\documentclass[11pt,a4paper]{article}
\pdfoutput=1

\usepackage{jcappub}
\usepackage{graphicx}
\usepackage{epsfig}
\usepackage{subfig}
\usepackage{comment}
\usepackage{slashed}
\usepackage{soul}
\usepackage{tabularx}
\usepackage{physics}
\usepackage[normalem]{ulem}
\usepackage{afterpage}
\usepackage{placeins}
\usepackage{booktabs}
\usepackage{float}
\usepackage{tensor}
\usepackage{xcolor}
\usepackage{lipsum}
\usepackage{dsfont}

\usepackage{cancel}

\usepackage{array}
\newcolumntype{C}[1]{>{\centering\let\newline\\\arraybackslash\hspace{0pt}}m{#1}}

\newcommand{\be}{\begin{equation}} 
\newcommand{\ee}{\end{equation}}
\newcommand{\bea}{\begin{equation}\begin{aligned}} 
\newcommand{\eea}{\end{aligned}\end{equation}}
\newcommand{\ber}{\begin{eqnarray}}
\newcommand{\ear}{\end{eqnarray}}

\def\lsim{\mathrel{\raise.3ex\hbox{$<$\kern-.75em\lower1ex\hbox{$\sim$}}}}
\def\gsim{\mathrel{\raise.3ex\hbox{$>$\kern-.75em\lower1ex\hbox{$\sim$}}}}

\newcommand{\m}{m_{\rm{P}}}
\newcommand{\G}{8\pi G}

\definecolor{DigitColor}{rgb}{0.5,0.5,0.5}

\definecolor{Red}{rgb}{1,0,0}
\definecolor{Blue}{rgb}{0,0,1}
\definecolor{Green}{rgb}{0,1,0}

\allowdisplaybreaks
\usepackage{array}
\renewcommand{\arraystretch}{1.35}
\begin{document}

\title{\boldmath\centering Post-Newtonian Constraints on Scalar-Tensor Gravity}
\author[a]{Alexandros Karam,}
\author[b]{Samuel S\'anchez L\'opez,}
\author[c]{and Jos\'e Jaime Terente D\'{i}az}

\emailAdd{alexandros.karam@kbfi.ee}
\emailAdd{samuel.lopez@iap.fr}
\emailAdd{jterente@uc.pt}

\vspace{5cm}

\affiliation[a]{National Institute of Chemical Physics and Biophysics, Rävala 10, 10143 Tallinn, Estonia.}
\affiliation[b]{Institut d’Astrophysique de Paris, 98 bis boulevard Arago, F-75014 Paris, France.}
\affiliation[c]{Faculdade de Ci\^{e}ncias e Tecnologia and CFisUC, Departamento de F\'isica,\\ 
Universidade de Coimbra, Rua Larga, P-3004-516 Coimbra, Portugal.}

\abstract{Solar-System constraints on a general scalar-tensor theory with generic non-minimal coupling function, non-canonical kinetic function, and scalar potential, are investigated in both the metric and Palatini formalisms. A unified post-Newtonian treatment is developed, yielding analytical expressions for the effective scalar mass, the effective gravitational coupling, and the parametrised post-Newtonian parameters $\gamma$ and $\beta$. The results show explicitly how the choice of variational principle affects the weak-field phenomenology. Comparison with Solar-System observations, primarily the Cassini bound on $\gamma$, indicates that the observational impact of the formalism is strongly model dependent. Generic non-minimally coupled scalar fields may satisfy significantly weaker local bounds in the Palatini case because of stronger Yukawa suppression, whereas in Brans-Dicke gravity the differences are typically small and become appreciable only in restricted regions of parameter space. For the point-particle source considered here, Palatini $f(\hat{R})$ gravity reproduces the general-relativistic exterior post-Newtonian limit, unlike metric $f(R)$ gravity.}

\maketitle

\section{Introduction}
The late-time accelerated expansion of the Universe, first established through Type Ia supernovae observations \cite{SupernovaSearchTeam:1998fmf,SupernovaCosmologyProject:1998vns} and subsequently corroborated by a wide range of cosmological probes~\cite{Planck:2018vyg,DESI:2025zgx,Huterer:2017buf}, remains one of the central open problems in modern cosmology. Although the $\Lambda$CDM model provides an excellent phenomenological description of the currently available data, the cosmological constant problem and the lack of a satisfactory understanding of the origin of dark energy within fundamental physics continue to motivate the study of dynamical alternatives to vacuum energy \cite{Clifton:2011jh,Joyce:2014kja}. Amongst the most thoroughly studied possibilities are quintessence \cite{Wetterich:1987fm,Ratra:1987rm} and, more generally, scalar-tensor theories, in which cosmic acceleration is sourced by an evolving scalar degree of freedom that may also modify gravity on large scales \cite{Boisseau:2000pr,Esposito-Farese:2000pbo,Gannouji:2006jm, Jarv:2008eb,Rossi:2019lgt}. In recent years, non-minimally coupled dark-energy scenarios have received renewed attention \cite{Wang:2025znm,Wolf:2024stt, Ye:2024ywg,Ye:2024zpk,Ferrari:2025egk,Pan:2025psn,Tiwari:2024gzo, Wolf:2025jed,SanchezLopez:2025uzw}, both because of their rich theoretical structure and because they may help accommodate late-time cosmological observations, including recent DESI results \cite{DESI:2025zgx}, while remaining subject to nontrivial local constraints \cite{Rossi:2019lgt,Braglia:2020iik}.

Modified gravity theories with a non-minimally coupled scalar field provide a natural framework in which such scenarios may be investigated. In these theories, the scalar field couples directly to curvature through a function multiplying the Ricci scalar, with the result that both the effective gravitational coupling and the scalar-field dynamics are modified \cite{Brans:1961sx,Clifton:2011jh}. An important subtlety is that, in the presence of non-minimal couplings or non-linear curvature terms, the choice of the variational principle has non-trivial physical consequences. In the metric formalism, the connection is assumed from the outset to be the Levi-Civita connection of the metric, whereas in the Palatini formalism the metric and connection are varied independently, and the form of the connection is determined only at the level of the field equations \cite{Palatini:1919ffw,Ferraris:1982wci}. While the two approaches are equivalent for the Einstein-Hilbert action with minimally coupled matter, they generally lead, in more general scalar-tensor or $f(R)$ theories, to inequivalent field equations, scalar dynamics, cosmological perturbations, and observational predictions \cite{Fan:2015rha,Koivisto:2005yk,Koivisto:2005yc,Kubota:2020ehu,Jarv:2014hma,Kozak:2018vlp,Jarv:2020qqm,Jarv:2024krk,Lopez:2025gfu}. For this reason, the Palatini formulation has become an important arena for modified-gravity phenomenology, with applications ranging from inflation and cosmological perturbations to quintessence, late-time acceleration, and alternative geometric realisations of dark energy \cite{Dimopoulos:2020pas,Verner:2020gfa,Dimopoulos:2021xld,Dimopoulos:2022tvn,Dimopoulos:2022rdp,Antoniadis:2022cqh,Dimopoulos:2025fuq,TerenteDiaz:2023kgc,SanchezLopez:2023ixx}.

Any viable scalar-tensor account of cosmic acceleration must nevertheless remain compatible with the exceptionally stringent weak-field constraints derived from Solar-System experiments. The appropriate framework for confronting such theories with local observations is provided by the parametrised post-Newtonian (PPN) formalism, in which deviations from General Relativity are encoded in a small set of parameters, most notably $\gamma$ and $\beta$, which characterise, respectively, the amount of spatial curvature produced per unit mass and the degree of source-independent non-linear gravitational self-interaction \cite{Will:2014kxa,Will:2018bme,Nordtvedt:1970uv}. The post-Newtonian regime of scalar-tensor gravity has been investigated in considerable generality, including general scalar-tensor theories, homogeneous gravitating spheres, screened theories, Horndeski theories, teleparallel analogues, and both metric and Palatini versions of $f(R)$ gravity \cite{Jarv:2014laa,Hohmann:2013rba,Hohmann:2017qje,Zhang:2016njn,Emtsova:2019qsl,Hohmann:2015kra,Olmo:2005hc,Olmo:2005hd,Harko:2011nh,Lu:2020eux,Lu:2019qgk,Dyadina:2019yon,Toniato:2021vmt}. Related analyses have also been carried out for generalised Brans-Dicke models in the metric and Palatini formalisms, as well as for hybrid metric-Palatini gravity and for exact post-Newtonian parameters of strongly gravitating bodies \cite{Lu:2019qgk,Lu:2020eux,Dyadina:2019yon,Chehab:2026vdx}. From the observational side, the most important local probes include Lunar Laser Ranging, Cassini time-delay measurements, Mercury perihelion precession, Earth-orbit tests, and modern planetary ephemerides \cite{Williams:2004qba,Bertotti:2003rm,Verma:2013ata,Genova:2018mjp,Fienga:2019uds,Fienga:2022bns,Scharer:2014kya,Zhang:2023nil}. These bounds constrain not only the PPN parameters themselves, but also the existence of Yukawa-type fifth forces, the effective gravitational coupling relevant for Cavendish-type measurements, and possible temporal variation of Newton's constant\footnote{Related constraints on scalar couplings to matter have also been obtained in strong-gravity orbital systems and through geodetic/frame-dragging observables (see \textit{e.g.} Refs.~\cite{Benisty:2021cmq,Benisty:2023dkn}).} \cite{Martin:2005bp, Adam:2025kve,Adelberger:2003zx, Barrow:1996kc,Esposito-Farese:2000pbo,Esposito-Farese:2004azw,Faraoni:2004qd,Nguyen:2024fpw}. While this weak-field phenomenology has been explored extensively in the metric formalism, its Palatini counterpart remains comparatively less developed, especially in the context of non-minimally coupled quintessence and late-time cosmology.

The present work is concerned with Solar-System constraints on scalar-tensor theories of gravity with non-minimal couplings, with particular emphasis placed on the differences between the metric and Palatini formalisms. We begin with a general Jordan-frame action containing an arbitrary non-minimal coupling function, kinetic function, and scalar potential. We then derive the field equations and construct the post-Newtonian expansion in a unified way for both formulations of gravity, following and extending the general scalar-tensor PPN analyses of Refs.~\cite{Jarv:2014laa,Hohmann:2013rba,Hohmann:2017qje}. Analytical expressions are obtained for the effective scalar mass, the effective gravitational coupling, and the PPN parameters $\gamma$ and $\beta$, thereby highlighting how the dependence on the independent connection modifies the weak-field phenomenology in the Palatini case relative to the metric one. On the basis of these results, the theory is confronted with Solar-System constraints, in particular the Cassini bound on $\gamma$~\cite{Bertotti:2003rm} and the MESSENGER and related constraints on $\beta$~\cite{Genova:2018mjp,doi:10.1126/science.1218809}, and the corresponding allowed regions of parameter space are determined. As illustrative examples, non-minimally coupled quintessence, Brans-Dicke gravity, and $f(R)$ gravity are considered. The extent to which current local tests can distinguish between their metric and Palatini realisations is assessed. 

The paper is organised as follows. In Sec.~\ref{sec:theory} we present the general scalar-tensor framework under consideration and derive the field equations in both the metric and Palatini formalisms. In Sec.~\ref{sec:ppn}, the parametrised post-Newtonian expansion is introduced and the weak-field equations relevant for Solar-System tests are established. In Sec.~\ref{subsubsec:ppnparams}, we obtain analytical expressions for the PPN parameters $\gamma$ and $\beta$, along with the effective scalar mass and the effective gravitational coupling, making explicit the differences between the two variational formulations. In Sec.~\ref{sec:constraints}, these results are confronted with Solar-System observations and the corresponding constraints on the parameter space are analysed. In particular, Sec.~\ref{sec:observables} is devoted to the relevant observables, namely the Shapiro time delay, the light-deflection angle, and Mercury's perihelion advance, whereas in Sec.~\ref{sec:apps} we employ our unified framework to examine non-minimally coupled quintessence, Brans-Dicke theory, and $f(R)$ gravity. Sec.~\ref{sec:conclusions} contains a summary of the main results together with a discussion of their implications. For completeness, App.~\ref{app:solution-connection} features the solution to the connection field equations, App.~\ref{sec:appendix-postnewtonian-details} presents the details of the post-Newtonian calculations, and App.~\ref{sec:appendix-geodesics} provides the derivation of the photon trajectory used in the discussion of light-propagation observables. In addition, App.~\ref{App:fifthforces} derives the perihelion-precession rate arising from the Yukawa contribution at Newtonian order, while App.~\ref{sec:appendix-f(R)} demonstrates the equivalence of metric and Palatini $f(R)$ gravity with metric Brans-Dicke gravity, alongside the derivation of the parametrised post-Newtonian parameters in Palatini $f(\hat{R})$ gravity.

Throughout this work, $G$ denotes Newton's gravitational constant, which has dimensions of inverse mass squared, and $\m=(8\pi G)^{-1/2}=2.44\times 10^{18}$GeV is the reduced Planck mass. The metric signature is taken to be the mostly positive one, namely $(-+++)$. Moreover, whenever `$x$' appears as the argument of a function, it is to be understood as referring to the spacetime coordinates $(t,\mathbf{x})$.

\section{\label{sec:theory}Scalar-Tensor Gravity: Metric and Palatini Formalisms}

\subsection{Action and Field Equations}
The action functional is taken to be\footnote{\label{first-footnote-in-your-face}As noted below Eq.~\eqref{eq:Newton-effective}, in scalar-tensor theories the gravitational coupling $G/A(\Phi)$, evaluated at the present cosmological epoch, does not coincide with the Newton constant inferred from Cavendish-type measurements, denoted by $G_{\textrm{eff}}$. Such experiments probe the local gravitational attraction between two test masses.}
\begin{equation}
    \label{eq:full_action}S = \frac{1}{2}\int \textrm{d}^4x \sqrt{-g}\left[\frac{A(\Phi)}{\G}\hat{R}-B(\Phi) g^{\mu\nu}\hat{\nabla}_{\mu}\Phi \hat{\nabla}_{\nu}\Phi -2V(\Phi)\right]+S_{\textrm{m}}\left[g_{\mu\nu},\Psi\right],
\end{equation}
where the Ricci scalar $\hat{R}$ depends on the inverse metric $g^{\mu\nu}$, the connection $\hat{\Gamma}^{\alpha}_{\mu\nu}$, and the first spacetime derivative of the latter. In the Palatini formulation of gravity, $\hat{\Gamma}^{\alpha}_{\mu\nu}$ is treated as an independent geometrical field and is not assumed a priori to coincide with the Levi-Civita (LC) connection $\Gamma^{\alpha}_{\mu\nu}$ of the metric $g_{\mu\nu}$, given by
\begin{equation}
    \Gamma^{\alpha}_{\mu\nu}\equiv \frac{1}{2}g^{\alpha\beta}\left(\partial_{\mu}g_{\beta\nu}+\partial_{\nu}g_{\beta\mu}-\partial_{\beta} g_{\mu\nu}\right),
\end{equation}
which is torsion-free, $\Gamma^{\alpha}_{\mu\nu} = \Gamma^{\alpha}_{\nu\mu}$, and metric-compatible, $\partial_{\alpha}g_{\mu\nu} = \Gamma^{\lambda}_{\alpha\mu} g_{\lambda\nu}+\Gamma^{\lambda}_{\alpha\nu} g_{\mu\lambda}$. This freedom reflects the fact that, in a generally covariant theory, the metric and the affine connection represent a priori distinct geometric structures: the metric encodes lengths and causal relations, while the connection determines parallel transport and covariant differentiation. Therefore, in this work, hatted geometric quantities, such as $\hat{R}$, are constructed from $\hat{\Gamma}^{\alpha}_{\mu\nu}$, whereas unhatted ones are defined with respect to the LC connection of $g_{\mu\nu}$. Only after varying the action does one determine the relation between $\hat{\Gamma}^{\alpha}_{\mu\nu}$ and $\Gamma^{\alpha}_{\mu\nu}$. Nevertheless, a general decomposition can always be introduced in the form
\begin{equation}
    \label{eq:arb-connection}\hat{\Gamma}^{\alpha}_{\mu\nu} = \Gamma^{\alpha}_{\mu\nu} +\tensor{\kappa}{^{\alpha}_{\mu\nu}}~,
\end{equation}
where $\tensor{\kappa}{^{\alpha}_{\mu\nu}}$ is the so-called distortion tensor\footnote{While $\hat{\Gamma}^{\alpha}_{\mu\nu}$ and $\Gamma^{\alpha}_{\mu\nu}$ do not transform as tensors under diffeomorphisms, their difference does define a tensor under such transformations \cite{Carroll:1997ar}.} \cite{BeltranJimenez:2016wxw,Shimada:2018lnm,Iosifidis:2023eom}. 

In addition, the action in Eq.~\eqref{eq:full_action} is assumed to depend on a dynamical scalar field $\Phi(t,\mathbf{x})$, non-minimally coupled to the Ricci scalar through the dimensionless coupling function $A(\Phi)>0$, and endowed with non-trivial kinetic and potential functions, $B(\Phi)$ and $V(\Phi)$, respectively. $\hat{\nabla}_{\mu}$ denotes the covariant derivative associated with the arbitrary connection $\hat{\Gamma}^{\alpha}_{\mu\nu}$. Since this derivative acts on the scalar field $\Phi$, one has $\hat{\nabla}_{\mu}\Phi = \nabla_{\mu}\Phi = \partial_{\mu}\Phi$, and therefore $g^{\mu\nu}\hat{\nabla}_{\mu}\Phi \hat{\nabla}_{\nu}\Phi = (\partial \Phi)^2$. Finally, $\Psi$ collectively denotes all matter fields with action $S_{\textrm{m}}$, which are taken to be minimally coupled to gravity. In particular, the connection $\hat{\Gamma}^{\alpha}_{\mu\nu}$ does not couple directly to the matter sector, so that the theory lies within the minimal class of metric-affine models \cite{Vitagliano:2010sr,Iosifidis:2021pta,JimenezCano:2021rlu}.

The field equations follow from requiring the full action to be stationary under independent variations with respect to the inverse metric, the connection, the scalar field, and the matter fields:
\begin{eqnarray}
    \label{eq:metric-field-eqs-Palatini} \hat{G}_{(\mu\nu)} &=& \frac{\G}{A}\left(T_{\mu\nu}+T^{\Phi}_{\mu\nu}\right),\\
    \label{eq:connection-field-eqs}\tensor{\kappa}{^{\alpha}_{\mu\nu}}+\tensor{\kappa}{_{\mu\nu}^{\alpha}}-\tensor{\kappa}{_{\mu\sigma}^{\sigma}}\delta_{\nu}^{\alpha}-\tensor{\kappa}{^{\sigma}_{\sigma\nu}}\delta_{\mu}^{\alpha} &=& \frac{A^{'}}{A}\left(\delta_{\mu}^{\sigma}\delta_{\nu}^{\alpha}-\delta_{\mu}^{\alpha}\delta_{\nu}^{\sigma}\right) \partial_{\sigma}\Phi~,\\
    \label{eq:sf-field-eqs-Palatini}B\Box \Phi +\frac{1}{2}B^{'}(\partial \Phi)^2-V^{'}+\frac{A^{'}}{16\pi G}\hat{R}&=&0~,\\
    \frac{\delta S_{\textrm{m}}[g_{\mu\nu},\Psi]}{\delta \Psi} &=& 0~,
\end{eqnarray}
where $T_{\mu\nu}$ is the energy-momentum tensor of the matter fields, defined by
\begin{equation}
    T_{\mu\nu}\equiv -\frac{2}{\sqrt{-g}}\frac{\delta S_{\textrm{m}}[g_{\mu\nu},\Psi]}{\delta g^{\mu\nu}}~,
\end{equation}
while $T^{\Phi}_{\mu\nu}$ denotes the canonical counterpart associated with the scalar field $\Phi$,
\begin{equation}
    T^{\Phi}_{\mu\nu}\equiv B\partial_{\mu}\Phi \partial_{\nu}\Phi -g_{\mu\nu}\left[\frac{1}{2}B(\partial \Phi)^2+V\right].
\end{equation}
$\hat{G}_{(\mu\nu)}\equiv \hat{R}_{(\mu\nu)}-\frac{1}{2}g_{\mu\nu}\hat{R}$ is the symmetrised Einstein tensor, and $\hat{R}_{(\mu\nu)}$ the symmetrised Ricci tensor. The latter, whether symmetric or not, is a function only of the independent connection and its first spacetime derivative, namely
\begin{equation}
    \label{eq:def-Riccitensor}\hat{R}_{\mu\nu} \equiv \partial_{\alpha}\hat{\Gamma}^{\alpha}_{\nu\mu}-\partial_{\nu}\hat{\Gamma}^{\alpha}_{\alpha\mu}+\hat{\Gamma}^{\alpha}_{\alpha\lambda}\hat{\Gamma}^{\lambda}_{\nu\mu}-\hat{\Gamma}^{\alpha}_{\nu\lambda}\hat{\Gamma}^{\lambda}_{\alpha\mu}~.
\end{equation}
$\Box\equiv g^{\mu\nu}\nabla_{\mu}\nabla_{\nu}$ is the d'Alembertian, and primes denote derivatives with respect to $\Phi$. The non-Riemannian contributions associated with the arbitrary connection enter the scalar-field equation through the Ricci scalar, and thus via the non-minimal coupling. The connection equations \eqref{eq:connection-field-eqs}, written in terms of the distortion tensor, are algebraic and can be readily solved, yielding (see App.~\ref{app:solution-connection} for details)
\begin{align}
    \label{eq:metric-connectiononshell}&A G_{\mu\nu} = \G\left(T_{\mu\nu}+T^{\Phi}_{\mu\nu}\right) +\nabla_{\mu}\partial_{\nu}A -g_{\mu\nu}\Box A -\frac{3}{2A}\left(\partial_{\mu}A\partial_{\nu}A-\frac{1}{2}g_{\mu\nu}\partial_{\sigma}A\partial^{\sigma}A\right), \\
    \label{eq:sf-connectiononshell}&B\Box \Phi +\frac{1}{2}B^{'}(\partial\Phi)^2 -V^{'} + \frac{A^{'}}{16\pi G}R = \frac{3A^{'}}{16\pi GA}\left(\Box A -\frac{1}{2A}\partial_{\sigma}A\partial^{\sigma}A\right).
\end{align}

Had we derived the metric and scalar field equations by imposing the LC connection before performing the independent variations with respect to the inverse metric and the scalar field, we would instead have obtained \cite{Clifton:2011jh,Faraoni:2004pi,Koivisto:2005yc}
\begin{align}
    &A G_{\mu\nu} =\G\left(T_{\mu\nu}+T^{\Phi}_{\mu\nu}\right) +\nabla_{\mu}\partial_{\nu}A-g_{\mu\nu}\Box A~,\\
    &B\Box \Phi +\frac{1}{2}B^{'}(\partial \Phi)^2-V^{'}+\frac{A^{'}}{16\pi G}R = 0~.
\end{align}
One may then group together the equations arising from the Palatini formalism, in which the metric and connection are varied independently before deriving the field equations, and those of the so-called `metric' formalism, where the LC connection is assumed at the level of the action: 
\begin{align}
    \label{eq:final_metric-eqs}&A G_{\mu\nu} = \G\left(T_{\mu\nu}+T^{\Phi}_{\mu\nu}\right) +\nabla_{\mu}\partial_{\nu}A-g_{\mu\nu}\Box A-\delta_{\textrm{P}}\frac{3}{2A}\left(\partial_{\mu}A\partial_{\nu}A-\frac{1}{2}g_{\mu\nu}\partial_{\sigma}A\partial^{\sigma}A\right),\\
    \label{eq:final_sf-eqs}&B\Box \Phi +\frac{1}{2}B^{'}(\partial \Phi)^2-V^{'}+\frac{A^{'}}{16\pi G}R = \delta_{\textrm{P}}\frac{3A^{'}}{16\pi GA}\left(\Box A -\frac{1}{2A}\partial_{\sigma}A\partial^{\sigma}A\right),
\end{align}
such that $\delta_{\textrm{P}} = 1$ for the Palatini formulation and $\delta_{\textrm{P}} = 0$ for the metric case. By virtue of the contracted Bianchi identity, $\nabla_{\nu}\tensor{G}{_{\mu}^{\nu}}=0$, together with the properties of the LC connection and the field equations above, one can show that the energy-momentum tensor of the matter sector is covariantly conserved with respect to the LC connection in both formalisms \cite{Koivisto:2005yk}:
\begin{equation}
    \nabla_{\nu}\tensor{T}{_{\mu}^{\nu}} = 0~.
\end{equation} 

Since this form is more convenient for the analysis of first-order post-Newtonian effects in the next section, we take the trace of the metric field equations \eqref{eq:final_metric-eqs} and express the Ricci scalar in terms of the traces of the energy-momentum tensors $T_{\mu\nu}$ and $T_{\mu\nu}^{\Phi}$. The same is done for the scalar-field equation \eqref{eq:final_sf-eqs}. The final field equations read
\begin{align}
    \label{eq:def-fieldeqs-metric}&AR_{\mu\nu} = 8\pi G\left[T_{\mu\nu}+T^{\Phi}_{\mu\nu}-\frac{1}{2}g_{\mu\nu}\left(T+T^{\Phi}\right)\right]+\nabla_{\mu}\partial_{\nu}A+\frac{1}{2}g_{\mu\nu}\Box A -\delta_{\textrm{P}}\frac{3}{2A}\partial_{\mu}A\partial_{\nu}A~,\\
    \label{eq:def-fieldeqs-scalar}&B\Box \Phi +\frac{1}{2}B^{'}(\partial \Phi)^2-V^{'}-\frac{A^{'}}{2A}\left(T+T^{\Phi}\right) = -\left(1-\delta_{\textrm{P}}\right)\frac{3A^{'}}{16\pi G A}\Box A~,
\end{align}
where the trace $T^{\Phi}$ is
\begin{equation}
    T^{\Phi} = -B(\partial \Phi)^2-4V~.
\end{equation}
It can be seen that, in the Palatini formulation, the non-minimal coupling function $A$ enters the scalar-field equation only via the traces of the energy-momentum tensors. This is expected because a conformal rescaling that brings the action into Einstein-frame form (see footnote~\ref{footnoteConfandProj}) does not modify the Ricci tensor $\hat{R}_{\mu\nu}$, which depends solely on the connection, as already noted. Moreover, the Palatini formalism does not give rise to additional second-order derivatives of the scalar field $\Phi$ in the metric field equations.

\section{Post-Newtonian Parameters}
\subsection{\label{sec:ppn}Parametrised Post-Newtonian Formalism}
We focus on the first-order post-Newtonian (PN) approximation, denoted `1PN'; further details are provided in App.~\ref{sec:appendix-postnewtonian-details}. In this regime, matter moves slowly compared to the speed of light in vacuum, which is set to unity in this work, and the gravitational field is weak, so that the spacetime geometry remains close to the Minkowski metric $\eta_{\mu\nu}=\textrm{diag}(-1,1,1,1)$ \cite{Weinberg:1972kfs,Will:2018bme}. The metric tensor $g_{\mu\nu}$ is then expanded about $\eta_{\mu\nu}$ as  
\begin{equation}
    \label{eq:expansion-metric-Minkowski}g_{\mu\nu}(x) \simeq \eta_{\mu\nu} + h_{\mu\nu}(x)~,
\end{equation}
where the correction $h_{\mu\nu}$ is expressed in powers of the slow-motion expansion parameter, namely the magnitude of the fluid element's three-velocity, $v\equiv |\mathbf{v}|$ (in units where $c=1$):
\begin{equation}
    h_{\mu\nu} = h^{(1)}_{\mu\nu}+h_{\mu\nu}^{(2)} +...+\mathcal{O}(n)~.
\end{equation}
Here, $h^{(n)}_{\mu\nu}$ is of order $\mathcal{O}(n)$, \textit{i.e.}, it is the $n$th-order term in the velocity expansion, scaling as $v^n$.\footnote{\label{footnote:deltaeta}Notice that $v^2=\delta_{ij}v^{i}v^{j}$, where $\delta_{ij}$ is the Kronecker delta; for spatial Cartesian indices, $\eta_{ij}=\delta_{ij}$.} The non-minimally coupled scalar degree of freedom $\Phi$ is likewise expanded about a fixed background value $\phi$ as \cite{Nordtvedt:1970uv,Hohmann:2013rba}
\begin{equation}
    \label{eq:expansion-FieldPhi-asymptoticBackground}\Phi(x) \simeq \phi + \varphi(x)~,
\end{equation}
where the correction $\varphi$ is ordered according to the aforementioned expansion parameter:
\begin{equation}
    \varphi = \varphi^{(1)}+\varphi^{(2)} +...+\mathcal{O}(n)~.
\end{equation}
The background scalar $\phi$ is of order $\mathcal{O}(0)$, as is the Minkowski metric in Eq.~\eqref{eq:expansion-metric-Minkowski}. Furthermore, the functions $A(\Phi)$, $B(\Phi)$, and $V(\Phi)$ are Taylor-expanded about $\phi$:
\begin{eqnarray}
    \label{eq:NMC-expansion}A(\Phi) &\simeq& A_0 + A_1\varphi+\frac{1}{2}A_2\varphi^2+...~,\\
    \label{eq:kinetic-expansion}B(\Phi) &\simeq& B_0 + B_1\varphi+\frac{1}{2}B_2\varphi^2+...~,\\
    \label{eq:VPot-expansion}V(\Phi) &\simeq& V_0 + V_1\varphi+\frac{1}{2}V_2\varphi^2+...~.
\end{eqnarray}
The Taylor coefficients are defined by the corresponding functions and their field derivatives evaluated at the fixed background value $\phi$. In particular, 
\begin{equation}
    A_0 \equiv A(\phi), \ \ \ A_1\equiv \left.\frac{\partial A}{\partial\Phi}\right|_{\Phi=\phi}, \ \ \  A_2\equiv \left.\frac{\partial^2 A}{\partial\Phi^2}\right|_{\Phi=\phi},~...~,
\end{equation} 
and similarly for $B(\Phi)$ and $V(\Phi$). These coefficients are therefore of order $\mathcal{O}(0)$.

A consistent Newtonian limit, denoted `0PN', is time-reversal invariant, in the sense that Newtonian gravity is a conservative force field. It further assumes a quasi-static gravitational field, such that any time dependence arises solely through the motion and evolution of the matter sources. Thus, terms of order $\mathcal{O}(1)$ in the expansions above may be neglected at lowest order. As a result, the 0PN limit is characterised by\footnote{See Refs.~\cite{Toniato:2017wmk,Toniato:2021vmt}, where the authors additionally require that the gravitational constant appearing in the Poisson equation, Eq.~\eqref{eq:O2-h00}, be exclusively the `bare' one, namely $G$, rather than an effective one. Under this more restrictive condition, the scalar-field expansion is truncated one order higher.}  
\begin{equation}
    h_{00} = h_{00}^{(2)} + \mathcal{O}(3)~, \ \ \ \varphi= \varphi^{(2)}+\mathcal{O}(3)~,
\end{equation}
while the remaining metric corrections do not contribute in this limit. Retaining the conservative, quasi-static PN counting at the next order, the first-order post-Newtonian effects are described by\footnote{\label{avertecomento}In the 1PN limit, the spatial metric is required only through $\mathcal{O}(2)$. This can be seen by expanding the action of a test particle moving along a trajectory $\gamma^{\mu}$ parametrised by the coordinate time $t$. Because the 1PN limit corresponds to the fourth order in velocity, the term $h_{ij}^{(4)}$ is excluded: one has $h_{ij}\dot{\gamma}^{i}\dot{\gamma}^{j}$, with $|\dot{\gamma}^{i}|\sim\mathcal{O}(1)$, so an $\mathcal{O}(4)$ contribution in $h_{ij}$ would enter only beyond 1PN. See Chap.~24 of Ref.~\cite{CANTATA:2021asi}.}
\begin{equation}
    \label{eq:terms-retained-each-order}h_{00} = h^{(2)}_{00} +h_{00}^{(4)}+\mathcal{O}(5)~, \ \ \ h_{0i} = h^{(3)}_{0i} + \mathcal{O}(5)~, \ \ \ h_{ij} = h^{(2)}_{ij} +\mathcal{O}(4)~,
\end{equation}
where $h_{0i} = h_{i0}$ by symmetry, and
\begin{equation}
    \varphi = \varphi^{(2)}+\varphi^{(4)}+\mathcal{O}(5)~.
\end{equation}
Accordingly, Eqs.~\eqref{eq:NMC-expansion}--\eqref{eq:VPot-expansion} become
\begin{eqnarray}
    \label{eq:NMC-expansion1}A(\Phi) &=& A_0 + A_1\left(\varphi^{(2)}+\varphi^{(4)}\right)+\frac{1}{2}A_2(\varphi^{(2)})^2+\mathcal{O}(5)~,\\
    \label{eq:kinetic-expansion1}B(\Phi) &=& B_0 + B_1\left(\varphi^{(2)}+\varphi^{(4)}\right)+\frac{1}{2}B_2(\varphi^{(2)})^2+\mathcal{O}(5)~,\\
    \label{eq:VPot-expansion1}V(\Phi) &=& V_0 + V_1\left(\varphi^{(2)}+\varphi^{(4)}\right)+\frac{1}{2}V_2(\varphi^{(2)})^2+\mathcal{O}(5)~.
\end{eqnarray}

The energy-momentum tensor of the matter sources is assumed to take on the perfect-fluid form, consistent with local isotropy in the fluid rest frame and negligible dissipative effects:
\begin{equation}
    \label{eq:EMT-perfect-fluid-first-appearance}T^{\mu\nu} = \left[\rho\left(1+\varepsilon\right) + p\right] u^{\mu}u^{\nu}+ p g^{\mu\nu}~, 
\end{equation}
where $\varepsilon$ denotes the specific internal energy per unit rest mass, measured in the local rest frame of the fluid, and $\rho$ is the corresponding rest-mass density.\footnote{This $\rho$ is the Newtonian source entering the Poisson equation, namely the leading-order contribution to the energy density in the PN expansion.} The momentum flux in that frame is given by the isotropic pressure $p$. The four-velocity of the fluid has components $u^{\mu} = u^{0}(1,v^{i})$, such that $v^{i}=u^{i}/u^{0}$. Consistency of the matter dynamics in that limit requires $\rho$ and $\varepsilon$ to be of order $\mathcal{O}(2)$, and therefore $\rho \varepsilon\sim \mathcal{O}(4)$. Additionally, for a non-relativistic fluid, $p$ is of order $\mathcal{O}(4)$, so that one has the scaling relation $p\sim \rho v^2$ \cite{Will:2014kxa} (see footnote~\ref{explain-PN-counting-pressure}).

As can be seen, the 1PN limit is obtained by retaining terms up to $\mathcal{O}(4)$ in the relevant expansions. It should be noted that no further assumptions about time-reversal symmetry or quasi-staticity at higher PN orders are required for the present analysis, since the associated effects do not enter at first PN order. At order $\mathcal{O}(0)$, the static limit, the metric and scalar field equations, Eqs.~\eqref{eq:def-fieldeqs-metric} and \eqref{eq:def-fieldeqs-scalar} respectively, read
\begin{eqnarray}
    \label{eq:first-trivial-eq}0 &=& 8\pi G A_0 V_0 \eta_{\mu\nu}~,\\
    \label{eq:more-trivial-if-possible}A_0V_1-2A_1 V_0 &=& 0~.
\end{eqnarray}
Since $A_0\neq 0$, Eq.~\eqref{eq:first-trivial-eq} implies $V_0=0$; otherwise, the Poisson equation would not be recovered, and hence neither would the Newtonian limit (see Eq.~\eqref{eq:O2-h00}). Eq.~\eqref{eq:more-trivial-if-possible} then yields $V_1=0$. As a result, the potential function reduces to
\begin{equation}
    \label{eq:potential-expansion}V(\Phi) = \frac{1}{2}V_2(\varphi^{(2)})^2+\mathcal{O}(5)~.
\end{equation} 
As indicated in footnote~\ref{footnote:negligible-order-O5}, the symbol $\mathcal{O}(5)$, or more generally the first neglected order, will be omitted for simplicity.

\subsubsection{0PN and 1PN Field Equations}
At order $\mathcal{O}(2)$, corresponding to the Newtonian limit, the metric field equations are\footnote{As explained in App.~\ref{sec:appendix-postnewtonian-details}, time derivatives, denoted by overdots, count as one additional order by virtue of the assumption of quasi-staticity. The same appendix also provides a detailed derivation of the metric and energy-momentum tensor components, together with other geometric and matter quantities.}
\begin{align}
    \label{eq:O2-h00}&\partial_i \partial^{i}h_{00}^{(2)} = -\frac{1}{A_0}\left(8\pi G \rho -A_1\partial_i \partial^{i}\varphi^{(2)}\right),\\
    \nonumber&\partial^{k}\left(\partial_i h_{kj}^{(2)}+\partial_j h_{ki}^{(2)}-\partial_k h_{ij}^{(2)}\right)+\partial_i \partial_j\left( h_{00}^{(2)}-\delta^{ks}h_{ks}^{(2)}-2\frac{A_1}{A_0}\varphi^{(2)}\right) =\\
    \label{eq:O2-metric-ij}&\phantom{----------------------}=\frac{1}{A_0}\left(8\pi G \rho +A_1\partial_k \partial^{k}\varphi^{(2)}\right)\delta_{ij}~,
\end{align}
while the scalar-field equation reads
\begin{equation}
    \label{eq:sf-eq-O2}A_0\left(B_0 \partial_i \partial^{i}\varphi^{(2)}-V_2\varphi^{(2)}\right)+\frac{1}{2}A_1 \rho = -\left(1-\delta_{\textrm{P}}\right)\frac{3A_1^2}{16\pi G}\partial_i \partial^{i}\varphi^{(2)}~.
\end{equation}
From the gauge condition \eqref{eq:linearised-gauge-Donder-modified} at order $\mathcal{O}(2)$, one obtains
\begin{equation}
    \label{eq:constraint-gauge-O2}\partial^{j}h_{ij}^{(2)}+\frac{1}{2}\partial_i \left(h_{00}^{(2)}-\delta^{ks}h_{ks}^{(2)}\right) = \frac{A_1}{A_0}\partial_i \varphi^{(2)}~,
\end{equation}
and inserting this into Eq.~\eqref{eq:O2-metric-ij} gives 
\begin{equation}
    \partial^{k}\left(\partial_i h_{kj}^{(2)}-\partial_j h_{ki}^{(2)}+\partial_k h_{ij}^{(2)}\right)=-\frac{1}{A_0}\left(8\pi G \rho +A_1\partial_k \partial^{k}\varphi^{(2)}\right)\delta_{ij}~.
\end{equation}
Separating the terms into their symmetric and antisymmetric parts under the exchange of indices $i$ and $j$, we find $\partial^{k}\left(\partial_i h_{kj}^{(2)}-\partial_j h_{ki}^{(2)}\right)=0$, and
\begin{equation}
    \label{eq:theotherequationatO2}\partial_k \partial^{k} h_{ij}^{(2)} = -\frac{1}{A_0}\left(8\pi G \rho + A_1\partial_k \partial^{k}\varphi^{(2)}\right)\delta_{ij}~.
\end{equation}
At that order, the difference between the metric and Palatini formalisms is encoded exclusively in the scalar-field equation. At order $\mathcal{O}(3)$, the metric field equations reduce to\footnote{As made explicit in Eq.~\eqref{eq:terms-retained-each-order}, the term $h^{(2)}_{0i}$ is excluded by PN counting, since $h_{0i}$ is sourced by the mass current $\rho v_i$, rather than by $\rho$. It is therefore one velocity order higher than $h_{00}$, and starts at order $\mathcal{O}(3)$.}
\begin{equation}
    \label{eq:metric-O3}A_0\left(\partial^{j}\dot{h}_{ij}^{(2)}-\partial_j \partial^{j}h_{0i}^{(3)} +\partial_i \partial^{j}h_{0j}^{(3)}-\delta^{jk}\partial_i \dot{h}^{(2)}_{jk}\right) = -16\pi G \rho v_i +2A_1\partial_i \dot{\varphi}^{(2)}~, 
\end{equation}
and we again find no dependence on $\delta_{\textrm{P}}$, meaning that the above equation is the same in both formalisms. From the gauge condition \eqref{eq:linearised-gauge-Donder-modified} we extract the following constraint at $\mathcal{O}(3)$: 
\begin{equation}
    \label{eq:constraint-gauge-O3}\frac{1}{2}\left(\dot{h}^{(2)}_{00}+\delta^{ij}\dot{h}^{(2)}_{ij}\right)-\partial^{i}h_{0i}^{(3)} = -\frac{A_1}{A_0}\dot{\varphi}^{(2)}~,
\end{equation}
which, together with Eq.~\eqref{eq:constraint-gauge-O2}, simplifies Eq.~\eqref{eq:metric-O3} to
\begin{equation}
    \partial_j \partial^{j}h_{0i}^{(3)} = \frac{16\pi G}{A_0} \rho v_i~.
\end{equation}

At order $\mathcal{O}(4)$, only the $00$ component of the metric field equations is required for the purposes of the present work (see footnote~\ref{avertecomento}):
\begin{align}
    \nonumber&A_0\left(\partial_i\partial^{i} h_{00}^{(4)}-2\partial^{i}\dot{h}_{0i}^{(3)}+\delta^{ij}\ddot{h}_{ij}^{(2)}-\partial_i h^{(2)ij}\partial_j h_{00}^{(2)}-h^{(2)ij}\partial_i \partial_j h_{00}^{(2)}+\frac{1}{2}\partial_i h_{00}^{(2)}\partial^{i} h_{00}^{(2)}+\right.\\
    \nonumber&\left.+\frac{1}{2}\delta^{jk}\partial_i h_{jk}^{(2)}\partial^{i}h_{00}^{(2)}\right)=-8\pi G \left[\rho\left(\varepsilon+5v^2-h_{00}^{(2)}+\frac{A_1}{A_0}\varphi^{(2)}\right)-V_2(\varphi^{(2)})^2\right]-\\
    \nonumber&-A_1\left[3\ddot{\varphi}^{(2)}+2\varphi^{(2)}\partial_i \partial^{i}h_{00}^{(2)}+h_{00}^{(2)}\partial_i \partial^{i}\varphi^{(2)}+\partial^{j}\varphi^{(2)}\partial^{i}h_{ij}^{(2)}+h^{(2)ij}\partial_i \partial_j \varphi^{(2)}+\right.\\
    \nonumber&\left.+\frac{3}{2}\partial^{i}h_{00}^{(2)}\partial_i \varphi^{(2)}-\frac{1}{2}\partial^{k}\varphi^{(2)}\delta^{ij}\partial_k h_{ij}^{(2)}-\left(\frac{A_1}{A_0}+\frac{A_2}{A_1}\right)\varphi^{(2)}\partial_i \partial^{i} \varphi^{(2)}-\frac{A_2}{A_1}\partial_i \varphi^{(2)}\partial^{i}\varphi^{(2)}-\right.\\
    &\phantom{-------------------------------}\left.-\partial_i \partial^{i}\varphi^{(4)}\right],
\end{align}
where $\delta_{\textrm{P}}$ again plays no explicit role. Using Eqs.~\eqref{eq:O2-h00}, \eqref{eq:constraint-gauge-O2}, and \eqref{eq:constraint-gauge-O3}, the above equation can be rewritten as
\begin{align}
    \nonumber&A_0\left(\partial_i \partial^{i}h_{00}^{(4)}-\ddot{h}^{(2)}_{00}+h_{00}^{(2)}\partial_i \partial^{i}h_{00}^{(2)}-h^{(2)ij}\partial_i \partial_j h_{00}^{(2)}+\partial_i h_{00}^{(2)}\partial^{i}h_{00}^{(2)}\right)=-8\pi G \left[\rho\left(\varepsilon +5v^2\right)-\right.\\
    \nonumber&\left.-V_2(\varphi^{(2)})^2\right]+A_1\left[\partial_i \partial^{i}\varphi^{(4)}-\ddot{\varphi}^{(2)}+\frac{A_2}{A_1}\varphi^{(2)}\partial_i \partial^{i}\varphi^{(2)} -h^{(2)ij}\partial_i \partial_j \varphi^{(2)}+\left(\frac{A_2}{A_1}-\frac{A_1}{A_0}\right)\times\right.\\
    \label{eq:metric-field-eqs-O4-00-final}&\phantom{-----------------------}\left.\times\partial_i \varphi^{(2)}\partial^{i}\varphi^{(2)}-\varphi^{(2)}\partial_i \partial^{i}h_{00}^{(2)}\right].
\end{align}
Lastly, because the term $\partial_i \partial^{i}\varphi^{(4)}$ appears, the scalar field equation \eqref{eq:def-fieldeqs-scalar} must also be considered at $\mathcal{O}(4)$, yielding
\begin{align}
    \nonumber&A_0 \left[B_0\left(\partial_i \partial^{i}\varphi^{(4)}-\ddot{\varphi}^{(2)}-h^{(2)ij}\partial_i \partial_j \varphi^{(2)}-\frac{1}{2}\partial_{i}h_{00}^{(2)}\partial^i \varphi^{(2)}-\partial^{j}\varphi^{(2)}\partial^{i}h_{ij}^{(2)}+\frac{1}{2}\partial^{k}\varphi^{(2)}\delta^{ij}\partial_k h_{ij}^{(2)}\right)+\right.\\
    \nonumber&\left.+B_1\left(\varphi^{(2)}\partial_i \partial^{i}\varphi^{(2)}+\frac{1}{2}\partial_i \varphi^{(2)}\partial^{i}\varphi^{(2)}\right)-V_2\varphi^{(4)}-\frac{1}{2}V_3(\varphi^{(2)})^2\right]+A_1B_0\left(\varphi^{(2)}\partial_i \partial^{i}\varphi^{(2)}+\right.\\
    \nonumber&\left.+\frac{1}{2}\partial_i \varphi^{(2)}\partial^{i}\varphi^{(2)}\right)+\frac{1}{2}\rho\left[A_1\left(\varepsilon-3v^2\right)+A_2\varphi^{(2)}\right]=-\left(1-\delta_{\textrm{P}}\right)\frac{3A_1^2}{16\pi G}\left(\partial_i \partial^{i}\varphi^{(4)}-\ddot{\varphi}^{(2)}+\right.\\
    \nonumber&\left.+2\frac{A_2}{A_1}\varphi^{(2)}\partial_i \partial^{i}\varphi^{(2)}-h^{(2)ij}\partial_i \partial_j \varphi^{(2)}+\frac{A_2}{A_1}\partial_i \varphi^{(2)}\partial^{i}\varphi^{(2)}-\partial^{j}\varphi^{(2)}\partial^{i}h_{ij}^{(2)}-\frac{1}{2}\partial_i h_{00}^{(2)}\partial^{i}\varphi^{(2)}+\right.\\
    &\phantom{---------------------------}\left.+\frac{1}{2}\partial^{k}\varphi^{(2)}\delta^{ij}\partial_k h_{ij}^{(2)}\right).
\end{align}
We may employ Eq.~\eqref{eq:constraint-gauge-O2} to simplify the above equation to
\begin{align}
    \nonumber&A_0 \left[B_0\left(\partial_i \partial^{i}\varphi^{(4)}-\ddot{\varphi}^{(2)}-h^{(2)ij}\partial_i \partial_j \varphi^{(2)}\right)+B_1\left(\varphi^{(2)}\partial_i \partial^{i}\varphi^{(2)}+\frac{1}{2}\partial_i \varphi^{(2)}\partial^{i}\varphi^{(2)}\right)-\right.\\
    \nonumber&\left.-V_2\varphi^{(4)}-\frac{1}{2}V_3(\varphi^{(2)})^2\right]+A_1B_0\left(\varphi^{(2)}\partial_i \partial^{i}\varphi^{(2)}-\frac{1}{2}\partial_i \varphi^{(2)}\partial^{i}\varphi^{(2)}\right)+\frac{1}{2}\rho\left[A_1\left(\varepsilon-3v^2\right)+\right.\\
    \nonumber&\left.+A_2\varphi^{(2)}\right]=-\left(1-\delta_{\textrm{P}}\right)\frac{3A_1^2}{16\pi G}\left[\partial_i \partial^{i}\varphi^{(4)}-\ddot{\varphi}^{(2)}+2\frac{A_2}{A_1}\varphi^{(2)}\partial_i \partial^{i}\varphi^{(2)}-h^{(2)ij}\partial_i \partial_j \varphi^{(2)}+\right.\\
    \label{eq:sf-eq-O(4)-final}&\phantom{-----------------------}\left.+\left(\frac{A_2}{A_1}-\frac{A_1}{A_0}\right)\partial_i \varphi^{(2)}\partial^{i}\varphi^{(2)}\right].
\end{align}

\subsection{\label{subsubsec:ppnparams}PPN Parameters $\gamma$ and $\beta$ in Metric and Palatini Formalisms}
With those equations, we can determine the parameters $\gamma$ and $\beta$ in the Parametrised Post-Newtonian (PPN) formalism. These quantify the space curvature produced per unit mass, and the degree of source-independent, non-linear gravitational self-interaction, respectively. Different theories of gravity can then be compared within the metric formalism by noting that $\gamma_{\textrm{GR}} = \beta_{\textrm{GR}} = 1$ \cite{Will:2014kxa}. Here, we derive analytical formulas that depend on the chosen formulation of gravity. Since the PPN parameters $\gamma$ and $\beta$ are read off from the weak-field metric outside the gravitating body, we restrict our attention to the exterior region of a static, spherically symmetric, non-rotating source. For present purposes, we further approximate the source by a point-like mass $M$, while noting that in some theories the exterior solution can retain dependence on the interior structure through matching conditions or effective scalar charges \cite{Olmo:2005hc,Olmo:2005hd}. The dependence on the formalism is encoded in $\delta_{\textrm{P}}$, as remarked previously. The non-vanishing metric components take on the following form:
\begin{eqnarray}
    \label{eq:g00-isotropic-PPN}g_{00} &\simeq& -1+2U_{\textrm{eff}}(r)-2\beta(r) U^2_{\textrm{eff}}(r)~,\\
    \label{eq:gij-isotropic-PPN}g_{ij} &\simeq& \left[1+2\gamma(r) U_{\textrm{eff}}(r)\right] \delta_{ij}~,
\end{eqnarray}
in Cartesian isotropic coordinates, where (see Eqs.~\eqref{eq:expansion-metric-Minkowski} and \eqref{eq:terms-retained-each-order}) 
\begin{equation}
    \label{eq:specifying-even-more}h_{00}^{(2)}(r) = 2U_{\textrm{eff}}(r)~, \ \ \ h_{ij}^{(2)}(r) = 2\gamma(r) U_{\textrm{eff}}(r) \delta_{ij}~, \ \ \ h_{00}^{(4)}(r) = -2\beta(r) U^2_{\textrm{eff}}(r)~.
\end{equation}
$r\equiv \sqrt{\delta_{ij}x^{i}x^{j}}$ is the radial coordinate, while $U_{\textrm{eff}}$ is an effective metric potential related to the Newtonian potential $U(r)=GM/r$ by
\begin{equation}
    \label{eq:Newton-effective}U_{\textrm{eff}}(r) \equiv G_{\textrm{eff}}(r)\frac{M}{r}=\frac{G_{\textrm{eff}}(r)}{G}U(r)~.
\end{equation}
As mentioned in the previous subsection, in particular in footnote~\ref{avertecomento}, only the $00$ component of the metric field equations is needed at $\mathcal{O}(4)$ within the present treatment, since $h_{ij}^{(4)}$ would contribute only beyond the 1PN limit. As a matter of fact, spherical symmetry, together with PPN power counting, implies that, at the order relevant here, the spatial metric is fully described by a single conformal factor, thereby excluding anisotropic terms.  

\subsubsection{The Parameter $\gamma$}
Given the rest-mass density for a point-mass source $\rho(r) = \frac{M}{4\pi r^2}\delta(r)$, with $\delta(r)$ denoting the Dirac delta distribution, and the Laplacian operator $\nabla^2 \equiv \partial_i \partial^{i} = \partial_r^2+\frac{2}{r} \partial_r$, we first solve the equation for $\varphi^{(2)}(r)$ (see Eq.~\eqref{eq:sf-eq-O2}):
\begin{equation}
    \label{eq:sf-eq-O2-isotropic}\left[B_0+\frac{3A_1^2}{16\pi G A_0}\left(1-\delta_{\textrm{P}}\right)\right]\nabla^2 \varphi^{(2)}(r) -V_2\varphi^{(2)}(r) = -\frac{A_1}{8\pi A_0}\frac{M}{r^2}\delta(r)~.
\end{equation}
Assuming $A_0, B_0>0$, and $V_2>0$,\footnote{\label{wearenotdoneyetMrMarshall}As reported in Ref.~\cite{Wolf:2025jed}, cosmological observations seem to favour $V_2<0$, in contrast to the exponential potential often employed in quintessence scenarios \cite{Barreiro:1999zs}. Because the prefactor of the gradient term is positive in both formalisms, as explained further in Sec.~\ref{sec:apps}, the scalar field would then exhibit not a monotonic exponential suppression, but rather an oscillatory modulation.} the solution is
\begin{equation}
    \varphi^{(2)}(r) = \frac{2A_1}{16\pi G A_0B_0 +3A_1^2(1-\delta_{\textrm{P}})}U(r) e^{-m_{\varphi}r}~,
\end{equation}
where $m_{\varphi}$ is the effective mass of the scalar field $\varphi$, 
\begin{equation}
    \label{eq:effective-mass-squared}m_{\varphi}^2 \equiv \frac{16\pi G A_0V_2}{16\pi G A_0 B_0+3A_1^2(1-\delta_{\textrm{P}})}~.
\end{equation}
We imposed $\varphi\to 0$ as $r\rightarrow \infty$ to discard the growing solution. The remaining constant of integration was then fixed by the delta-function matching condition at the origin, obtained by integrating the field equation over a small sphere around $r=0$. The screening\footnote{In this work, the word ``screening'' is used in a specific sense to refer to Yukawa suppression. Indeed, the effective mass $m_\varphi$ of the scalar field perturbation is a constant that does not depend on the ambient matter density. Therefore, other types of mechanisms, such as chameleon \cite{Khoury:2003aq,Khoury:2003rn}, symmetron \cite{Hinterbichler:2010es,Hinterbichler:2011ca}, or Vainshtein \cite{Vainshtein:1972sx} screening, lie beyond the scope of the present work.} length is given by $1/m_{\varphi}$, and depends on the formalism through $\delta_{\textrm{P}}$. Moreover, in the Palatini formulation ($\delta_{\textrm{P}} =1$), Newton's constant drops out of the effective mass of the field, which is instead determined by the ratio $V_2/B_0$; in this sense, $m_{\varphi}$ takes on the same form as in General Relativity (GR) with a non-canonical scalar field. In neither formulation does $m^2_{\varphi}$ become negative. However, the field value of $\varphi^{(2)}$ is negative if $A_1<0$. For $A_0>0$, $B_0>0$, and $V_2>0$, the effective mass in the Palatini case is always larger than in the metric one, unless $A_1=0$, in which case both masses coincide. Accordingly, the Palatini screening length is shorter, and the scalar field is therefore more strongly suppressed at large distances. 

In view of Eq.~\eqref{eq:specifying-even-more}, one of the metric field equations at order $\mathcal{O}(2)$, namely Eq.~\eqref{eq:O2-h00}, can be written as
\begin{equation}
    \label{eq:effective-Poisson-eq}\nabla^2 U_{\textrm{eff}}(r) = -\frac{U(r)}{A_0}\left\{\left[1+\frac{A_1^2}{16\pi G A_0 B_0+3A_1^2(1-\delta_{\textrm{P}})}\right]\frac{\delta(r)}{r}-\frac{A_1^2m^2_{\varphi}e^{-m_{\varphi}r}}{16\pi G A_0 B_0+3A_1^2(1-\delta_{\textrm{P}})}\right\}, 
\end{equation}
and its solution is
\begin{equation}
    \label{asdfiabwerawer}
    U_{\textrm{eff}}(r) = \frac{U(r)}{A_0}\left[1+\frac{A_1^2e^{-m_{\varphi}r}}{16\pi G A_0 B_0+3A_1^2(1-\delta_{\textrm{P}})}\right]+U_0~,
\end{equation}
where the additive constant $U_0$ is set to zero by imposing the boundary condition $U_{\textrm{eff}}(r)\rightarrow 0$ as $r\rightarrow \infty$. As before, the delta-source matching (flux) condition at the origin fixes the remaining constant of integration. Comparing the above expression with Eq.~\eqref{eq:Newton-effective}, we obtain $G_{\textrm{eff}}(r)$:
\begin{equation}
    \label{eq:effective-G-PPN}G_{\textrm{eff}}(r) = \frac{G}{A_0}\left[1+\frac{A_1^2e^{-m_{\varphi}r}}{16\pi G A_0 B_0+3A_1^2(1-\delta_{\textrm{P}})}\right].
\end{equation}
Contrary to the effective mass of $\varphi^{(2)}$, the effective gravitational constant differs from its GR counterpart with a non-canonical scalar field when $\delta_{\textrm{P}}=1$. In addition to the screened Poisson equation \eqref{eq:effective-Poisson-eq}, we can derive the PPN parameter $\gamma$ from the remaining metric field equation at order $\mathcal{O}(2)$, Eq.~\eqref{eq:theotherequationatO2}, together with Eq.~\eqref{eq:gij-isotropic-PPN}. Proceeding in the same way as in the previous cases, and imposing the same boundary conditions, we arrive at
\begin{equation}
    \label{asdofubaiwerawer}
    \gamma(r) =1- \frac{2A_1^2 e^{-m_{\varphi}r}}{16\pi G A_0 B_0+3A_1^2(1-\delta_{\textrm{P}})+A_1^2 e^{-m_{\varphi}r}}~.
\end{equation}
The right-hand side of Eq.~\eqref{eq:theotherequationatO2}, after substituting $h_{ij}(r)$ from Eq.~\eqref{eq:specifying-even-more}, matches that of Eq.~\eqref{eq:effective-Poisson-eq}, except that the signs in front of the corresponding $A_1^2$ terms in the numerators are reversed. As can again be seen, $\gamma$ in the Palatini formalism differs from its GR value, $\gamma_{\textrm{GR}}=1$, even though the effective masses coincide in GR (supplemented by a scalar field with a non-canonical kinetic term). 

\subsubsection{The Parameter $\beta$}
To determine $\beta(r)$, the scalar-field equation at order $\mathcal{O}(4)$, Eq.~\eqref{eq:sf-eq-O(4)-final}, must first be solved. We remind the reader that a point-like mass is assumed, meaning that the pressure $p$ and the internal energy $\varepsilon$ vanish. Because the calculations are performed in the rest frame of the source, the velocity term $v$ also vanishes. We neglect the term $\rho \varphi^{(2)}$ as well, since we are interested only in the contribution from the pure gravitational self-interaction present in vacuum, given that $\beta$ measures the non-linearity of the gravitational field itself, as noted at the beginning of this subsection. Accordingly, in order to determine $\varphi^{(4)}$, we use Eq.~\eqref{eq:sf-eq-O2-isotropic} with the right-hand side set to zero, corresponding to the vacuum solution for $r>0$. Consequently, the scalar field equation becomes
\begin{align}
    \nonumber&\left[B_0+\frac{3A_1^2}{16\pi G A_0}(1-\delta_{\textrm{P}})\right]\nabla^2 \varphi^{(4)}(r)-V_2\varphi^{(4)}(r) = \left\{\frac{1}{2}V_3-m^2_{\varphi}\left[B_1+\frac{A_1}{A_0}B_0+\right.\right.\\
    \nonumber&\left.\left.+\frac{3A_1 A_2}{8\pi G A_0}(1-\delta_{\textrm{P}})\right]\right\}[\varphi^{(2)}(r)]^2+2m^2_{\varphi}\left[B_0+\frac{3A_1^2}{16\pi G A_0}(1-\delta_{\textrm{P}})\right]\gamma(r) U_{\textrm{eff}}(r) \varphi^{(2)}(r)+\\
    &\phantom{--------}+\frac{1}{2}\left[\frac{A_1}{A_0}B_0-B_1-\frac{3A_1^2}{8\pi G A_0}\left(\frac{A_2}{A_1}-\frac{A_1}{A_0}\right)(1-\delta_{\textrm{P}})\right][\partial_r \varphi^{(2)}(r)]^2~.
\end{align}
The solution is
\begin{equation}
    \label{eq:sol-no-homogeneous-varphi4}\varphi^{(4)}(r) =\frac{\kappa e^{-m_{\varphi}r}}{r} -\frac{1}{2m_{\varphi}r}\left[e^{-m_{\varphi}r}\int^{r}_{0}\textrm{d}\tilde{r}~S(\tilde{r})\tilde{r}e^{m_{\varphi}\tilde{r}}+e^{m_{\varphi}r}\int_r^{\infty}\textrm{d}\tilde{r}~S(\tilde{r})\tilde{r}e^{-m_{\varphi}\tilde{r}}\right],
\end{equation}
where $\kappa$ is a constant of integration, and $S(r)$ reads
\begin{equation}
    S(r) \equiv \left[c_1 +\left(c_2+c_3\right) e^{-m_{\varphi}r}+c_4\frac{\left(1+m_{\varphi}r\right)^2}{r^2} e^{-m_{\varphi}r}\right] U^2(r)e^{-m_{\varphi}r}~.   
\end{equation}
The constant coefficients are 
\begin{align}
    &c_1 \equiv \frac{4A_1m^2_{\varphi}}{[16\pi G A_0 B_0+3A_1^2(1-\delta_{\textrm{P}})]A_0}~,\\
    &c_2 \equiv \frac{64\pi G A_0 A_1^2}{\left[16\pi G A_0 B_0+3A_1^2(1-\delta_{\textrm{P}})\right]^3}\left\{\frac{1}{2}V_3-m^2_{\varphi}\left[B_1+\frac{A_1}{A_0}B_0+\frac{3A_1A_2}{8\pi G A_0}(1-\delta_{\textrm{P}})\right]\right\},\\
    &c_3 \equiv -\frac{4A_1^3 m^2_{\varphi}}{\left[16\pi G A_0 B_0+3A_1^2(1-\delta_{\textrm{P}})\right]^2A_0}~,\\
    &c_4 \equiv \frac{32\pi G A_0A_1^2}{\left[16\pi G A_0 B_0+3A_1^2(1-\delta_{\textrm{P}})\right]^3}\left[\frac{A_1}{A_0}B_0-B_1-\frac{3A_1^2}{8\pi G A_0}\left(\frac{A_2}{A_1}-\frac{A_1}{A_0}\right)(1-\delta_{\textrm{P}})\right].
\end{align}
Because our interest is restricted to the vacuum solution for $r>0$, the lower limit of integration in Eq.~\eqref{eq:sol-no-homogeneous-varphi4} should not be taken to $0$. In the Green-function representation, the first integral in Eq.~\eqref{eq:sol-no-homogeneous-varphi4} contributes only through an effective Yukawa charge, which diverges in the point-particle limit \cite{Hohmann:2013rba}. We therefore introduce a cutoff $\epsilon>0$, writing 
\begin{equation}
    \varphi^{(4)}(r) =\frac{\kappa e^{-m_{\varphi}r}}{r} -\frac{1}{2m_{\varphi}r}\left[e^{-m_{\varphi}r}\int^{r}_{\epsilon}\textrm{d}\tilde{r}~S(\tilde{r})\tilde{r}e^{m_{\varphi}\tilde{r}}+e^{m_{\varphi}r}\int_r^{\infty}\textrm{d}\tilde{r}~S(\tilde{r})\tilde{r}e^{-m_{\varphi}\tilde{r}}\right],
\end{equation}
and, upon carrying out the integration, the Yukawa monopole term reads
\begin{align}
    \nonumber&\frac{e^{-m_\varphi r}}{r}\left[\kappa + \frac{G^2M^2}{2m_\varphi}\left(c_1 \ln{m_{\varphi}\epsilon}+(c_2+c_3)\text{Ei}(-m_{\varphi}\epsilon)-c_4\frac{e^{-m_{\varphi} \epsilon}(1+3m_{\varphi}\epsilon)}{2\epsilon^2}-\right.\right.\\
    \label{aoisdbfasdf}&\left.\left.\phantom{--------------------------}-c_4\frac{m_\varphi^2}{2}\text{Ei}(-m_{\varphi}\epsilon)\right)\right],
\end{align}
where $\textrm{Ei}(-x)$ denotes the exponential integral, 
\begin{equation}
    \textrm{Ei}(-x) \equiv -\int^{\infty}_{x}\textrm{d}t ~\frac{e^{-t}}{t}~.
\end{equation}
We then impose that the scalar charge vanish at order $\mathcal{O}(4)$. This is consistent with removing any explicit source or contact (boundary-injection) term at that order.\footnote{The same reasoning applies to the later calculation of $h_{00}^{(4)}(r)$.} $\varphi^{(4)}$ is then given by 
\begin{align}
    \nonumber&\varphi^{(4)}(r) =\frac{c_1}{2m_{\varphi}}U^2(r) r\left[e^{m_{\varphi}r}\textrm{Ei}(-2m_{\varphi}r)-e^{-m_{\varphi}r}\ln(m_{\varphi}r)\right]+\frac{1}{2}\left(\frac{c_2+c_3}{m_{\varphi}}-\frac{c_4 m_{\varphi}}{2}\right)U^2(r)r\times\\
    &\phantom{----------}\times\left[e^{m_{\varphi}r}\textrm{Ei}(-3m_{\varphi}r)-e^{-m_{\varphi}r}\textrm{Ei}(-m_{\varphi}r)\right]+\frac{1}{2}c_4 U^2(r) e^{-2m_{\varphi}r}~.
\end{align}

Once $\varphi^{(4)}$ has been determined, $\beta(r)$ can be obtained from Eq.~\eqref{eq:metric-field-eqs-O4-00-final} by dropping out terms involving source-dependent contributions, the pressure, the internal energy, and the velocity, as in the case of $\varphi^{(4)}$ for reasons already discussed. The equation then reads (see Eq.~\eqref{eq:specifying-even-more})
\begin{align}
    \nonumber&\nabla^2 h_{00}^{(4)}(r) = \frac{A_1}{A_0}\left[m^2_{\varphi}\varphi^{(4)}(r)+S(r)\right]+\left[\frac{8\pi G}{A_0}V_2+\frac{A_1}{A_0}\left(\frac{A_2}{A_1}-\frac{A_1}{A_0}\right)m^2_{\varphi}\right]\left[\varphi^{(2)}(r)\right]^2-\\
    &\phantom{-----}-2\frac{A_1}{A_0}m^2_{\varphi}U_{\textrm{eff}}(r)\varphi^{(2)}(r)-4\left[\partial_r U_{\textrm{eff}}(r)\right]^2+\frac{A_1}{A_0}\left(\frac{A_2}{A_1}-\frac{A_1}{A_0}\right)\left[\partial_r \varphi^{(2)}(r)\right]^2~.
\end{align}
Imposing asymptotic decay, the solution is
\begin{equation}
    h^{(4)}_{00}(r)=\frac{\tilde{\kappa}}{r}-\frac{1}{r}\int^{r}_{0}\textrm{d}\tilde{r}~F(\tilde{r})\tilde{r}^2-\int^{\infty}_r \textrm{d}\tilde{r}~F(\tilde{r}) \tilde{r}~, 
\end{equation}
where $\tilde{\kappa}$ is an integration constant, and $F(r)$ is given by
\begin{align}
    \nonumber&F(r) =\left\{c_5r\left[e^{m_{\varphi}r}\textrm{Ei}(-2m_{\varphi}r)-e^{-m_{\varphi}r}\ln(m_{\varphi}r)\right]+c_6r\left[e^{m_{\varphi}r}\textrm{Ei}(-3m_{\varphi}r)-e^{-m_{\varphi}r}\textrm{Ei}(-m_{\varphi}r)\right]+\right.\\
    &\left.\phantom{-----}+c_7e^{-2m_{\varphi}r}+c_8\frac{1+m_{\varphi}r}{r^2}e^{-m_{\varphi}r}+c_{9}\frac{(1+m_{\varphi}r)^2}{r^2}e^{-2m_{\varphi}r}-\frac{4}{A_0^2r^2}\right\}U^2(r)~, 
\end{align}
and
\begin{align}
    &c_5 \equiv \frac{A_1m_{\varphi}}{2A_0}c_1~,\\
    &c_6 \equiv \frac{A_1m_{\varphi}}{2A_0}\left(c_2+c_3-\frac{c_4m^2_{\varphi}}{2}\right),\\
    &c_7 \equiv \frac{A_1}{A_0}\left(c_2+c_3+\frac{c_4m_{\varphi}^2}{2}\right)+\frac{4A_1^2}{[16\pi G A_0 B_0+3A_1^2(1-\delta_{\textrm{P}})]^2}\left[\frac{8\pi G}{A_0}V_2+\frac{A_1}{A_0}\left(\frac{A_2}{A_1}-2\frac{A_1}{A_0}\right)m^2_{\varphi}\right],\\
    &c_8 \equiv -\frac{8A_1^2}{[16\pi G A_0 B_0+3A_1^2(1-\delta_{\textrm{P}})]A_0^2}~,\\
    &c_9 \equiv \frac{A_1}{A_0}\left\{c_4+\frac{4A_1^2}{[16\pi G A_0 B_0 +3A_1^2(1-\delta_{\textrm{P}})]^2}\left(\frac{A_2}{A_1}-2\frac{A_1}{A_0}\right)\right\}.
\end{align}
Proceeding in the same way as for $\varphi^{(4)}$, the fourth-order solution reads
\begin{align}
    h_{00}^{(4)}(r)=U^2(r)&\Bigg\{\frac{c_5 r}{m_{\varphi}^2}\left[e^{m_{\varphi} r}\text{Ei}(-2m_{\varphi}r)-e^{-m_{\varphi}r}\ln{(m_{\varphi}r)}-2e^{-m_{\varphi} r}-2m_{\varphi} r \text{Ei}(-m_{\varphi}r)\right]+\nonumber\\
    &+\frac{c_6 r}{m_{\varphi}^2}\left[e^{m_{\varphi} r}\text{Ei}(-3 m_{\varphi} r) - e^{-m_{\varphi} r}\text{Ei}(-m_{\varphi} r)-e^{-2 m_{\varphi} r} - 2m_{\varphi} r \text{Ei}(-2m_{\varphi} r)\right]+\nonumber\\
    &+c_7 r^2\left[\text{Ei}(-2 m_{\varphi} r)+\frac{e^{-2 m_{\varphi} r}}{2 m_{\varphi} r}\right]+\frac{c_8}{2}\left[(1-m_{\varphi} r)e^{-m_{\varphi} r}-m_{\varphi}^2r^2 \text{Ei}(-m_{\varphi} r)\right]+\nonumber\\
    &\phantom{-------}+c_9\left[\frac{1-m_{\varphi}r}{2}e^{-2 m_{\varphi} r}-m_{\varphi}^2 r^2 \text{Ei}(-2 m_{\varphi} r)\right]-\frac{2}{A_0^2}\Bigg\}.
\end{align}
Using $h_{00}^{(4)}(r)=-2\beta(r)U^2_{\textrm{eff}}(r)$ and Eq.~\eqref{eq:Newton-effective}:
\begin{align}
    \nonumber&\beta(r) \frac{G^2_{\textrm{eff}}(r)}{G^2} = \frac{1}{A_0^2}+c_5 \left\{\left[1+\frac{1}{2}\ln(m_{\varphi}r)\right]e^{-m_{\varphi}r}+m_{\varphi}r\textrm{Ei}(-m_{\varphi}r)-\frac{e^{m_{\varphi}r}}{2}\textrm{Ei}(-2m_{\varphi}r)\right\}\frac{r}{m_{\varphi}^2}+\\
    \nonumber&+c_6\left\{m_{\varphi}r\textrm{Ei}(-2m_{\varphi}r)+\frac{1}{2}\left[e^{-2m_{\varphi}r}+e^{-m_{\varphi}r}\textrm{Ei}(-m_{\varphi}r)-e^{m_{\varphi}r}\textrm{Ei}(-3m_{\varphi}r)\right]\right\}\frac{r}{m_{\varphi}^2}-\\
    \nonumber&-\frac{c_7}{2} \left[\textrm{Ei}(-2m_{\varphi}r)+\frac{e^{-2m_{\varphi}r}}{2m_{\varphi}r}\right]r^2-\frac{c_8}{4} \left[(1-m_{\varphi}r)e^{-m_{\varphi}r}-m^2_{\varphi}r^2 \textrm{Ei}(-m_{\varphi}r)\right]-\\
    &\phantom{-----------------}-\frac{c_9}{2} \left[\frac{1-m_{\varphi}r}{2}e^{-2m_{\varphi}r}-m^2_{\varphi}r^2 \textrm{Ei}(-2m_{\varphi}r)\right].
\end{align}
Hence, we obtain $\beta(r)$: 
\begin{align}
    \nonumber&\beta(r) = 1+\frac{G^2}{G_{\textrm{eff}}^2(r)}\left\{c_5 \left\{\left[1+\frac{1}{2}\ln(m_{\varphi}r)\right]e^{-m_{\varphi}r}+m_{\varphi}r\textrm{Ei}(-m_{\varphi}r)-\frac{e^{m_{\varphi}r}}{2}\textrm{Ei}(-2m_{\varphi}r)\right\}\frac{r}{m^2_{\varphi}}+\right.\\
    \nonumber&\left.+c_6\left\{m_{\varphi}r\textrm{Ei}(-2m_{\varphi}r)+\frac{1}{2}\left[e^{-2m_{\varphi}r}+e^{-m_{\varphi}r}\textrm{Ei}(-m_{\varphi}r)-e^{m_{\varphi}r}\textrm{Ei}(-3m_{\varphi}r)\right]\right\}\frac{r}{m_{\varphi}^2}-\right.\\
    \nonumber&\left.-\frac{c_7}{2} \left[\textrm{Ei}(-2m_{\varphi}r)+\frac{e^{-2m_{\varphi}r}}{2m_{\varphi}r}\right]r^2+\frac{c_8}{4} \left[e^{-m_{\varphi}r}+m_{\varphi}r \textrm{Ei}(-m_{\varphi}r)\right]m_{\varphi}r+\right.\\
    &\left.\phantom{------------}+\frac{c_9}{4}\left[e^{-2m_{\varphi}r}+2m_{\varphi}r \textrm{Ei}(-2m_{\varphi}r)\right]m_{\varphi}r-\frac{c_{10}}{4}e^{-2m_{\varphi}r}\right\},
\end{align}
or, after rearranging the expression so as to group terms according to their radial dependence, and noting that $c_8/4 = -c_5/m_{\varphi}^3$: 
\begin{align}
    \nonumber&\beta(r)-1=\frac{G^2}{G^2_{\textrm{eff}}(r)}\left\{c_5\frac{r}{2m_{\varphi}^2}\left[e^{-m_{\varphi}r}\ln(m_{\varphi}r)-e^{m_{\varphi}r}\textrm{Ei}(-2m_{\varphi}r)\right]-\frac{c_{10}}{4}e^{-2m_{\varphi}r}+\right.\\
    \nonumber&\left.+c_6\frac{r}{2m_{\varphi}^2}\left[e^{-m_{\varphi}r}\textrm{Ei}(-m_{\varphi}r)-e^{m_{\varphi}r}\textrm{Ei}(-3m_{\varphi}r)\right]+\left(\frac{c_6}{m_{\varphi}^3}-\frac{c_7}{2m^2_{\varphi}}+\frac{c_9}{2}\right)m_{\varphi}r\times\right.\\
    \label{eq:beta-not-yet-simplified}&\left.\phantom{---------------------}\times\left[\frac{e^{-2m_{\varphi}r}}{2}+m_{\varphi}r\textrm{Ei}(-2m_{\varphi}r)\right]\right\},
\end{align}
where $c_{10}$ is defined as
\begin{equation}
    c_{10} \equiv c_9 +\frac{4A_1^4}{[16\pi G A_0 B_0+3A_1^2(1-\delta_{\textrm{P}})]^2A_0^2}~.
\end{equation}
Upon making the Taylor coefficients $A_0$, $A_1$, and so on explicit in Eq.~\eqref{eq:beta-not-yet-simplified}, the PPN parameter $\beta$ reads
\begin{align}
    \nonumber&\beta(r)-1 =-\frac{G^2 A_1^3 e^{-2m_{\varphi}r}}{G_{\textrm{eff}}^2(r)[16\pi G A_0 B_0+3A_1^2(1-\delta_{\textrm{P}})]^2A_0}\left[\frac{16\pi G A_0 B_0}{16\pi G A_0 B_0+3A^2_1(1-\delta_{\textrm{P}})}\left(\frac{A_2}{A_1}-\frac{A_1}{A_0}\right)+\right.\\
    \nonumber&\left.+\frac{m^2_{\varphi}}{2V_2}\left(\frac{A_1}{A_0} B_0-B_1\right)\right]-\frac{G^2A_1^2 m_{\varphi}r}{G^2_{\textrm{eff}}(r)[16\pi G A_0 B_0 +3A_1^2(1-\delta_{\textrm{P}})]A_0^2}\left\{\frac{e^{-2m_{\varphi}r}}{2}+(m_{\varphi}r+e^{m_{\varphi}r})\times\right.\\
    \nonumber&\left.\times\textrm{Ei}(-2m_{\varphi}r)-e^{-m_{\varphi}r}\ln(m_{\varphi}r)+\frac{A_1A_0}{16\pi G A_0 B_0+3A_1^2(1-\delta_{\textrm{P}})}\left[\frac{V_3}{2V_2}-\frac{A_1}{A_0}\left(1+\frac{5m_{\varphi}^2}{4V_2}B_0\right)-\right.\right.\\
    &\left.\left.-\frac{3m^2_{\varphi}}{4V_2}B_1-\frac{3A_1^2(1-\delta_{\textrm{P}})}{2[16\pi G A_0 B_0+3A_1^2(1-\delta_{\textrm{P}})]}\left(\frac{A_1}{A_0}+3\frac{A_2}{A_1}\right)\right]\left[e^{m_{\varphi}r}\textrm{Ei}(-3m_{\varphi}r)-e^{-m_{\varphi}r}\textrm{Ei}(-m_{\varphi}r)\right]\right\}.
\end{align}
For strong Yukawa suppression ($m_{\varphi}r\gg 1$), where the Yukawa correction becomes negligible, $\beta$ is approximated as
\begin{align}
    \nonumber&\beta(r) -1\simeq \frac{G^2e^{-m_{\varphi}r}}{2G_{\textrm{eff}}^2(r)}\left\{\frac{c_5}{m_{\varphi}^3}\left[m_{\varphi}r\ln(m_{\varphi}r)+\frac{1}{2}-\frac{1}{4m_{\varphi}r}\right]-\right.\\
    &\left.\phantom{------------------}-\left[\frac{c_6}{3m_{\varphi}^3}+\frac{1}{2}\left(\frac{c_7}{m_{\varphi}^2}-c_9\right)+c_{10}\right]\frac{e^{-m_{\varphi}r}}{2}\right\},
\end{align}
and it can be seen that the coefficients containing the derivatives of the potential, $V_2$ and $V_3$, are subleading, namely those in which $A_2$ and $B_1$ also appear. 

We next translate the PPN parameters derived above into observational bounds, allowing for a direct comparison with Solar-System measurements and a contrast between the metric and Palatini formalisms. After assessing their agreement with Solar-System data, we delineate the allowed parameter space and highlight the phenomenological differences between the two approaches. 

\section{\label{sec:constraints}Solar-System Constraints and Comparison of Formalisms}
In this section, we use bounds on the PPN parameters from Solar System experiments to determine whether the allowed parameter space is sensitive to the choice of the gravitational formalism. We begin by introducing a set of redefinitions that enable us to express the relevant quantities more compactly.

The Yukawa coupling strength is defined by
\begin{equation}
    \alpha \equiv \frac{A_1^2}{16\pi G A_0 B_0+3A_1^2(1-\delta_{\rm P})}~,
\end{equation}
while the kinetic fraction is introduced as
\begin{equation}
    \chi\equiv \frac{16\pi G A_0 B_0}{16\pi G A_0 B_0+3A^2_1(1-\delta_{\textrm{P}})}=\frac{m_\varphi^2 B_0}{V_2}=1-3\alpha(1-\delta_{\rm P})~,
\end{equation}
which quantifies the relative importance of the kinetic contribution $A_0 B_0$ in the denominator of the Yukawa coupling strength. In the Palatini formalism, one has $\chi=1$, independently of the specific theory under consideration. We also define three dimensionless derivative ratios: 
\begin{equation}
    \eta_1\equiv \frac{A_0 A_2}{A_1^2}~,\ \ \  \eta_2\equiv \frac{A_0 B_1}{A_1 B_0}~,\ \ \ \eta_3\equiv \frac{A_0 V_3}{A_1 V_2}~,
\end{equation}
and two radial functions:
\begin{eqnarray}
    F_1(r)&\equiv& \frac{e^{-2 m_\varphi r}}{2}+(m_\varphi r + e^{m_\varphi r})\text{Ei}(-2m_\varphi r) - e^{-m_\varphi r}\ln{(m_\varphi r)}~,\\
    F_2(r)&\equiv& e^{m_\varphi r}\text{Ei}(-3 m_\varphi r) - e^{-m_\varphi r}\text{Ei}(-m_\varphi r)~.
\end{eqnarray}
In the effectively massless regime $m_\varphi r\ll 1$, corresponding to a length scale associated with $m_{\varphi}$ much larger than $r$, we have
\begin{equation}
    F_1(r)\xrightarrow[ m_\varphi r \ll 1 ]{}\frac{1}{2}+\gamma_{\rm E}+\ln{2}~, \ \ \ F_2(r)\xrightarrow[ m_\varphi r \ll 1 ]{}\ln{3}~,
\end{equation}
where $\gamma_{\rm E}$ is the Euler-Mascheroni constant. In the opposite limit, $m_\varphi r\gg 1$,
\begin{equation}
    F_1(r)\xrightarrow[ m_\varphi r \gg 1 ]{}0~,\ \ \ \text{and} \ \ \ F_2(r)\xrightarrow[ m_\varphi r \gg 1 ]{}0~.
\end{equation}
With these definitions, the effective Newton's constant reads
\begin{equation}
    \label{asdfijabsdfae}
    G_{\rm eff}(r)=\frac{G}{A_0}\left(1+\alpha e^{-m_\varphi r}\right),
\end{equation}
while the light-deflection and nonlinearity parameters read
\begin{equation}
    \label{asdfbahierawer}
    \gamma(r)=\frac{1-\alpha e^{-m_\varphi r}}{1+\alpha e^{-m_\varphi r}}~,
\end{equation}
and
\begin{equation}
    \label{adifbjahwerwaer}
    \beta(r)-1=\frac{\alpha^2 \chi e^{-2 m_\varphi r}}{\left(1+\alpha e^{-m_\varphi r}\right)^2}\left(\frac{1}{2}-\eta_1+\frac{\eta_2}{2}\right)-\frac{\alpha m_\varphi r}{\left(1+\alpha e^{-m_\varphi r}\right)^2}\left[F_1(r)+\alpha C F_2(r)\right],
\end{equation}
respectively. $C$ is 
\begin{equation}
    C\equiv \frac{\eta_3}{2}-1-\frac{5}{4}\chi-\frac{3}{4}\chi \eta_2 - \frac{3}{2}\alpha (1-\delta_{\rm P})(1+3\eta_1)~.
\end{equation}

Eqs.~\eqref{asdfijabsdfae}--\eqref{adifbjahwerwaer} take the same form in both the metric and Palatini formalisms. Consequently, any dependence on the gravitational formalism can only enter via the mapping between the underlying coefficients $\{A_0, A_1, A_2, B_0, B_1, V_2,V_3\}$ and the redefined quantities $\{m_\varphi, \alpha, \chi, \eta_1, \eta_2,\eta_3\}$. One must therefore impose constraints directly on the former rather than on the latter. 

\subsection{\label{sec:observables}Constraining Scalar-Tensor Gravity with Solar-System Observables}
The observables relevant for constraining the PN expansion are briefly reviewed in this section. In particular, the focus is placed on the Shapiro time delay, the deflection of light, and the perihelion precession of Mercury. The interested reader is referred to Refs.~\cite{Will:2014kxa,Will:2018bme} for classic treatises on experimental tests of gravity.

\subsubsection{Shapiro Time Delay and Light Deflection}
The Shapiro time delay $\delta t$ and the light-deflection angle $\delta \theta$ can be obtained from the photon trajectory. For a general action \eqref{eq:full_action}, the latter is derived in App.~\ref{sec:appendix-geodesics} for a point-like source of mass $M$ located at the origin, described by the density $\rho(r)=M\delta(r)/(4\pi r^2)$. The trajectory reads
\begin{equation}
    \label{asdoabnwerawer}
    x^i(t)=x_0^i+k^i(t-t_0)+k^i\delta x_{\parallel}+\delta x_{\bot}^i~,
\end{equation}
where
\begin{equation}
    \delta x_{\parallel} = -\frac{2GM}{A_0} \ln{\left(\frac{r(t)+k_jx^j_0+(t-t_0)}{r_0+k_jx_0^j}\right)}
\end{equation}
is a 1PN correction defined along the 0PN photon path, since $k_ik^i=1$, and 
\begin{equation}
    \delta x_{\bot}^i = -\frac{2GM}{A_0}\frac{b^i}{b^2}\left[r(t)-r_0-\frac{k_jx_0^j}{r_0}(t-t_0)\right]
\end{equation}
is a 1PN correction evaluated in the direction perpendicular to the 0PN photon trajectory, as $b_ik^i=0$. Eq.~\eqref{asdoabnwerawer} holds in both the metric and Palatini formalisms. To obtain $\delta t$, Eq.~\eqref{asdoabnwerawer} is contracted with $k^i$, which yields
\begin{equation}
    t-t_0=\abs{x^i(t)-x^i_0}+\frac{2GM}{A_0} \ln{\left(\frac{r(t)+k_jx^j_0+(t-t_0)}{r_0+k_jx_0^j}\right)}~,
\end{equation}
where $k_i(x^i(t)-x^i_0)=\abs{x^i(t)-x^i_0}+\mathcal{O}(4)$. For a signal sent from Earth, $(t_{\oplus},x_{\oplus}^i)$, to a spacecraft, $(t_{\rm sp},x_{\rm sp}^i)$, and back, the total travel time $t_{\rm tot}$ is then 
\begin{equation}
    t_{\rm tot} = 2\abs{x^i_{\oplus}-x^i_{\rm sp}}+\frac{4GM}{A_0}\ln{\left(\frac{r_\oplus+r_{\rm sp}+\abs{x^i_{\oplus}-x^i_{\rm sp}}}{r_\oplus+r_{\rm sp}-\abs{x^i_{\oplus}-x^i_{\rm sp}}}\right)}\simeq 2\abs{x^i_{\oplus}-x^i_{\rm sp}} +\frac{4GM}{A_0}\ln{\left(\frac{4r_\oplus r_{\rm sp}}{b^2}\right)}~,
\end{equation}
the second equality following from the standard superior-conjunction approximation $b\ll r_\oplus, r_{\rm sp}$ \cite{Will:2014kxa,Will:2018bme}. The Shapiro time delay is therefore
\begin{equation}
    \label{asdfiabwerrawef}
    \delta t = \frac{4GM}{A_0}\ln{\left(\frac{4r_\oplus r_{\rm sp}}{b^2}\right)}~.
\end{equation}
\begin{table}[t]
\centering
\renewcommand{\arraystretch}{1.22}
\begin{tabular}{|c| c| l| }
\hline
Quantity & Constraint & Experiment \\
\hline\hline
$\gamma-1$ &
$(2.1\pm 2.3)\times 10^{-5}$ ($1\sigma$) & Cassini \cite{Bertotti:2003rm} \\[2mm]
\hline
$\gamma-1$ &
$(-0.8\pm 1.2)\times 10^{-4}$ ($1\sigma$) & VLBI \cite{Lambert:2009xy,Titov:2010zn} \\[2mm]
\hline
&$\abs{\delta \phi} <
1.75\times 10^{-11}$rad/orbit ($1\sigma$ equiv.) & MESSENGER \cite{Park:2017zgd} \\[2mm]
\hline
\end{tabular}
\caption{Experimental constraints on the light-deflection parameter $\gamma -1$, derived from Cassini data (first row) and VLBI data (second row), and on Mercury's perihelion advance, $\delta\phi$, derived from MESSENGER data (third row).}
\label{tab:constraints}
\end{table}

We can obtain $\delta \theta$ in a similar way. The quantity of interest is the change in the direction of the photon trajectory, which is determined by the component of $\text{d}x^i(t)/\text{d}t$ perpendicular to the 0PN path. Equivalently, $\delta \theta$ is given by the magnitude of the transverse change in direction:
\begin{equation}
    q^i=\text{d}(\delta x_{\bot}^i)/\text{d}t~.
\end{equation}
Assuming that the photon is emitted at $(t_0,x^i_0)$ with $q^i(t_0)=0$, this is simply given by the norm of Eq.~\eqref{asdofbawerawer}, which we repeat here:
\begin{align}
    \label{qeaubfawefsdssdf}
    \delta \theta=\frac{2GM}{A_0 b}\abs{\frac{k_j x^j_0 + (t-t_0)}{r(t)}-\frac{k_jx_0^j}{r_0}}\simeq \frac{4GM}{A_0 b}~,
\end{align}
where the last step follows from taking the emission and observation events to lie in the asymptotic past and future, $t_0\to -\infty$ and $t\to +\infty$, respectively. In general, however, the initial direction of the photon is not known, so a different observable must be used in order to test GR through light deflection. In practice, the measured quantity is the angle between two known sources, first when the deflecting mass is far from the line of sight, and again when it lies close to it. That is, the observable is the differential deflection angle (see Ref.~\cite{Will:2018bme} for further details). Owing to the cancellation in Eq.~\eqref{adfiabweirawer}, the resulting expression for $\delta \theta$ takes on the same form as Eq.~\eqref{qeaubfawefsdssdf}, and we do not repeat it here.

Since Eqs.~\eqref{asdfiabwerrawef} and \eqref{qeaubfawefsdssdf} reproduce the GR results, up to the overall normalisation by $A_0$, one might be tempted to conclude that Shapiro-delay and light-deflection experiments cannot constrain the details of the scalar-tensor theory \cite{Huang:2024gvi}. However, we remind the reader that the bare gravitational coupling constant $G$ appearing in the action is not directly measured in Cavendish-type experiments (see footnote~\ref{first-footnote-in-your-face}). Rather, the locally measured gravitational parameter is inferred from orbital dynamics and ephemeris fits, assuming a force law of the form
\begin{equation}
    F_r=-m \partial_r U_{\rm eff}(r)=-\frac{(G_{\rm eff}(r)-rG_{\rm eff}'(r))Mm}{r^2}\equiv -\frac{G_{\rm dyn}(r)Mm}{r^2}~,
\end{equation}
where
\begin{equation}
    G_{\rm dyn}(r)=\frac{G}{A_0}\left[1+(1+m_\varphi r)\alpha e^{-m_\varphi r}\right].
\end{equation}
Taking into account that $G_{\rm eff}(r)$ varies negligibly across Solar-System scales (see Ref.~\cite{Jarv:2017npl} and references therein), one has $G_{\rm dyn}(r)\simeq G_{\rm eff}(r)$. In this way, the quantity actually measured is $\mu\equiv G_{\rm dyn}(\tilde{r})M\simeq G_{\rm eff}(\tilde{r})M$, where $\tilde{r}$ is a characteristic scale that depends on how the Keplerian mass of the object is determined \cite{Dyadina:2021paa}. Since the Shapiro time delay measured by Cassini is sourced by the Sun, we follow Ref.~\cite{Alsing:2011er} and take $\tilde{r}=1$AU, as this is the scale associated with the determination of the Sun's Keplerian mass. Thus, using Eq.~\eqref{asdfijabsdfae} in the time-delay expression, we obtain
\begin{equation}
    \delta t = \frac{4 \mu_{\odot}}{1+\alpha e^{- m_\varphi \tilde{r}}}\ln{\left(\frac{4r_\oplus r_{\rm sp}}{b^2}\right)}=[1+\gamma(\tilde{r})]2\mu_{\odot}\ln{\left(\frac{4r_\oplus r_{\rm sp}}{b^2}\right)}~,
\end{equation}
with $\mu_{\odot}\simeq G_{\rm eff}(\tilde{r})M_{\odot}$. The resulting expression takes the standard form used in Ref.~\cite{Bertotti:2003rm} and can now be used to constrain the PPN parameter $\gamma$.

\subsubsection{Perihelion Advance of Mercury}
The last observable considered here is Mercury's perihelion advance, $\delta \phi$. Starting from the 1PN Lagrangian for a test particle of mass $m$ (see App.~\ref{App:fifthforces}),
\begin{align}
    \frac{L}{m}&=-1+U_{\rm eff}(r)+\frac{1}{2}\left(\dot{r}^2+r^2\dot{\phi}^2\right)-\left[\beta(r)-\frac{1}{2}\right]U_{\rm eff}^2(r)+\nonumber\\
    &\phantom{------------}+\left[\gamma(r) + \frac{1}{2}\right]U_{\rm eff}(r)\left(\dot{r}^2+r^2\dot{\phi}^2\right)+\frac{1}{8}\left(\dot{r}^2+r^2\dot{\phi}^2\right)^2~,\label{aisdfbaiyhsdfawer}
\end{align}
together with the radial dependence of $\gamma(r)$, $\beta(r)$, and $U_{\rm eff}(r)$, suggests that the secular perihelion precession depends on the interaction scale rather than on a single constant PPN parameter. The MESSENGER mission provided the high-precision radiometric tracking data that underpin modern determinations of Mercury’s orbit \cite{doi:10.1126/science.1218809}. However, in order to place an observational constraint on $\beta$, an ephemeris fit would have to be repeated within the modified theory \cite{Fienga:2023ocw} using the MESSENGER data, in analogy with the analysis performed in Ref.~\cite{Park:2017zgd}, where GR is assumed. Only then, with an updated constraint on $\beta$, would it be possible to derive rigorous 1PN bounds on the parameter space of the theory. This should be contrasted with the time-delay case, where the general calculation of the observable yields a GR-like expression that can be applied directly to the Cassini data. For these reasons, we also refrain from evaluating $\beta$ at a reference radius and imposing the GR-derived constraint on it, as was done in Ref.~\cite{Hohmann:2013rba}. 

A full 1PN computation of $\delta \phi$ for the general action \eqref{eq:full_action}, together with an ephemeris fit, lies beyond the scope of the present work.\footnote{See, however, Ref.~\cite{Alves:2023cuo} for a computation of the periastron advance in the particular case of massive Brans-Dicke gravity in the large-mass limit. There, the authors emphasise that the perihelion advance must be treated directly rather than being encoded solely in the standard effective $\beta$. For related phenomenological analyses of Yukawa-modified two-body dynamics and orbital-precession constraints, see also Ref.~\cite{Benisty:2022txp}.} Nevertheless, an approximate constraint on the theory can still be obtained. Indeed, Eq.~\eqref{aisdfbaiyhsdfawer} shows that a perihelion shift arises already at 0PN from the Yukawa contribution to $U_{\rm eff}(r)$. The perihelion advance per orbit is given by (see App.~\ref{App:fifthforces})
\begin{equation}
    \label{asdfiuabewwer}
        \delta \phi =  \pi \alpha (m_{\varphi}p)^2 e^{-m_\varphi p}~,
\end{equation}
with 
\begin{equation}
    p=a(1-e^2)~,
\end{equation}
where $a$ is the semimajor axis and $e$ is the eccentricity. For Mercury's orbit, one has $a=0.387$AU, and $e=0.206$. The 0PN perihelion advance per orbit in Eq.~\eqref{asdfiuabewwer} is required to be smaller than the uncertainty in the GR-based estimate of Mercury's total perihelion precession reported in Ref.~\cite{Park:2017zgd}. The quoted uncertainty is $0.0015''/\mathrm{century}$, corresponding to $1.75\times 10^{-11}\mathrm{rad}/\mathrm{orbit}$. We emphasise that this does not constitute a strict bound, since the quoted GR-based value for Mercury's perihelion precession is inferred from a signal that is also correlated with $\gamma$, $\beta$, and $J_{2\odot}$.

\subsection{\label{sec:apps}Parameter Space}
\begin{figure}[h]
         \centering
         \includegraphics[width=0.7\textwidth]{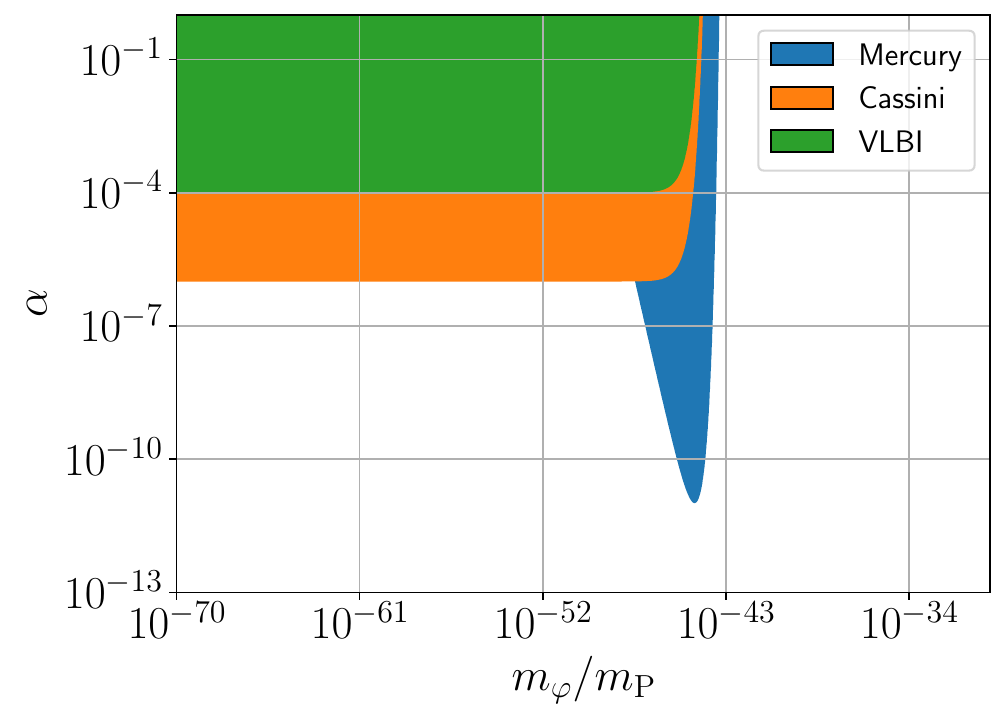}
     \caption{Excluded regions in the $(m_{\varphi},\alpha)$ plane obtained from the Cassini (orange) and VLBI (green) constraints on $\gamma(\tilde{r})$, together with the approximate bound from Mercury's perihelion advance (blue). The effective scalar-field mass $m_\varphi$ is given in units of the reduced Planck mass, with $\tilde{r}=1$AU.}
     \label{fig:malphacountour}
\end{figure}
Fig.~\ref{fig:malphacountour} shows the excluded regions in the $(m_\varphi,\alpha)$ plane from the Cassini and VLBI constraints on $\gamma(\tilde r)$, along with the approximate bound from Mercury's perihelion advance, as summarised in Table~\ref{tab:constraints}. The Cassini bound is the most stringent constraint on $\gamma(\tilde{r})$, requiring $\alpha \lesssim 10^{-6}$ for sufficiently light scalars with $m_\varphi \lesssim 10^{-45}\m$. For larger masses, Yukawa suppression rapidly renders the Cassini and VLBI bounds ineffective. The approximate Mercury bound, by contrast, excludes an intermediate mass band centred around $m_\varphi \sim 10^{-45}\m$, where the Yukawa contribution to the perihelion advance is maximal. In that region, one finds approximately $\alpha \lesssim 10^{-11}$, whereas away from this mass range the Yukawa coupling strength becomes effectively unconstrained.

In what follows, we explore the available parameter space in light of Solar-System experiments. In principle, $\gamma$ can be constrained through measurements of both $\delta t$ and $\delta \theta$. However, the former yield constraints that are stronger by roughly two orders of magnitude (see Fig.~\ref{fig:malphacountour}). We therefore restrict our attention to the results from the Cassini mission. The approximate bound on $\delta\phi$ from Mercury's perihelion advance will be discussed separately. 

Before exploring the available parameter space, we make a few comments that simplify the analysis. Any viable theory must satisfy $A_0>0$, so that $A(\Phi)\hat{R}$ yields the correct sign in the gradient term in the Newtonian limit (see Eq.~\eqref{eq:O2-h00}). The absence of ghosts in the scalar sector requires the effective kinetic coefficient to be positive, namely $16\pi G A_0 B_0 + 3A_1^2(1-\delta_{\textrm{P}}) > 0$ \cite{Jarv:2024krk,Jarv:2025qgo}. In the Palatini formalism, this reduces to $B_0>0$, whereas in the metric formalism $B_0$ may be negative, provided it is not too negative. We do not consider the $B_0<0$ branch here (see Ref.~\cite{Ballardini:2023mzm} for a cosmological analysis of this branch), and instead restrict attention to the subclass with $B_0>0$, for which $\alpha>0$ (unless $A_1=0$, corresponding to the trivial minimally coupled case), and $\chi>0$. Finally, as mentioned in Sec.~\ref{subsubsec:ppnparams} (see also footnote~\ref{wearenotdoneyetMrMarshall}), we consider non-tachyonic potentials with $V_2>0$, which implies $m_{\varphi}^2>0$.

With the above assumptions, Eq.~\eqref{asdfbahierawer} shows that the light-bending parameter $\gamma(\tilde{r})$ increases monotonically from $(1-\alpha)/(1+\alpha)$ to $1$ as $V_2$ increases from 0 to $\infty$, for fixed $A_0$, $A_1$, and $B_0$, in both the metric and Palatini formalisms. Since $\alpha>0$, it follows that $\gamma(\tilde{r})<1$, and therefore it is sufficient to consider only the lower Cassini bound, $\gamma_{\rm Cassini}^{\rm lower}=1-0.2\times 10^{-5}$. If 
\begin{equation}
    \frac{1-\alpha}{1+\alpha}>\gamma_{\rm Cassini}^{\rm lower}~,
\end{equation}
then all values of $V_2$ are allowed. However, if
\begin{equation}
    \frac{1-\alpha}{1+\alpha}<\gamma_{\rm Cassini}^{\rm lower}~,
\end{equation}
then, for fixed $(A_0,A_1,B_0)$, there exists a minimum value $V_2^{\rm min}$ such that all values $V_2<V_2^{\rm min}$ are excluded, since in that case $\gamma(\tilde{r})<\gamma_{\rm Cassini}^{\rm lower}$. This minimum value is given by
\begin{equation}
    V_2^{\rm min} = \frac{B_0}{\chi \tilde{r}^2}\left[\ln{\left(\frac{\alpha(1+\gamma_{\rm Cassini}^{\rm lower})}{1-\gamma_{\rm Cassini}^{\rm lower}}\right)}\right]^2~.
\end{equation}

In what follows, we focus on a few cases of interest and identify the excluded regions in parameter space by specifying $V_2^{\rm min}$. The situation is slightly different for Mercury's perihelion advance, $\delta \phi$. Indeed, Eq.~\eqref{asdfiuabewwer} shows that $\delta\phi$ vanishes in both the light- and heavy-mass limits, and attains a maximum at $m_\varphi p=2$. In other words, for fixed $A_0$, $A_1$, and $B_0$, $\delta \phi$ does not vary monotonically as $V_2$ increases from 0 to $\infty$. Instead, the condition $m_\varphi p=2$ identifies the mass scale about which the exclusion band is centred, leading to an excluded interval $V_{2,-}<V_2<V_{2,+}$, rather than to a single lower bound as in the Cassini case. At the maximum, imposing the bound yields
\begin{equation}
    4\pi\alpha e^{-2} < 1.75\times 10^{-11}~,
\end{equation}
where $e$ denotes Euler's number. This implies
\begin{equation}
    \alpha \lesssim 10^{-11}~,
\end{equation}
in agreement with Fig.~\ref{fig:malphacountour}. A complete parameter scan incorporating the bound on $\delta \phi$ would require introducing two threshold surfaces instead of one, while still relying on an approximate ephemeris-based constraint, rather than on a dedicated fit within the modified theory. Since the main robust conclusions of this work are already captured by the Cassini constraint on $\gamma(\tilde{r})$, we restrict ourselves to a qualitative discussion of the Mercury bound in the $(m_\varphi,\alpha)$ plane.

\subsubsection{\label{sec:NMC-check}Non-Minimal Coupling}

\begin{figure}[h]
         \centering
         \includegraphics[width=0.9\textwidth]{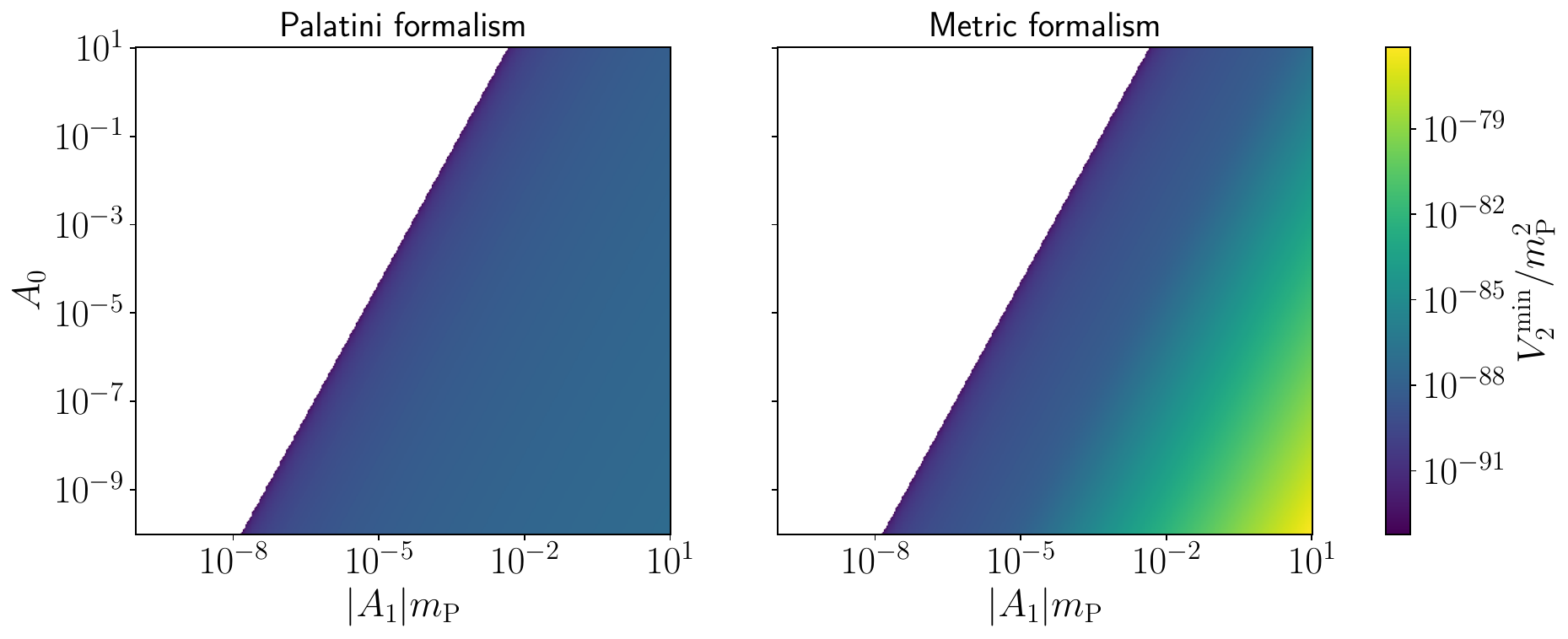}
     \caption{Minimum value of $V_2$ (heat map) below which the Cassini bound is violated, shown as a function of $(A_0,A_1)$ in the Palatini (left) and metric (right) formalisms, for the case of a non-minimally coupled scalar field. The white region is unconstrained, in the sense that the Cassini bound is satisfied for all values of $V_2$. Since both the mass and the Yukawa coupling are symmetric under $A_1\to -A_1$, we plot $\abs{A_1}$ on the $x$-axis.}
     \label{fig:NMCscan}
\end{figure}
Motivated by recent studies pointing to a non-minimal coupling of the quintessence field in light of recent cosmological data~\cite{Wolf:2024stt,Ye:2024ywg,Ye:2024zpk,Ferrari:2025egk,Pan:2025psn,Tiwari:2024gzo,Wolf:2025jed,Myrzakulov:2025jpk,Wang:2025znm,SanchezLopez:2025uzw}, we focus on the case of a scalar field with a canonical kinetic term and a non-minimal coupling $A(\Phi)$, which we leave unspecified. This corresponds to setting $B_0=1$ and $B_1=0$, so that the effective mass and Yukawa coupling in the metric formalism are given by
\begin{equation}
    m_\varphi = \sqrt{\frac{16 \pi G A_0 V_2}{16 \pi G A_0 + 3 A_1^2}}~, \ \ \ \text{and} \ \ \ \alpha = \frac{A_1^2}{16\pi G A_0 + 3 A_1^2}~,
\end{equation}
while in the Palatini formalism they become
\begin{equation}
    \hat{m}_\varphi = \sqrt{V_2}~, \ \ \ \text{and} \ \ \ \hat{\alpha} = \frac{A_1^2}{16\pi G A_0}~.
\end{equation}
Fig.~\ref{fig:NMCscan} shows $V_2^{\rm min}$ for each pair $(A_0,A_1)$ in both Palatini and metric formalisms (note $A_1$ has dimensions of inverse mass). Since both the mass and the Yukawa coupling depend evenly on $A_1$, we display $\abs{A_1}$ on the $x$-axis. In the central region of the figure, the values of $V_2^{\rm min}$ are comparable in the two cases. By contrast, for sufficiently large $\abs{A_1}$ and small $A_0$, the minimum allowed value of $V_2$ is significantly larger in the metric formalism than in the Palatini formalism, in some regions by as much as roughly ten orders of magnitude. In this sense, the Palatini formulation admits a wider allowed region in the $(A_0,A_1,V_2)$ parameter space under the Cassini bound. This can be understood by noting that, in the limit $A_1^2\gg 16\pi G A_0$, the Yukawa couplings and masses satisfy $\hat{\alpha}\gg \alpha$ and $\hat{m}_{\varphi}\gg  m_{\varphi}$, respectively. Since both quantities enter the PPN parameter $\gamma$ via the combination $\alpha e^{- m_\varphi \tilde{r}}$, the much stronger exponential suppression in the Palatini case dominates over the enhancement in the coupling. In other words, because the screening length is much shorter in the Palatini formalism in this limit, $V_2$ can span a wider range of values while still satisfying the Cassini bound.

\subsubsection{\label{sec:BD-check}Brans-Dicke Theory}
\begin{figure}[h]
         \centering
         \includegraphics[width=0.9\textwidth]{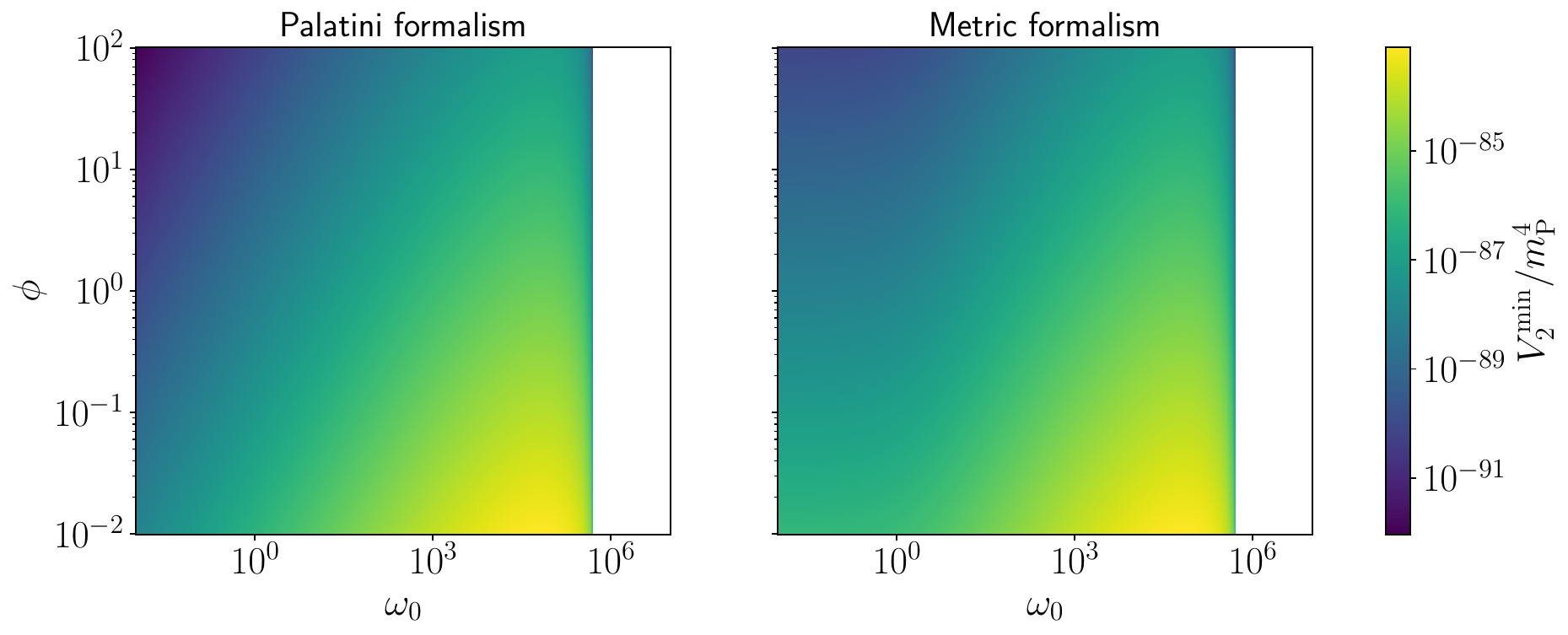}
     \caption{Minimum value of $V_2$ (heat map) below which the Cassini bound is violated, as a function of $(\phi,\omega_0)$ in the Palatini (left) and metric (right) formalisms for Brans-Dicke gravity. The white region is unconstrained, meaning that the Cassini bound is satisfied for all values of $V_2$.}
     \label{fig:BDGscan}
\end{figure}

We consider Brans-Dicke gravity, for which $A(\Phi)=\Phi$ and $B(\Phi)=\omega(\Phi)/(8\pi G \Phi)$. In this theory, both $\Phi$ and $\omega(\Phi)$ are dimensionless. Since $A_1=1$, the effective mass and Yukawa coupling in the metric formalism are
\begin{equation}
    \label{asdjfnbaihbwefawerr}
    m_\varphi = \sqrt{\frac{16 \pi G \phi V_2}{2\omega_0 + 3}}~, \ \ \ \text{and} \ \ \ \alpha = \frac{1}{2\omega_0 + 3}~,
\end{equation}
while in the Palatini formalism they are
\begin{equation}
    \hat{m}_\varphi = \sqrt{\frac{16 \pi G \phi V_2}{2\omega_0 }}~, \ \ \ \text{and} \ \ \ \hat{\alpha} = \frac{1}{2\omega_0 }~.
\end{equation}
We note that $V_2$ now has dimensions $[V_2]=M^4$, since $\Phi$ is dimensionless.

Fig.~\ref{fig:BDGscan} shows $V_2^{\rm min}$ for each pair $(\phi,\omega_0)$ in both the Palatini and metric formalisms. The two panels are nearly identical, differing appreciably only for small values of $\omega_0$. This is expected, since for large $\omega_0$ one has $\alpha\simeq \hat{\alpha}$, and likewise for $\chi$, so that the corresponding values of $V_2^{\rm min}$ become nearly equal. Moreover, the white region corresponds to the part of parameter space in which the massless-limit prediction already satisfies the Cassini bound, so that the experiment imposes no lower bound on $V_2$. Since this condition depends only on $\omega_0$ through $\alpha$, the transition is independent of $\phi$. 

In Brans-Dicke gravity, the Cassini bound has limited power to distinguish between the two formalisms, except at small $\omega_0$. In that region, the figure shows that $V_2^{\rm min}$ is smaller in the Palatini formalism, meaning that for fixed $(\phi,\omega_0)$ the allowed range of $V_2$ is wider in the Palatini case. Finally, we note that Yukawa suppression allows $\omega_0$ to take values much smaller than the usually quoted $\omega_0\gtrsim 10^4$ \cite{Will:2014kxa} for the effectively massless case, provided that $V_2$ is sufficiently large.

\subsubsection{$f(R)$ Gravity}
Turning now to $f(R)$ gravity, it is well known~\cite{Sotiriou:2008rp} that $f(R)$ theories are dynamically equivalent to Brans-Dicke gravity for particular choices of the function $\omega(\Phi)$. In particular, metric $f(R)$ gravity is equivalent to Brans-Dicke gravity in the metric formalism with $\omega(\Phi)=0$, while Palatini $f(\hat{R})$ gravity is equivalent to Brans-Dicke gravity in the metric formalism with $\omega(\Phi)=-3/2$. Since setting $\omega_0=-3/2$ directly causes most expressions to diverge, the corresponding quantities in the PPN formalism must be re-derived; this is done in detail in App.~\ref{sec:appendix-f(R)}. 

\begin{figure}[h]
         \centering
         \includegraphics[width=0.45\textwidth]{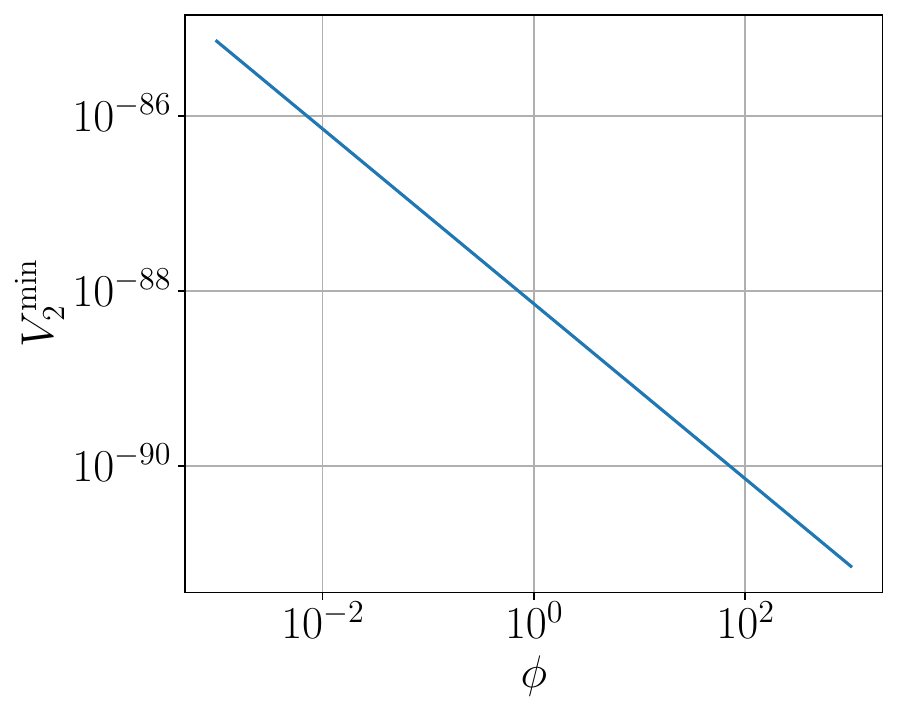}
     \caption{Minimum value of $V_2$ below which the Cassini bound is violated, as a function of $\phi$, in metric $f(R)$ gravity.}
     \label{fig:f(R)scan}
\end{figure}

In our setup, Palatini $f(\hat{R})$ gravity reduces at 1PN order to the GR values of the PPN parameters. This should not be interpreted as a general result, but rather a consequence of our idealised point-particle source in the exterior limit, on an asymptotically flat background. For extended bodies, the non-trivial relation between $\Phi$ and the matter sources leads to matter-sensitive quantities, as shown in Ref.~\cite{Olmo:2005hd} (see also Refs.~\cite{Toniato:2019rrd,Toniato:2021vmt}). In particular, the analysis can depend on the regularity of the matter profile and on the matching conditions across matter boundaries. Addressing these issues for finite-radius matter distributions would require a dedicated study and is left for future work. We therefore focus exclusively on metric $f(R)$ gravity. The effective mass and Yukawa coupling are then
\begin{equation}
    m_\varphi = \sqrt{\frac{16 \pi G \phi V_2}{3}}~, \ \ \ \text{and} \ \ \ \alpha = \frac{1}{3}~.
\end{equation}
Since $\omega_0=0$, the parameter space is two-dimensional, with coordinates $\{\phi,V_2\}$. In Fig.~\ref{fig:f(R)scan} we show $V_2^{\rm min}$ as a function of $\phi$. 

\section{\label{sec:conclusions}Summary and Conclusions}
In this work, we investigated Solar-System constraints on a general scalar-tensor theory with non-minimal couplings in both the metric and Palatini formalisms. Beginning from a Jordan-frame action characterised by arbitrary non-minimal coupling, kinetic, and potential functions, the corresponding field equations were derived in Sec.~\ref{sec:theory}, and the post-Newtonian expansion was constructed within a unified framework in the subsequent section. In this manner, analytical expressions were obtained for the effective scalar mass, the effective gravitational coupling, and the parametrised post-Newtonian parameters $\gamma$ and $\beta$, thereby making explicit the way in which the dependence on the variational principle enters the weak-field phenomenology. In particular, we showed that, at first post-Newtonian order, the scalar field equations depend explicitly on the underlying formalism, with the result, presented in Sec.~\ref{subsubsec:ppnparams}, that the Palatini case generally features a shorter screening length than its metric counterpart.

These results were subsequently confronted with Solar-System observables in Secs.~\ref{sec:observables} and \ref{sec:apps}, with primary emphasis placed on the Cassini bound on the light-deflection parameter $\gamma$, while the advance of Mercury's perihelion was incorporated as a complementary probe. We found that the observational consequences associated with the choice of formalism are model dependent. For a generic non-minimally coupled scalar field endowed with a canonical kinetic term, analysed in Sec.~\ref{sec:NMC-check}, a substantially broader region of parameter space can be accommodated within the Palatini formulation than within the metric one, especially in the regime of small $A_0$ and large $|A_1|$, where the scalar interaction is more strongly Yukawa-suppressed in the Palatini case. In this regime, the shorter screening length associated with the Palatini effective mass may compensate for the enhanced coupling, thereby relaxing the Cassini bound on the parameters of the scalar-field potential. In the opposite limit $m_\varphi r\ll 1$, the Yukawa suppression is ineffective, and fifth-force constraints reduce to strong bounds on the coupling $\alpha$.

In Brans-Dicke gravity, considered in Sec.~\ref{sec:BD-check}, by contrast, the differences between the two formalisms were found to be less pronounced. The available parameter space becomes nearly indistinguishable in both formalisms for large values of the Brans-Dicke parameter $\omega_0$, while appreciable deviations arise only for small $\omega_0$. Even in this case, however, values of $\omega_0$ lying well below the usual effectively massless bound remain admissible as a consequence of Yukawa suppression, provided that the scalar potential is sufficiently steep. 

In $f(R)$ gravity, discussed in Sec.~\ref{sec:apps}, the distinction between the two variational formulations is stark. Metric $f(R)$ gravity, being dynamically equivalent to metric Brans-Dicke gravity with $\omega_0=0$, inherits a non-trivial scalar-mediated Yukawa correction and is therefore constrained by the Cassini bound through a lower limit on the curvature of the scalar potential, encoded in $V_2$. Palatini $f(\hat{R})$ gravity, by contrast, corresponds to the special non-dynamical case $\omega_0=-3/2$, for which the scalar field is determined algebraically by the matter distribution rather than propagating as an independent degree of freedom. Within the point-particle, asymptotically flat exterior setup adopted here, this leads to the general-relativistic post-Newtonian limit, with $\gamma=\beta=1$. In this sense, the usual Yukawa fifth-force bounds on $(\alpha,m_\varphi)$ are not applicable, although other constraints, such as those arising from matter-profile dependence, may still apply. This absence of Yukawa fifth-force constraints is not due to a screening mechanism, but rather to the fact that the scalar field does not propagate in vacuum.

Overall, our analysis provides a unified post-Newtonian framework for the comparison of the metric and Palatini formalisms in scalar-tensor gravity, and clarifies the way Solar-System experiments constrain non-minimal couplings in the presence of a scalar potential. The results presented here are applicable not only to Brans-Dicke models and non-minimally coupled quintessence, but also, more broadly, to late-time dark-energy scenarios in which the scalar degree of freedom remains relevant on cosmological scales, while being locally screened through Yukawa suppression. More generally, our results show that local tests of gravity may, in principle, discriminate between the metric and Palatini realisations of scalar-tensor theories, although such discrimination is expected to be significant only within a restricted region of the parameter space. It would be interesting in future work to combine the local bounds derived here with cosmological data, so that a more comprehensive assessment may be made as to whether the metric and Palatini versions of non-minimally coupled dark-energy models can be distinguished observationally across both astrophysical and cosmological regimes.

\vspace{10mm}

\textbf{Acknowledgements}

\vspace{4mm}

\noindent
A.~K. is supported by the Estonian Research Council grants TARISTU24-TK10, TARISTU24-TK3, and the Center of Excellence program TK202 “Foundations of the Universe”.
S.~S.~L. would like to thank the Indo-French Centre for the Promotion of Advanced Research (IFCPAR/CEFIPRA) for support of the proposal 6704-4 titled `Testing flavors of the early universe beyond vanilla models with cosmological observations’ under the Collaborative Scientific Research Programme. J.~J.~T.~D. acknowledges the financial support provided by FCT-Fundação para a Ciência e Tecnologia (FCT), I.P., through the Strategic Funding UID/04650/2025 and UID/04564/2025 and national funds with DOI identifiers 10.54499 /2023.11681.PEX, 10.54499/2024.00249.CERN funded by measure RE-C06-i06.m02-``Reinfor-cement of funding for International Partnerships in Science, Technology and Innovation'' of the Recovery and Resilience Plan–RRP, within the framework of the financing contract signed between the Recover Portugal Mission Structure (EMRP) and the Foundation for Science and Technology I.P. (FCT), as an intermediate beneficiary, as well as the advanced computing projects 2024.00249.CERN.F1. 

\appendix
\section{\label{app:solution-connection}Solution of the Connection Field Equations}
In this appendix, we solve the connection field equations \eqref{eq:connection-field-eqs} and show how the metric and scalar-field equations, Eqs.~\eqref{eq:metric-connectiononshell} and \eqref{eq:sf-connectiononshell}, respectively, can be obtained in terms of geometric quantities depending only on the Levi-Civita (LC) connection. By writing the cyclic permutations of the free indices $\alpha\mu\nu$ of Eq.~\eqref{eq:connection-field-eqs}, subtracting two of them, and adding the third, the following expression is obtained (see Ref.~\cite{TerenteDiaz:2024uxb} for further details):
\begin{equation}
    \label{eq:sol-kappa-distortion}\tensor{\kappa}{^{\alpha}_{\mu\nu}} = W_{\mu}\delta_{\nu}^{\alpha}-W_{\nu}\delta_{\mu}^{\alpha}+g_{\mu\nu}g^{\alpha\beta}V_{\beta}~,
\end{equation}
where
\begin{eqnarray}
    W_{\mu} &\equiv& \frac{1}{2}\left(\tensor{\kappa}{_{\mu\sigma}^{\sigma}}-\tensor{\kappa}{^{\sigma}_{\sigma\mu}}\right) +\frac{A^{'}}{A}\partial_{\mu}\Phi~,\\
    V_{\mu} &\equiv& \frac{1}{2}\left(\tensor{\kappa}{_{\mu\sigma}^{\sigma}}+\tensor{\kappa}{^{\sigma}_{\sigma\mu}}\right).
\end{eqnarray}
We recall that primes denote derivatives with respect to $\Phi$. Taking traces of Eq.~\eqref{eq:sol-kappa-distortion}, we arrive at
\begin{eqnarray}
    \tensor{\kappa}{^{\sigma}_{\mu\sigma}} &=& V_{\mu}+3W_{\mu}~,\\
    \tensor{\kappa}{^{\sigma}_{\sigma\mu}} &=& V_{\mu}-3W_{\mu}~,\\
    \tensor{\kappa}{_{\mu\sigma}^{\sigma}} &=& 4V_{\mu}~,
\end{eqnarray}
and consequently $W_{\mu} = V_{\mu} = -\frac{A^{'}}{2A}\partial_{\mu}\Phi$. Thus
\begin{equation}
    \tensor{\kappa}{^{\alpha}_{\mu\nu}} = \frac{A^{'}}{2A}\left(\delta_{\mu}^{\alpha}\partial_{\nu}\Phi  -\delta_{\nu}^{\alpha}\partial_{\mu}\Phi  -g_{\mu\nu}g^{\alpha\beta}\partial_{\beta}\Phi\right),
\end{equation}
which is not symmetric in its covariant indices and therefore does not define a metric-compatible connection. However, the torsion induced by this asymmetry can be removed by means of a projective transformation of the form
\begin{equation}
    \label{eq:projective-trans}\tensor{\kappa}{^{\alpha}_{\mu\nu}}(x) \rightarrow \tensor{\kappa}{^{\alpha}_{\mu\nu}}(x)+\delta_{\nu}^{\alpha}C_{\mu}(x)~,
\end{equation}
provided that $C_{\mu}\equiv \frac{A^{'}}{A}\partial_{\mu}\Phi$. 

It can be seen that both the connection field equations \eqref{eq:connection-field-eqs} and the full action in Eq.~\eqref{eq:full_action} are invariant under a generic projective transformation with arbitrary vector $C_{\mu}$.\footnote{For a projective transformation of the form $\hat{\Gamma}^{\alpha}_{\mu\nu}\rightarrow \hat{\Gamma}^{\alpha}_{\mu\nu}+\delta_{\nu}^{\alpha}C_{\mu}$, one finds (see Eq.~\eqref{eq:def-Riccitensor})
\begin{equation}
    \hat{R}_{\mu\nu}\rightarrow \hat{R}_{\mu\nu}+\partial_{\mu}C_{\nu}-\partial_{\nu}C_{\mu}~. 
\end{equation}
Hence, $\hat{R}\equiv g^{\mu\nu}\hat{R}_{\mu\nu}\rightarrow \hat{R}$, and the action \eqref{eq:full_action} remains unchanged, since the independent connection appears only through the Ricci scalar.} We therefore arrive at the equally valid solution
\begin{equation}
    \label{eq:Weyl-conn-trans-LC}\hat{\Gamma}^{\alpha}_{\mu\nu}= \Gamma^{\alpha}_{\mu\nu}+\frac{A^{'}}{2A}\left(\delta_{\mu}^{\alpha}\partial_{\nu}\Phi +\delta_{\nu}^{\alpha}\partial_{\mu}\Phi-g_{\mu\nu}g^{\alpha\beta}\partial_{\beta}\Phi\right),
\end{equation}
which is torsion-free but exhibits non-vanishing non-metricity. This connection is an example of a Weyl connection \cite{Yuan:2013cv}, which can be recast as the LC connection of a conformally rescaled metric via $\tilde{g}_{\mu\nu}(x) \equiv A[\Phi(x)] g_{\mu\nu}(x)$.\footnote{\label{footnoteConfandProj}In fact, when the metric is rescaled in the action \eqref{eq:full_action} in this way, the resulting gravitational sector becomes that of the Einstein-Hilbert action of General Relativity, where the connection is known to coincide with the LC one up to a projective transformation \cite{Dadhich:2012htv,Bernal:2016lhq}. We display the explicit argument in the conformal rescaling formula because such a transformation is not a coordinate transformation \cite{Dabrowski:2008kx}, just as the projective transformations \eqref{eq:projective-trans} are not.} Using Eq.~\eqref{eq:def-Riccitensor}, the definition of the Einstein tensor $\hat{G}_{\mu\nu}\equiv \hat{R}_{\mu\nu}-\frac{1}{2}g_{\mu\nu}\hat{R}$, and the connection in Eq.~\eqref{eq:Weyl-conn-trans-LC}, we obtain
\begin{equation}
    \label{eq:rel-Einstein-tensor-metric-hat}\hat{G}_{\mu\nu} = G_{\mu\nu}-\nabla_{\mu}\partial_{\nu}\ln A +\frac{1}{2}\partial_{\mu}\ln A \partial_{\nu}\ln A+\frac{1}{2}g_{\mu\nu}\left(2\Box \ln A +\frac{1}{2}\partial_{\sigma}\ln A \partial^{\sigma} \ln A\right),
\end{equation}
where the Weyl vector is $\frac{1}{2}\partial_{\mu}\ln A = \frac{A^{'}}{2A} \partial_{\mu}\Phi$ \cite{BeltranJimenez:2014iie}. Since this vector is Abelian, $\hat{G}_{\mu\nu}$ is automatically symmetric, $\hat{G}_{\mu\nu} = \hat{G}_{(\mu\nu)}$. 

Replacing $\hat{G}_{(\mu\nu)}$ from Eq.~\eqref{eq:rel-Einstein-tensor-metric-hat} in Eq.~\eqref{eq:metric-field-eqs-Palatini}, we recover Eq.~\eqref{eq:metric-connectiononshell}. As for the Ricci scalar in the scalar-field equation \eqref{eq:sf-field-eqs-Palatini}, it can be substituted by
\begin{equation}
    \label{aidfbawerawer}
    \hat{R} = R-3\left(\Box \ln A+\frac{1}{2}\partial_{\sigma}\ln A\partial^{\sigma}\ln A\right).
\end{equation}
Hence, one arrives at Eq.~\eqref{eq:sf-connectiononshell}. We remind the reader that unhatted geometric quantities are constructed from the LC connection.    

\section{\label{sec:appendix-postnewtonian-details}The Post-Newtonian Approximation: Detailed Calculations}
For completeness, we provide below detailed calculations of the metric and energy-momentum tensor components, along with other relevant quantities. 

The normalisation condition for the four-velocity $u^{\mu}$ entering the energy-momentum tensor in Eq.~\eqref{eq:EMT-perfect-fluid-first-appearance}, $g_{\mu\nu}u^{\mu}u^{\nu}=-1$, implies\footnote{\label{footnote:negligible-order-O5}For brevity, we omit `$\mathcal{O}(5)$', or whatever the first neglected order may be, in subsequent equations to avoid repetition. Recall that, in the present work, no assumptions concerning time-reversal symmetry or the quasi-staticity of the gravitational field are required at higher orders in the post-Newtonian approximation, since they do not affect the first-order result.}
\begin{eqnarray}
    u^{0}&\simeq& 1+\frac{1}{2}\left[v^2+h_{00}^{(2)}+h_{00}^{(4)}+2h_{0i}^{(3)} v^{i}+h_{ij}^{(2)}v^{i}v^{j}+\frac{3}{4}\left(v^2+h^{(2)}_{00}\right)^2\right],\\
    u^{i}&\simeq&v^{i}\left[1+\frac{1}{2}\left(v^2+h_{00}^{(2)}\right)\right].
\end{eqnarray}
The covariant components are then given by (Latin indices are lowered and raised by $\delta_{ij}$ and $\delta^{ij}$, respectively)
\begin{eqnarray}
    u_0 &\simeq& -1-\frac{1}{2}\left[v^2-h_{00}^{(2)}-h_{00}^{(4)}+h_{ij}^{(2)} v^{i} v^{j}+v^4-\frac{1}{4}\left(v^2-h_{00}^{(2)}\right)^2\right],\\
    u_i &\simeq& v_i \left[1+\frac{1}{2}\left(v^2+h_{00}^{(2)}\right)\right]+h_{0i}^{(3)}+h_{ij}^{(2)} v^{j}~.
\end{eqnarray}
Hence, we can determine the covariant components of the symmetric energy-momentum tensor $T_{\mu\nu}$ to be\footnote{\label{explain-PN-counting-pressure}The PPN counting $p\sim \rho v^2$ follows from the Euler equation along with the assignment $U\sim \mathcal{O}(2)$ for the Newtonian potential, which implies $\nabla p \sim \rho \nabla U \sim \rho v^2/L$ \cite{Anber:2009qp,Will:2014kxa}.}
\begin{eqnarray}
    \label{eq:T00-postNewt}T_{00}&\simeq& \rho\left(1+\varepsilon +v^2-h_{00}^{(2)}\right),\\
    T_{0i}&\simeq&-\rho v_i~, \\
    \label{eq:Tij-postNewt}T_{ij}&\simeq& \rho\left(v_i v_j +v^2 \delta_{ij}\right).
\end{eqnarray}

Additionally, for our purposes, we need the Ricci tensor components, constructed from the Levi-Civita connection, and the Christoffel symbols appearing in the d'Alembertian $\Box$ in Eqs.~\eqref{eq:final_metric-eqs} and \eqref{eq:final_sf-eqs}). We first write the components of the inverse metric (recall that $\eta_{ij} = \delta_{ij}$; see footnote~\ref{footnote:deltaeta}): 
\begin{eqnarray}
    g^{00}&\simeq& -1-h_{00}^{(2)}-h_{00}^{(4)}-\left(h^{(2)}_{00}\right)^2,\\
    g^{0i}&\simeq& \delta^{ij}h_{0j}^{(3)}~,\\
    g^{ij}&\simeq& \delta^{ij}-h^{(2)ij}-h^{(4)ij}+\delta_{ks}h^{(2)ik}h^{(2)js}~,
\end{eqnarray}
where $h^{ij}=\delta^{ik}\delta^{jm}h_{km}$. These are then used to calculate the Christoffel symbols:
\begin{align}
    &\Gamma^{0}_{00} \simeq -\frac{1}{2}\dot{h}_{00}^{(2)}~, \label{eq:Christ-000}\\
    &\Gamma^{0}_{0i} \simeq -\frac{1}{2}\left(\partial_i h_{00}^{(2)}+\partial_i h_{00}^{(4)}+h_{00}^{(2)}\partial_i h_{00}^{(2)}\right),\\
    &\Gamma^{0}_{ij} \simeq \frac{1}{2}\left(\dot{h}^{(2)}_{ij}-\partial_i h_{0j}^{(3)}-\partial_j h_{0i}^{(3)}\right),\\
    &\Gamma^{i}_{00} \simeq -\frac{1}{2}\partial^{i}h_{00}^{(2)}+\delta^{ij}\dot{h}_{0j}^{(3)}-\frac{1}{2}\left(\partial^{i}h_{00}^{(4)}-h^{(2)ij}\partial_j h_{00}^{(2)}\right),\\
    &\Gamma^{i}_{0j} \simeq \frac{1}{2}\delta^{ik}\left(\dot{h}_{kj}^{(2)}-\partial_k h_{0j}^{(3)}+\partial_j h_{0k}^{(3)}\right),\\
    \nonumber&\Gamma^{k}_{ij} \simeq\frac{1}{2}\delta^{ks}\left(\partial_i h_{sj}^{(2)}+\partial_j h_{si}^{(2)}-\partial_s h_{ij}^{(2)}+\partial_i h_{sj}^{(4)}+\partial_j h_{si}^{(4)}-\partial_s h_{ij}^{(4)}\right)-\\
    \label{eq:Christ-kij}&\phantom{--------------------}-\frac{1}{2}h^{(2)ks}\left(\partial_i h_{sj}^{(2)}+\partial_j h_{si}^{(2)}-\partial_s h_{ij}^{(2)}\right), 
\end{align} 
where we keep in mind the symmetry of the lower indices. Overdots denote derivatives with respect to the coordinate time $t$. We note that these quantities are of order $\mathcal{O}(1)$, in accordance with the assumption of quasi-staticity of the gravitational field, meaning that $\partial_{0} \sim v \partial_i$. Using the inverse metric components, and Eqs.~\eqref{eq:T00-postNewt}--\eqref{eq:Tij-postNewt}, we may compute the trace of the matter energy-momentum tensor, $T\equiv g^{\mu\nu}T_{\mu\nu}$, up to $\mathcal{O}(4)$: 
\begin{equation}
    T \simeq -\rho\left(1+\varepsilon-3v^2\right).
\end{equation} 
With the Christoffel symbols at hand, one can readily compute the Ricci tensor components from Eq.~\eqref{eq:def-Riccitensor} for the Levi-Civita connection, recalling that it is torsion-free and metric-compatible, and that the Ricci tensor $R_{\mu\nu}$ is symmetric: 
\begin{align}
    \nonumber&R_{00} \simeq-\frac{1}{2}\partial_i \partial^{i}h_{00}^{(2)}+\partial^{i}\dot{h}_{0i}^{(3)}-\frac{1}{2}\partial_i \partial^{i}h_{00}^{(4)}-\frac{1}{2}\delta^{ij}\ddot{h}^{(2)}_{ij}+\frac{1}{2}\partial_i h^{(2)ij}\partial_j h_{00}^{(2)}+\frac{1}{2}h^{(2)ij}\partial_i \partial_j h_{00}^{(2)}- \\
    &\phantom{--------------------}-\frac{1}{4}\partial_i h_{00}^{(2)} \partial^{i}h_{00}^{(2)}-\frac{1}{4} \delta^{jk}\partial_i h_{jk}^{(2)}\partial^{i}h_{00}^{(2)}~,\\
    &R_{0i} \simeq \frac{1}{2}\partial^{j}\dot{h}^{(2)}_{ij}-\frac{1}{2}\partial_j \partial^{j} h_{0i}^{(3)}+\frac{1}{2}\partial_i \partial^{j}h_{0j}^{(3)}-\frac{1}{2}\delta^{jk}\partial_i \dot{h}^{(2)}_{jk}~,\\
    \nonumber&R_{ij} \simeq \frac{1}{2}\left(\partial_i\partial^{k} h_{kj}^{(2)}+\partial_j \partial^{k} h_{ki}^{(2)}-\partial_k \partial^{k}h_{ij}^{(2)}\right)+\frac{1}{2}\partial_i \partial_j h_{00}^{(2)} -\frac{1}{2}\delta^{ks}\partial_i \partial_j h_{ks}^{(2)}+\frac{1}{2}\left(\partial_i \partial^{k}h_{kj}^{(4)}+\right.\\
    \nonumber&\left.+\partial_j \partial^{k}h_{ki}^{(4)}-\partial_k \partial^{k} h_{ij}^{(4)}\right)+\frac{1}{2}\partial_i \partial_j h_{00}^{(4)} -\frac{1}{2}\delta^{ks}\partial_i \partial_j h_{ks}^{(4)} +\frac{1}{2}\left(\ddot{h}^{(2)}_{ij}-\partial_i \dot{h}^{(3)}_{0j}-\partial_j \dot{h}^{(3)}_{0i}\right)+\\
    \nonumber&+\frac{1}{2}\partial_i h_{ks}^{(2)}\partial_j h^{(2)ks}+\frac{1}{4}\partial_i h_{00}^{(2)}\partial_j h_{00}^{(2)}+\frac{1}{2}h_{00}^{(2)}\partial_i \partial_j h_{00}^{(2)}+\frac{1}{2}h^{(2)ks}\partial_i \partial_j h_{ks}^{(2)} +\frac{1}{4}\delta^{ks}\partial^{r}h_{ks}^{(2)}\left(\partial_i h_{rj}^{(2)}+\right.\\
    \nonumber&\left.+\partial_j h_{ri}^{(2)}-\partial_r h_{ij}^{(2)}\right)-\frac{1}{2}h^{(2)ks}\left(\partial_i \partial_k h_{sj}^{(2)}+\partial_j \partial_k h_{si}^{(2)}-\partial_k \partial_s h_{ij}^{(2)}\right)-\frac{1}{4}\partial^{k}h_{00}^{(2)}\left(\partial_i h_{kj}^{(2)}+\partial_j h_{ki}^{(2)}-\right.\\
    \nonumber&\left.-\partial_k h_{ij}^{(2)}\right)-\frac{1}{2}\partial_k h^{(2)ks}\left(\partial_i h_{sj}^{(2)}+\partial_j h_{si}^{(2)}-\partial_s h_{ij}^{(2)}\right)-\frac{1}{4}\delta^{kp}\delta^{sr}\left(\partial_i h_{ps}^{(2)}+\partial_s h_{pi}^{(2)}-\partial_p h_{is}^{(2)}\right)\times \\
    &\phantom{-----------------------}\times \left(\partial_j h_{rk}^{(2)}+\partial_k h_{rj}^{(2)}-\partial_r h_{jk}^{(2)}\right).
\end{align}

Given that we assume a covariant theory of gravity, the field equations are invariant under arbitrary smooth coordinate transformations \cite{Weinberg:1972kfs}. A gauge choice is therefore possible in order to simplify the Ricci tensor components above, as well as the equations more generally. Under the transformation $x^{\mu}\rightarrow x^{\mu}+\xi^{\mu}(x)$, the metric perturbation transforms linearly as
\begin{equation}
    h_{\mu\nu}(x)\rightarrow h_{\mu\nu}(x)-\partial_{\mu}\xi_{\nu}(x)-\partial_{\nu} \xi_{\mu}(x)~.
\end{equation}
It follows that, whereas $\xi_i \sim \mathcal{O}(2)$, $\xi_0 \sim \mathcal{O}(3)$ because $\partial_0 \sim \mathcal{O}(1)$, as already noted below Eq.~\eqref{eq:Christ-kij}. This means that $h_{00}^{(2)}$ is gauge-invariant, as expected, since it corresponds to the Newtonian potential. The so-called harmonic gauge condition in linearised General Relativity (GR) is \cite{Carroll:1997ar} 
\begin{equation}
    \label{eq:linearised-gauge-Donder}\partial^{\nu}\left(h_{\mu\nu}-\frac{1}{2}\eta_{\mu\nu}\eta^{\sigma\rho}h_{\sigma\rho}\right)=0~,
\end{equation}
meaning that the components of the gauge vector $\xi^{\mu}$ are chosen so that this condition is satisfied after the gauge transformation. This condition ensures that the metric field equations of GR, namely the Einstein field equations, are hyperbolic, thereby making the initial-value (Cauchy) problem well posed \cite{Reula:1998ty,Choquet-Bruhat:2014hta}. In the scalar-tensor theory under consideration, Eq.~\eqref{eq:full_action}, the condition is different in the Jordan frame, but it takes on its standard form in the Einstein frame (see footnote~\ref{footnoteConfandProj}). Pulling this condition back to the Jordan frame by means of a conformal rescaling (see below Eq.~\eqref{eq:Weyl-conn-trans-LC}) gives
\begin{equation}
    \label{eq:linearised-gauge-Donder-modified}g^{\alpha\beta}\Gamma^{\mu}_{\alpha\beta} = g^{\mu\nu}\partial_{\nu}\ln A~.
\end{equation}
This is the modified Donder (harmonic) gauge condition \cite{Fujii:2003pa,Steinwachs:2011zs,Copeland:2021qby}. In GR, $g^{\alpha\beta}\Gamma^{\mu}_{\alpha\beta}=0$, and its linearised version is Eq.~\eqref{eq:linearised-gauge-Donder}.

\section{\label{sec:appendix-geodesics}Photon Trajectories in Scalar-Tensor Gravity}
Here, the trajectory of a photon is derived for the general scalar-tensor action \eqref{eq:full_action}, valid in both the metric and Palatini formalisms, in the field of a point mass $M$ located at the origin, $\rho(r)=M\delta^{(3)}(x,y,z)$. Since the matter action is minimally coupled to gravity and is independent of the Palatini connection, test bodies follow geodesics of the LC connection of the metric. The theory is therefore metric in the sense of Ref.~\cite{Will:2014kxa}. For a photon, the geodesic equation is
\begin{equation}
    \label{asdifuabwer}
    \frac{\text{d}^2x^{\mu}}{\text{d}\lambda^2}+\Gamma^{\mu}_{\alpha\beta}\frac{\text{d}x^\alpha}{\text{d}\lambda}\frac{\text{d}x^\beta}{\text{d}\lambda}=0~,
\end{equation}
where $\lambda$ is an affine parameter along the worldline $x^\mu(\lambda)$. The spatial component of this equation, expressed in terms of the coordinate time $t$, reads
\begin{equation}   
    \label{asdoifuabwerawer}
    \frac{\text{d}^2x^i}{\text{d}t^2}=\left(\Gamma^0_{\mu\nu}\frac{\text{d}x^i}{\text{d}t}-\Gamma^i_{\mu\nu}\right)\frac{\text{d}x^\mu}{\text{d}t}\frac{\text{d}x^\nu}{\text{d}t}~.
\end{equation}
To obtain this equation, the time component of Eq.~\eqref{asdifuabwer} has been used to eliminate derivatives with respect to $\lambda$ from the spatial component. We will also make use of the null-trajectory condition:
\begin{equation}
    g_{\mu\nu}\frac{\text{d}x^\mu}{\text{d}t}\frac{\text{d}x^\nu}{\text{d}t}=0~.
\end{equation}

For null propagation, the 0PN trajectory is of order $\mathcal{O}(0)$, while the 1PN correction is of order $\mathcal{O}(2)$. Accordingly, terms of order $\mathcal{O}(4)$, \textit{i.e.} in the 2PN limit, are neglected in what follows. Using Eqs.~\eqref{eq:g00-isotropic-PPN}--\eqref{eq:gij-isotropic-PPN}, the null condition yields
\begin{equation}
    \label{adsifbawera}
    u^2=\frac{1-2U_{\rm eff}}{1+2\gamma U_{\rm eff}}~.
\end{equation}
As for Eq.~\eqref{asdoifuabwerawer}, we use Eqs.~\eqref{eq:Christ-000}--\eqref{eq:Christ-kij} to recast it in the form
\begin{equation}
    \label{asdifufbuaiweawe}
    \frac{\text{d}u^i}{\text{d}t}=\delta^{ij}\partial_j U_{\rm eff}+u^2\delta^{ij}\partial_j(\gamma U_{\rm eff})-2u^i u^k \partial_k[(1+\gamma) U_{\rm eff}]~,
\end{equation}
where $u^i=\text{d}x^i/\text{d}t$. Expanding Eq.~\eqref{adsifbawera} to 1PN order gives
\begin{equation}
    \label{asdfiaubserawer}
    u^i=\left[1-(1+\gamma)U_{\rm eff}\right]n^i~,
\end{equation}
with $n^i\equiv u^i/u$ denoting a unit vector. Substituting this expression into Eq.~\eqref{asdifufbuaiweawe}, one obtains
\begin{equation}
    \label{aidbfaisdfawer}
    \frac{\text{d}n^i}{\text{d}t}=\delta^{ij}\partial_j[(1+\gamma)U_{\rm eff}]-n^i n^k \partial_k[(1+\gamma)U_{\rm eff}]~.
\end{equation}
We solve this equation order by order by writing 
\begin{equation}
    \label{adfiuaubwewerrwe}
    n^i=k^i+q^i~,
\end{equation}
where $k^i$ corresponds to 0PN and $q^i$ to 1PN. At 0PN, Eq.~\eqref{aidbfaisdfawer} reads
\begin{equation}
    \frac{\text{d}k^i}{\text{d}t}=0~,
\end{equation}
showing that $k^i$ is indeed constant and $\delta_{ij}k^ik^j=1$. Eq.~\eqref{asdfiaubserawer} then becomes
\begin{equation}
    \frac{\text{d}x^i}{\text{d}t}=k^i~,
\end{equation}
from which it follows that
\begin{equation}
    \label{asdfiuaubwerawer}
    x^i(t)=x^i_0 + k^i(t-t_0)~.
\end{equation}
In this way, $k^i$ represents the wavevector of the unperturbed photon trajectory, emitted from $x_0^i$ at the initial time $t_0$. At 1PN, Eq.~\eqref{aidbfaisdfawer} reads
\begin{equation}
    \frac{\text{d}q^i}{\text{d}t}=\left(\delta^{ij}-k^i k^j \right)\partial_j[(1+\gamma)U_{\rm eff}]~.
\end{equation}
Although both $\gamma$ and $U_{\rm eff}$ depend on the coordinates in principle (see Eqs.~\eqref{asdofubaiwerawer} and~\eqref{asdfiabwerawer}), straightforward algebra shows that their combination simplifies to
\begin{equation}
    (1+\gamma)U_{\rm eff}=(1+\gamma)G_{\rm eff}\frac{M}{r}=\frac{2 G M}{A_0 r}~,
\end{equation}
so that
\begin{equation}
    \label{adfiabweirawer}
    \partial_i[(1+\gamma)U_{\rm eff}]=-\frac{2G M}{A_0 r^3}\delta_{ij}x^j=-\frac{2G M}{A_0 r^3}\delta_{ij}\left[x^j_0 + k^j(t-t_0)\right],
\end{equation}
where we have used Eq.~\eqref{asdfiuaubwerawer} in the last step. Putting everything together, the equation to be solved is
\begin{equation}
    \label{asdfiaubwer}
    \frac{\text{d}q^i}{\text{d}t}=-\frac{2G M}{A_0 r^3}b^i~,
\end{equation}
with
\begin{equation}
    b^i\equiv x_0^i-k^i(k_j x^j_0)
\end{equation}
denoting the impact-parameter vector relative to the point-mass source at the origin, which satisfies $k_ib^i=0$. Defining $y(t)$ via
\begin{equation}
    r(t)=\sqrt{\delta_{ij}x^i x^j}=\sqrt{b^2 + \left[k^ix_0^i+(t-t_0)\right]^2}\equiv \sqrt{b^2 + y(t)^2}~,
\end{equation}
we can integrate Eq.~\eqref{asdfiaubwer} as
\begin{align}
    q^i&=-\frac{2GM}{A_0}b^i\int_{t_0}^t\frac{\text{d}t'}{\left[b^2+y^2(t')\right]^{3/2}}=-\frac{2GM}{A_0}b^i\int_{y_0}^y\frac{\text{d}y'}{\left[b^2+(y')^2\right]^{3/2}}=\nonumber\\
    &\phantom{---------}=-\frac{2GM}{A_0}\frac{b^i}{b^2}\left(\frac{y(t)}{r(t)}-\frac{y_0}{r_0}\right)=-\frac{2GM}{A_0}\frac{b^i}{b^2}\left(\frac{k_j x^j(t)}{r(t)}-\frac{k_jx_0^j}{r_0}\right),\label{asdofbawerawer}
\end{align}
where $y_0=k^ix_0^i$. 

Finally, upon substituting Eq.~\eqref{adfiuaubwewerrwe} into Eq.~\eqref{asdfiaubserawer}, one obtains
\begin{equation}
    u^i=\left[1-(1+\gamma)U_{\rm eff}\right]k^i+q^i=\left[1-\frac{2GM}{A_0 r}\right]k^i+q^i~.
\end{equation}
Using Eq.~\eqref{asdofbawerawer} and performing the straightforward integrations yield the general 1PN photon trajectory
\begin{align}
    &x^i(t)=x_0^i+k^i(t-t_0)-\nonumber\\
    \phantom{--}&-\frac{2GM}{A_0} k^i\ln{\left(\frac{r(t)+k_jx^j_0+(t-t_0)}{r_0+k_jx_0^j}\right)}-\frac{2GM}{A_0}\frac{b^i}{b^2}\left[r(t)-r_0-\frac{k_jx_0^j}{r_0}(t-t_0)\right],
\end{align}
where
\begin{equation}
    r_0\equiv r(t_0)~.
\end{equation}

\section{\label{App:fifthforces}Timelike Geodesics and Fifth Forces in Scalar-Tensor Gravity} 
In this appendix we derive the perihelion precession rate used in the main text. To this end, the action for a test particle of mass $m$ is first expanded as
\begin{align}
    S&=-m\int \text{d}\tau =-m\int \text{d}t\sqrt{-g_{\mu\nu}\frac{\text{d}x^\mu}{\text{d}t}\frac{\text{d}x^\nu}{\text{d}t}}=\nonumber\\
    &\phantom{-------------}=-m\int \text{d}t\sqrt{1-2U_{\rm eff}+2\beta U_{\rm eff}^2-u^2-2u^2\gamma U_{\rm eff}}=\nonumber\\
    &\phantom{--}=-m\int\text{d}t\left[1-U_{\rm eff}-\frac{u^2}{2}+\left(\beta-\frac{1}{2}\right)U_{\rm eff}^2-\left(\gamma + \frac{1}{2}\right)U_{\rm eff}u^2-\frac{u^4}{8}+\mathcal{O}(u^6)\right],
\end{align}
where we have used the metric components \eqref{eq:g00-isotropic-PPN} and \eqref{eq:gij-isotropic-PPN}. The 1PN Lagrangian takes on the form
\begin{equation}
    \frac{L}{m}=-1+U_{\rm eff}+\frac{u^2}{2}-\left(\beta-\frac{1}{2}\right)U_{\rm eff}^2+\left(\gamma + \frac{1}{2}\right)U_{\rm eff}u^2+\frac{u^4}{8}~.
\end{equation}
Equivalently, the calculation may be carried out in spherical coordinates, using the metric\footnote{Here, and in Eq.~\eqref{aisdfbaiyhsdfawer}, $\phi$, denoting the azimuthal angle, should not be confused with the $\mathcal{O}(0)$ term in the PPN expansion of $\Phi$.}
\begin{equation}
    \text{d}s^2=\left(-1+2U_{\rm eff}-2\beta U_{\rm eff}^2\right)\text{d}t^2+\left(1+2\gamma U_{\rm eff}\right)\left(\text{d}r^2 + r^2\text{d}\theta^2 + r^2\sin^2\theta \text{d}\phi^2\right).
\end{equation}
Since $\gamma(r)$, $\beta(r)$, and $U_{\rm eff}(r)$ are functions of the radial coordinate $r$ only, the problem remains spherically symmetric, and angular momentum is conserved. One may therefore choose the orbital plane to be $\theta=\pi/2$, which yields the Lagrangian 
\begin{align}
    \frac{L}{m}&=-1+U_{\rm eff}+\frac{1}{2}\left(\dot{r}^2+r^2\dot{\phi}^2\right)-\left(\beta-\frac{1}{2}\right)U_{\rm eff}^2+\left(\gamma + \frac{1}{2}\right)U_{\rm eff}\left(\dot{r}^2+r^2\dot{\phi}^2\right)+\nonumber\\
    &\phantom{---------------------------}+\frac{1}{8}\left(\dot{r}^2+r^2\dot{\phi}^2\right)^2.
\end{align}

At 0PN, the angular equation of motion yields
\begin{equation}
    l\equiv r^2\dot{\phi}=\text{const.}~,
\end{equation}
showing the conservation of the orbital angular momentum per unit mass. At the same order, the radial equation gives
\begin{equation}
    \ddot{r}-r\dot{\phi}^2=\frac{\text{d}U_{\rm eff}}{\text{d}r}~.
\end{equation}
Upon introducing the variable $w(\phi)\equiv 1/r(\phi)$, the radial equation becomes
\begin{equation}
    \frac{\text{d}^2w}{\text{d}\phi^2}+w=-\frac{1}{l^2 w^2}\frac{\text{d}U_{\rm eff}}{\text{d}r}\Bigg{|}_{r=1/w}~.
\end{equation}
Using Eq. \eqref{asdfiabwerawer}, we obtain
\begin{equation}
    \label{asdifbawefawerwer}
    \frac{\text{d}^2w}{\text{d}\phi^2}+w=\frac{GM}{A_0 l^2}\left[1+f(1/w)\right],
\end{equation}
where
\begin{equation}
    \label{asdfiujujabnwerawer}
    f(r)=(1+m_\varphi r)\alpha e^{-m_\varphi r}~.
\end{equation}
The function $f(r)$ encodes a Yukawa-like force that induces deviations from the pure Newtonian inverse-square law. Under the assumption $\abs{f(r)}\ll1$, Eq.~\eqref{asdifbawefawerwer} may be solved perturbatively. At leading order, we arrive at
\begin{equation}
    \frac{\text{d}^2w_0}{\text{d}\phi^2}+w_0=\frac{1}{p}~,
\end{equation}
with
\begin{equation}
    p\equiv \frac{A_0 l^2}{GM}~.
\end{equation}
The solution is
\begin{equation}
    w_0(\phi)=\frac{1}{p}\left[1 + e \cos{(\phi-\phi_0)}\right],
\end{equation}
where $e$ is the eccentricity of the unperturbed Keplerian orbit, and $G/A_0$ is the corresponding constant inverse-square coupling. In the unperturbed orbit, the quantity $p$ is related to $e$ by
\begin{equation}
    p=a(1-e^2)~,
\end{equation}
with $a$ the semimajor axis. Since only the leading correction to the orbital precession is of interest, it is sufficient to expand the Yukawa term about the characteristic radius $r=p$ of the unperturbed Kepler orbit, so that
\begin{equation}
    w=\frac{1}{p}+\delta w~.
\end{equation}
Expanding $f(1/w)$ about $w=1/p$ yields
\begin{equation}
    f(1/w)=f(p)-p^2 f'(p)\delta w + \mathcal{O}(\delta w^2)~.
\end{equation}
We then obtain Eq. \eqref{asdifbawefawerwer} at first order:
\begin{equation}
    \frac{\text{d}^2(\delta w)}{\text{d}\phi^2} + \left[1 + p f'(p)\right]\delta w = \frac{f(p)}{p}~,
\end{equation}
whose solution is
\begin{equation}
    \delta w = \frac{f(p)}{p(1+p f'(p))} + w_{\rm e}\cos{\left[\omega(\phi-\phi_0)\right]}~,
\end{equation}
with
\begin{equation}
    \label{asdofuubawerawer}
    \omega^2 = 1 + p f'(p)~.
\end{equation}

Collecting the above results, the first-order solution is
\begin{equation}
    w(\phi)=\frac{1}{p}\left[1+\frac{f(p)}{\omega^2}\right]+ w_{\rm e}\cos{\left[\omega(\phi-\phi_0)\right]}~.
\end{equation}
Perihelion occurs when the orbital radius $r(\phi)$ is minimal, that is, when $w(\phi)$ is maximal. This takes place at $\omega(\phi-\phi_0)=2\pi n$, with $n\in\mathds{Z}$. The angular separation between successive perihelia is therefore
\begin{equation}
    \Delta \phi=\frac{2\pi}{\omega}~.
\end{equation}
Hence, if $f'(p)=0$, then $\Delta \phi=2\pi$ and the orbit closes. If, on the other hand, $f'(p)\neq 0$, the orbit does not close, and the perihelion advance per orbit is given by 
\begin{equation}
    \delta \phi = \Delta \phi - 2\pi=\frac{2\pi (1-\omega)}{\omega}\simeq -\pi p f'(p)~,
\end{equation}
where Eq.~\eqref{asdofuubawerawer} has been Taylor-expanded about $\omega=1$, assuming only a small deviation from unity. Using Eq.~\eqref{asdfiujujabnwerawer}, one finally obtains
\begin{equation}
        \delta \phi =  \pi \alpha (m_{\varphi}p)^2 e^{-m_\varphi p}~.
\end{equation}
For a massless field, $m_\varphi = 0$, the Yukawa-induced precession vanishes.

\section{\label{sec:appendix-f(R)}$f(R)$ Gravity as a Scalar-Tensor Theory}
This appendix shows the equivalence of both metric and Palatini $f(R)$ gravity with metric Brans-Dicke gravity \cite{Sotiriou:2008rp}. 

Let us start with metric $f(R)$ gravity, defined by the action
\begin{equation}
    \label{asdifubawerawerwer}
    S=\frac{1}{16 \pi G}\int \text{d}^4x \sqrt{-g}f(R)+S_{\rm m}[g_{\mu\nu},\Psi]~,
\end{equation}
where $\Psi$ denotes the collection of matter fields. This action is dynamically equivalent to
\begin{equation}
    \label{asidufubawerasda}
    S=\frac{1}{16 \pi G}\int \text{d}^4x \sqrt{-g}\left[f(\psi)+f'(\psi)(R-\psi)\right]+S_{\rm m}[g_{\mu\nu},\Psi]~.
\end{equation}
Indeed, varying Eq. \eqref{asidufubawerasda} with respect to $\psi$ yields
\begin{equation}
    f''(\psi)(R-\psi)=0~.
\end{equation}
Provided that $f''(\psi)\neq 0$, this implies that $R=\psi$. Plugging this back into Eq.~\eqref{asidufubawerasda} recovers Eq.~\eqref{asdifubawerawerwer}. Now, after redefining $\psi$ as $\Phi = f'(\psi)$, Eq.~\eqref{asidufubawerasda} can be rewritten as
\begin{equation}
    \label{asdfbaipwbeqwerweqw}
    S=\frac{1}{16\pi G}\int \text{d}^4x \sqrt{-g}\left[\Phi R - V(\Phi)\right]+S_{\rm m}[g_{\mu\nu},\Psi]~,
\end{equation}
with the potential defined as
\begin{equation}
    V(\Phi)\equiv\Phi \psi(\Phi) - f[\psi(\Phi)]~.
\end{equation}
Eq.~\eqref{asdfbaipwbeqwerweqw} corresponds to the action of Brans-Dicke gravity in the metric formalism, with $A(\Phi)=\Phi$ and $B(\Phi)=0$, namely $\omega(\Phi)=0$. 

Palatini $f(\hat{R})$ gravity is defined by the action 
\begin{equation}
    S=\frac{1}{16 \pi G}\int \text{d}^4x \sqrt{-g}f(\hat{R})+S_{\rm m}[g_{\mu\nu},\Psi]~,
\end{equation}
where $\hat{R}$ is constructed from the independent connection $\hat{\Gamma}^\mu_{\alpha\beta}$ via $\hat{R}=g^{\mu\nu}\hat{R}_{\mu\nu}$, with $\hat{R}_{\mu\nu}$ depending only on $\hat{\Gamma}^\mu_{\alpha\beta}$. Following the same procedure as in the metric formalism, one obtains 
\begin{equation}
    S=\frac{1}{16\pi G}\int \text{d}^4x \sqrt{-g}\left[\Phi \hat{R} - V(\Phi)\right]+S_{\rm m}[g_{\mu\nu},\Psi]~.
\end{equation}
Using Eq.~\eqref{aidfbawerawer} to rewrite $\hat{R}$ in terms of $R$, and recalling that $\Box\ln{\Phi}=\left(\Box\Phi\right)/\Phi-(\partial\Phi)^2/\Phi^2$, we find, upon discarding the total derivative,
\begin{equation}
    \label{awidfbawhefawe}
    S=\frac{1}{16\pi G}\int \text{d}^4x \sqrt{-g}\left[\Phi R + \frac{3}{2\Phi}g^{\mu\nu}\partial_\mu \Phi \partial_\nu \Phi - V(\Phi)\right]+S_{\rm m}[g_{\mu\nu},\Psi]~,
\end{equation}
which corresponds to the action of Brans-Dicke gravity in the metric formalism, now with $A(\Phi)=\Phi$ and $B(\Phi)=-3/(16\pi G \Phi)$, namely $\omega(\Phi)=-3/2$. Variation of Eq.~\eqref{awidfbawhefawe} with respect to $g^{\mu\nu}$ and $\Phi$ yields the corresponding metric and scalar-field equations. Using the trace of the former in the latter, one arrives at \cite{Sotiriou:2008rp}
\begin{equation}
    \label{asdiabwierawerrawe}
    2 V(\Phi) - \Phi V'(\Phi) = 8\pi G T~,
\end{equation}
where $T$ is the trace of the energy-momentum tensor, derived from $S_{\rm m}$ in Eq.~\eqref{awidfbawhefawe}. Eq.~\eqref{asdiabwierawerrawe} shows that $\Phi$ is not a dynamical field, but is instead determined algebraically by the matter content of the theory through $T$. Palatini $f(\hat{R})$ gravity therefore constitutes a special case that must be treated separately from our general PPN analysis of scalar-tensor gravity. 

Using $A_0=\phi$, $A_1=1$, $B_0=-3/(16\pi G \phi)$, and $\delta_{\rm P}=1$ in Eq.~\eqref{eq:sf-eq-O2-isotropic}, the Laplacian term cancels. For the formal point-particle source $\rho(r)=\frac{M}{4\pi r^2}\delta(r)$, one then obtains
\begin{equation}
    \varphi^{(2)}(r)=\frac{M}{8\pi \phi V_2 r^2}\delta(r)~.
\end{equation}
$\varphi^{(2)}(r)$ vanishes for $r>0$ and therefore contributes only through a contact term localised at the origin, which we discard consistently with the rest of our treatment. Turning to the metric perturbations, Eqs.~\eqref{eq:O2-h00} and \eqref{eq:theotherequationatO2} then become
\begin{align}
    &\partial_i\partial^i h_{00}^{(2)} =  -\frac{2GM}{\phi r^2}\delta(r) + \frac{M}{8\pi \phi^2 V_2}\partial_i\partial^i \left(\frac{\delta(r)}{r^2}\right),\\
    &\partial_k\partial^k h_{ij}^{(2)} =  -\frac{2GM}{\phi r^2}\delta(r)\delta_{ij} - \frac{M}{8\pi \phi^2 V_2}\partial_k\partial^k \left(\frac{\delta(r)}{r^2}\right)\delta_{ij}~,
\end{align}
whose formal solutions are
\begin{align}
    &h_{00}^{(2)}=\frac{2 G M}{\phi r} + \frac{M}{8\pi\phi^2 V_2 r^2}\delta(r)~,\\
    &h_{ij}^{(2)}=\frac{2 G M}{\phi r}\delta_{ij} - \frac{M}{8\pi\phi^2 V_2 r^2}\delta(r)\delta_{ij}~,
\end{align}
where the integration constants have been fixed by imposing decay at infinity and the delta-source matching condition. As mentioned above, we exclude the second term in both solutions, since we are interested in the vacuum regime. Hence,
\begin{equation}
    G_{\rm eff}=\frac{G}{\phi}~,
\end{equation}
and
\begin{equation}
    \gamma = 1~.
\end{equation}
The scalar-field equation at $\mathcal{O}(4)$, Eq.~\eqref{eq:sol-no-homogeneous-varphi4}, with $\varphi^{(2)}=0$, gives 
\begin{equation}
    \varphi^{(4)}=0~.
\end{equation}
With this, the $00$ component of the metric field equations at order $\mathcal{O}(4)$ reads
\begin{equation}
    \partial_i\partial^i h_{00}^{(4)}=-\frac{4G^2 M^2}{\phi^2}\frac{1}{r^4}~,
\end{equation}
whose solution is 
\begin{equation}
    h_{00}^{(4)}=\frac{c_1}{r}-\frac{2G^2M^2}{\phi^2 r^2}~.
\end{equation}
Setting $c_1=0$, consistently with the removal of contact terms, we then compare with Eq.~\eqref{eq:g00-isotropic-PPN} and obtain
\begin{equation}
    \beta = 1~.
\end{equation}
Thus, in the point-particle limit and in the vacuum exterior region $r>0$, the contact terms localised at the origin do not contribute to the metric, yielding $G_{\rm eff}=G/\phi$, $\gamma = 1$, and $\beta=1$. This should not be taken as a general proof that Palatini $f(\hat{R})$ gravity is equivalent to GR at 1PN order, but only as the outcome of the idealised delta-source exterior limit in an asymptotically flat background. For extended bodies, the scalar field remains non-dynamical, but is still determined algebraically by the local matter distribution, so that the weak-field metric may depend non-trivially on matter-sensitive quantities, including the regularity of the density profile and the treatment of matter boundaries, as shown in Ref.~\cite{Olmo:2005hd} (see also Refs.~\cite{Toniato:2019rrd,Toniato:2021vmt}).

It is useful to make explicit why the Palatini $f(\hat R)$ scalar does not
generate a Yukawa correction in the exterior region. In the vacuum exterior of the point source considered here, $T=0$, Eq. \eqref{asdiabwierawerrawe} reduces to
\begin{equation}
    2V(\Phi)-\Phi V'(\Phi)=0~,
\end{equation}
which is an algebraic equation for $\Phi$ implying that
\begin{equation}
        \Phi=\Phi_0=\mathrm{const~.}
\end{equation}
This constant $\Phi_0$ is fixed to the vacuum branch compatible with the asymptotic value $\phi$ used in the PPN expansion. Since $\partial_\mu\Phi=0$, there is no scalar profile proportional to $e^{-m_\varphi r}/r$, and hence no Yukawa force mediated by the Palatini $f(\hat R)$ scalar. The usual fifth-force bounds on a propagating Yukawa mode therefore do not apply to this exterior, non-dynamical sector.

\bibliography{PalatiniPPN}

@article{Wolf:2025jed,
    author = "Wolf, William J. and Garc{\'\i}a-Garc{\'\i}a, Carlos and Anton, Theodore and Ferreira, Pedro G.",
    title = "{Assessing Cosmological Evidence for Nonminimal Coupling}",
    eprint = "2504.07679",
    archivePrefix = "arXiv",
    primaryClass = "astro-ph.CO",
    doi = "10.1103/jysf-k72m",
    journal = "Phys. Rev. Lett.",
    volume = "135",
    number = "8",
    pages = "081001",
    year = "2025"
}

@article{Fienga:2022bns,
    author = "Fienga, A. and Bernus, L. and Minazzoli, O. and Hees, A. and Bigot, L. and Herrera, C. and Mariani, V. and Di Ruscio, A. and Durante, D. and Mary, D.",
    title = "{INPOP Planetary ephemerides and applications in the frame of the BepiColombo mission including new constraints on the graviton mass and dilaton parameters}",
    eprint = "2211.04881",
    archivePrefix = "arXiv",
    primaryClass = "gr-qc",
    month = "11",
    year = "2022"
}

@article{Verma:2013ata,
    author = "Verma, Ashok and Fienga, Agnes and Laskar, Jacques and Manche, Herve and Gastineau, Mickael",
    title = "{Use of MESSENGER radioscience data to improve planetary ephemeris and to test general relativity}",
    eprint = "1306.5569",
    archivePrefix = "arXiv",
    primaryClass = "astro-ph.EP",
    doi = "10.1051/0004-6361/201322124",
    journal = "Astron. Astrophys.",
    volume = "561",
    pages = "A115",
    year = "2014"
}

@article{Williams:2004qba,
    author = "Williams, James G. and Turyshev, Slava G. and Boggs, Dale H.",
    title = "{Progress in lunar laser ranging tests of relativistic gravity}",
    eprint = "gr-qc/0411113",
    archivePrefix = "arXiv",
    doi = "10.1103/PhysRevLett.93.261101",
    journal = "Phys. Rev. Lett.",
    volume = "93",
    pages = "261101",
    year = "2004"
}

@article{Zhang:2016njn,
    author = "Zhang, Xing and Zhao, Wen and Huang, He and Cai, Yifu",
    title = "{Post-Newtonian parameters and cosmological constant of screened modified gravity}",
    eprint = "1603.09450",
    archivePrefix = "arXiv",
    primaryClass = "gr-qc",
    doi = "10.1103/PhysRevD.93.124003",
    journal = "Phys. Rev. D",
    volume = "93",
    number = "12",
    pages = "124003",
    year = "2016"
}

@article{Emtsova:2019qsl,
    author = "Emtsova, Elena D. and Hohmann, Manuel",
    title = "{Post-Newtonian limit of scalar-torsion theories of gravity as analogue to scalar-curvature theories}",
    eprint = "1909.09355",
    archivePrefix = "arXiv",
    primaryClass = "gr-qc",
    doi = "10.1103/PhysRevD.101.024017",
    journal = "Phys. Rev. D",
    volume = "101",
    number = "2",
    pages = "024017",
    year = "2020"
}

@article{Hohmann:2015kra,
    author = "Hohmann, Manuel",
    title = "{Parametrized post-Newtonian limit of Horndeski{\textquoteright}s gravity theory}",
    eprint = "1506.04253",
    archivePrefix = "arXiv",
    primaryClass = "gr-qc",
    doi = "10.1103/PhysRevD.92.064019",
    journal = "Phys. Rev. D",
    volume = "92",
    number = "6",
    pages = "064019",
    year = "2015"
}

@article{Toniato:2021vmt,
    author = "Toniato, J{\'u}nior D. and Rodrigues, Davi C.",
    title = "{Post-Newtonian {\ensuremath{\gamma}}-like parameters and the gravitational slip in scalar-tensor and f(R) theories}",
    eprint = "2106.12542",
    archivePrefix = "arXiv",
    primaryClass = "gr-qc",
    doi = "10.1103/PhysRevD.104.044020",
    journal = "Phys. Rev. D",
    volume = "104",
    number = "4",
    pages = "044020",
    year = "2021"
}

@article{Scharer:2014kya,
    author = {Sch{\"a}rer, Andreas and Ang{\'e}lil, Raymond and Bondarescu, Ruxandra and Jetzer, Philippe and Lundgren, Andrew},
    title = "{Testing scalar-tensor theories and parametrized post-Newtonian parameters in Earth orbit}",
    eprint = "1410.7914",
    archivePrefix = "arXiv",
    primaryClass = "gr-qc",
    doi = "10.1103/PhysRevD.90.123005",
    journal = "Phys. Rev. D",
    volume = "90",
    number = "12",
    pages = "123005",
    year = "2014"
}

@article{Jarv:2014laa,
    author = {J{\"a}rv, L. and Kuusk, P. and Saal, M. and Vilson, O.},
    editor = "Paal, E. and Kuusk, P. and Stolin, A.",
    title = "{Parametrizations in scalar-tensor theories of gravity and the limit of general relativity}",
    eprint = "1501.07781",
    archivePrefix = "arXiv",
    primaryClass = "gr-qc",
    doi = "10.1088/1742-6596/532/1/012011",
    journal = "J. Phys. Conf. Ser.",
    volume = "532",
    pages = "012011",
    year = "2014"
}

@article{Clifton:2011jh,
    author = "Clifton, Timothy and Ferreira, Pedro G. and Padilla, Antonio and Skordis, Constantinos",
    title = "{Modified Gravity and Cosmology}",
    eprint = "1106.2476",
    archivePrefix = "arXiv",
    primaryClass = "astro-ph.CO",
    doi = "10.1016/j.physrep.2012.01.001",
    journal = "Phys. Rept.",
    volume = "513",
    pages = "1--189",
    year = "2012"
}

@article{Hohmann:2017qje,
    author = {Hohmann, Manuel and Sch{\"a}rer, Andreas},
    title = "{Post-Newtonian parameters {\ensuremath{\gamma}} and {\ensuremath{\beta}} of scalar-tensor gravity for a homogeneous gravitating sphere}",
    eprint = "1708.07851",
    archivePrefix = "arXiv",
    primaryClass = "gr-qc",
    doi = "10.1103/PhysRevD.96.104026",
    journal = "Phys. Rev. D",
    volume = "96",
    number = "10",
    pages = "104026",
    year = "2017"
}

@article{Hohmann:2013rba,
    author = "Hohmann, Manuel and Jarv, Laur and Kuusk, Piret and Randla, Erik",
    title = "{Post-Newtonian parameters $\gamma$ and $\beta$ of scalar-tensor gravity with a general potential}",
    eprint = "1309.0031",
    archivePrefix = "arXiv",
    primaryClass = "gr-qc",
    doi = "10.1103/PhysRevD.88.084054",
    journal = "Phys. Rev. D",
    volume = "88",
    number = "8",
    pages = "084054",
    year = "2013",
    note = "[Erratum: Phys.Rev.D 89, 069901 (2014)]"
}

@article{Jarv:2014hma,
    author = {J{\"a}rv, Laur and Kuusk, Piret and Saal, Margus and Vilson, Ott},
    title = "{Invariant quantities in the scalar-tensor theories of gravitation}",
    eprint = "1411.1947",
    archivePrefix = "arXiv",
    primaryClass = "gr-qc",
    doi = "10.1103/PhysRevD.91.024041",
    journal = "Phys. Rev. D",
    volume = "91",
    number = "2",
    pages = "024041",
    year = "2015"
}

@article{Boisseau:2000pr,
    author = "Boisseau, B. and Esposito-Farese, Gilles and Polarski, D. and Starobinsky, Alexei A.",
    title = "{Reconstruction of a scalar tensor theory of gravity in an accelerating universe}",
    eprint = "gr-qc/0001066",
    archivePrefix = "arXiv",
    reportNumber = "CPT-99-P-3917",
    doi = "10.1103/PhysRevLett.85.2236",
    journal = "Phys. Rev. Lett.",
    volume = "85",
    pages = "2236",
    year = "2000"
}

@article{Rossi:2019lgt,
    author = "Rossi, Massimo and Ballardini, Mario and Braglia, Matteo and Finelli, Fabio and Paoletti, Daniela and Starobinsky, Alexei A. and Umilt{\`a}, Caterina",
    title = "{Cosmological constraints on post-Newtonian parameters in effectively massless scalar-tensor theories of gravity}",
    eprint = "1906.10218",
    archivePrefix = "arXiv",
    primaryClass = "astro-ph.CO",
    doi = "10.1103/PhysRevD.100.103524",
    journal = "Phys. Rev. D",
    volume = "100",
    number = "10",
    pages = "103524",
    year = "2019"
}

@article{SanchezLopez:2025uzw,
    author = "S{\'a}nchez L{\'o}pez, Samuel and Karam, Alexandros and Hazra, Dhiraj Kumar",
    title = "{Non-Minimally Coupled Quintessence in Light of DESI}",
    eprint = "2510.14941",
    archivePrefix = "arXiv",
    primaryClass = "astro-ph.CO",
    month = "10",
    year = "2025"
}

@article{Braglia:2020iik,
    author = "Braglia, Matteo and Ballardini, Mario and Emond, William T. and Finelli, Fabio and Gumrukcuoglu, A. Emir and Koyama, Kazuya and Paoletti, Daniela",
    title = "{Larger value for $H_0$ by an evolving gravitational constant}",
    eprint = "2004.11161",
    archivePrefix = "arXiv",
    primaryClass = "astro-ph.CO",
    doi = "10.1103/PhysRevD.102.023529",
    journal = "Phys. Rev. D",
    volume = "102",
    number = "2",
    pages = "023529",
    year = "2020"
}

@article{Kozak:2018vlp,
    author = "Kozak, Aleksander and Borowiec, Andrzej",
    title = "{Palatini frames in scalar{\textendash}tensor theories of gravity}",
    eprint = "1808.05598",
    archivePrefix = "arXiv",
    primaryClass = "hep-th",
    doi = "10.1140/epjc/s10052-019-6836-y",
    journal = "Eur. Phys. J. C",
    volume = "79",
    number = "4",
    pages = "335",
    year = "2019"
}

@article{Jarv:2020qqm,
    author = {J{\"a}rv, Laur and Karam, Alexandros and Kozak, Aleksander and Lykkas, Angelos and Racioppi, Antonio and Saal, Margus},
    title = "{Equivalence of inflationary models between the metric and Palatini formulation of scalar-tensor theories}",
    eprint = "2005.14571",
    archivePrefix = "arXiv",
    primaryClass = "gr-qc",
    doi = "10.1103/PhysRevD.102.044029",
    journal = "Phys. Rev. D",
    volume = "102",
    number = "4",
    pages = "044029",
    year = "2020"
}

@article{Jarv:2024krk,
    author = {J{\"a}rv, Laur and Karamitsos, Sotirios and Saal, Margus},
    title = "{Global portraits of nonminimal inflation: Metric and Palatini formalism}",
    eprint = "2401.12314",
    archivePrefix = "arXiv",
    primaryClass = "gr-qc",
    doi = "10.1103/PhysRevD.109.084073",
    journal = "Phys. Rev. D",
    volume = "109",
    number = "8",
    pages = "084073",
    year = "2024"
}

@article{Nguyen:2024fpw,
    author = "Nguyen, Hoang Ky and Chauvineau, Bertrand",
    title = "{Impact of star pressure on $\gamma $ in modified gravity beyond post-Newtonian approach}",
    eprint = "2404.00094",
    archivePrefix = "arXiv",
    primaryClass = "gr-qc",
    doi = "10.1140/epjc/s10052-024-13080-2",
    journal = "Eur. Phys. J. C",
    volume = "84",
    number = "7",
    pages = "710",
    year = "2024"
}

@article{Zhang:2023nil,
    author = "Zhang, Xing and Wang, Bo and Niu, Rui",
    title = "{Constraining the attractive fifth force in the general free scalar{\textendash}tensor gravity with solar system experiments}",
    eprint = "2305.06752",
    archivePrefix = "arXiv",
    primaryClass = "gr-qc",
    doi = "10.1140/epjc/s10052-024-12723-8",
    journal = "Eur. Phys. J. C",
    volume = "84",
    number = "4",
    pages = "381",
    year = "2024"
}

@article{Olmo:2005hc,
    author = "Olmo, Gonzalo J.",
    title = "{Post-Newtonian constraints on f(R) cosmologies in metric and Palatini formalism}",
    eprint = "gr-qc/0505135",
    archivePrefix = "arXiv",
    doi = "10.1103/PhysRevD.72.083505",
    journal = "Phys. Rev. D",
    volume = "72",
    pages = "083505",
    year = "2005"
}

@article{Olmo:2005hd,
    author = "Olmo, Gonzalo J.",
    title = "{Post-Newtonian constraints on f(R) cosmologies in Palatini formalism}",
    eprint = "gr-qc/0505136",
    archivePrefix = "arXiv",
    month = "5",
    year = "2005"
}

@article{Dimopoulos:2020pas,
    author = "Dimopoulos, Konstantinos and S{\'a}nchez L{\'o}pez, Samuel",
    title = "{Quintessential inflation in Palatini $f(R)$ gravity}",
    eprint = "2012.06831",
    archivePrefix = "arXiv",
    primaryClass = "gr-qc",
    doi = "10.1103/PhysRevD.103.043533",
    journal = "Phys. Rev. D",
    volume = "103",
    number = "4",
    pages = "043533",
    year = "2021"
}

@article{Dimopoulos:2022tvn,
    author = "Dimopoulos, Konstantinos and Karam, Alexandros and S{\'a}nchez L{\'o}pez, Samuel and Tomberg, Eemeli",
    title = "{Modelling Quintessential Inflation in Palatini-Modified Gravity}",
    eprint = "2203.05424",
    archivePrefix = "arXiv",
    primaryClass = "gr-qc",
    doi = "10.3390/galaxies10020057",
    journal = "Galaxies",
    volume = "10",
    number = "2",
    pages = "57",
    year = "2022"
}

@article{Dimopoulos:2022rdp,
    author = "Dimopoulos, Konstantinos and Karam, Alexandros and S{\'a}nchez L{\'o}pez, Samuel and Tomberg, Eemeli",
    title = "{Palatini R $^{2}$ quintessential inflation}",
    eprint = "2206.14117",
    archivePrefix = "arXiv",
    primaryClass = "gr-qc",
    doi = "10.1088/1475-7516/2022/10/076",
    journal = "JCAP",
    volume = "10",
    pages = "076",
    year = "2022"
}

@article{TerenteDiaz:2023kgc,
    author = "Terente D{\'\i}az, Jos{\'e} Jaime and Dimopoulos, Konstantinos and Kar{\v{c}}iauskas, Mindaugas and Racioppi, Antonio",
    title = "{Quintessence in the Weyl-Gauss-Bonnet model}",
    eprint = "2310.08128",
    archivePrefix = "arXiv",
    primaryClass = "gr-qc",
    doi = "10.1088/1475-7516/2024/02/040",
    journal = "JCAP",
    volume = "02",
    pages = "040",
    year = "2024"
}

@article{Dimopoulos:2021xld,
    author = "Dimopoulos, Konstantinos",
    title = "{Jointly modelling Cosmic Inflation and Dark Energy}",
    eprint = "2106.14966",
    archivePrefix = "arXiv",
    primaryClass = "gr-qc",
    doi = "10.1088/1742-6596/2105/1/012001",
    journal = "J. Phys. Conf. Ser.",
    volume = "2105",
    number = "1",
    pages = "012001",
    year = "2021"
}

@article{Dimopoulos:2025fuq,
    author = {Dimopoulos, Konstantinos and Dioguardi, Christian and H{\"u}tsi, Gert and Racioppi, Antonio},
    title = "{Quintessential inflation in Palatini F(R,~X) gravity}",
    eprint = "2503.21610",
    archivePrefix = "arXiv",
    primaryClass = "gr-qc",
    doi = "10.1140/epjp/s13360-025-07025-1",
    journal = "Eur. Phys. J. Plus",
    volume = "140",
    number = "11",
    pages = "1109",
    year = "2025"
}

@article{Koivisto:2005yk,
    author = "Koivisto, Tomi",
    title = "{Covariant conservation of energy momentum in modified gravities}",
    eprint = "gr-qc/0505128",
    archivePrefix = "arXiv",
    doi = "10.1088/0264-9381/23/12/N01",
    journal = "Class. Quant. Grav.",
    volume = "23",
    pages = "4289--4296",
    year = "2006"
}

@article{Harko:2011nh,
    author = "Harko, Tiberiu and Koivisto, Tomi S. and Lobo, Francisco S. N. and Olmo, Gonzalo J.",
    title = "{Metric-Palatini gravity unifying local constraints and late-time cosmic acceleration}",
    eprint = "1110.1049",
    archivePrefix = "arXiv",
    primaryClass = "gr-qc",
    doi = "10.1103/PhysRevD.85.084016",
    journal = "Phys. Rev. D",
    volume = "85",
    pages = "084016",
    year = "2012"
}

@article{Koivisto:2005yc,
    author = "Koivisto, Tomi and Kurki-Suonio, Hannu",
    title = "{Cosmological perturbations in the palatini formulation of modified gravity}",
    eprint = "astro-ph/0509422",
    archivePrefix = "arXiv",
    reportNumber = "HIP-2005-38-TH",
    doi = "10.1088/0264-9381/23/7/009",
    journal = "Class. Quant. Grav.",
    volume = "23",
    pages = "2355--2369",
    year = "2006"
}

@article{Kubota:2020ehu,
    author = "Kubota, Mio and Oda, Kin-Ya and Shimada, Keigo and Yamaguchi, Masahide",
    title = "{Cosmological Perturbations in Palatini Formalism}",
    eprint = "2010.07867",
    archivePrefix = "arXiv",
    primaryClass = "hep-th",
    doi = "10.1088/1475-7516/2021/03/006",
    journal = "JCAP",
    volume = "03",
    pages = "006",
    year = "2021"
}

@article{Will:2014kxa,
    author = "Will, Clifford M.",
    title = "{The Confrontation between General Relativity and Experiment}",
    eprint = "1403.7377",
    archivePrefix = "arXiv",
    primaryClass = "gr-qc",
    doi = "10.12942/lrr-2014-4",
    journal = "Living Rev. Rel.",
    volume = "17",
    pages = "4",
    year = "2014"
}

@article{Martin:2005bp,
    author = "Martin, Jerome and Schimd, Carlo and Uzan, Jean-Philippe",
    title = "{Testing for w{\ensuremath{<}} -1 in the solar system}",
    eprint = "astro-ph/0510208",
    archivePrefix = "arXiv",
    doi = "10.1103/PhysRevLett.96.061303",
    journal = "Phys. Rev. Lett.",
    volume = "96",
    pages = "061303",
    year = "2006"
}

@article{Nordtvedt:1970uv,
    author = "Nordtvedt, Jr., Kenneth",
    title = "{PostNewtonian metric for a general class of scalar tensor gravitational theories and observational consequences}",
    doi = "10.1086/150607",
    journal = "Astrophys. J.",
    volume = "161",
    pages = "1059--1067",
    year = "1970"
}

@article{Chehab:2026vdx,
    author = "Chehab, Thomas and Minazzoli, Olivier",
    title = "{On the numerical evaluation of the `exact' Post-Newtonian parameters in Brans-Dicke and Entangled Relativity theories}",
    eprint = "2602.11811",
    archivePrefix = "arXiv",
    primaryClass = "gr-qc",
    month = "2",
    year = "2026"
}

@article{Joyce:2014kja,
    author = "Joyce, Austin and Jain, Bhuvnesh and Khoury, Justin and Trodden, Mark",
    title = "{Beyond the Cosmological Standard Model}",
    eprint = "1407.0059",
    archivePrefix = "arXiv",
    primaryClass = "astro-ph.CO",
    doi = "10.1016/j.physrep.2014.12.002",
    journal = "Phys. Rept.",
    volume = "568",
    pages = "1--98",
    year = "2015"
}

@article{Barrow:1996kc,
    author = "Barrow, John D. and Parsons, Paul",
    title = "{The Behavior of cosmological models with varying G}",
    eprint = "gr-qc/9607072",
    archivePrefix = "arXiv",
    reportNumber = "SUSSEX-AST-96-7-5",
    doi = "10.1103/PhysRevD.55.1906",
    journal = "Phys. Rev. D",
    volume = "55",
    pages = "1906--1936",
    year = "1997"
}

@article{Esposito-Farese:2000pbo,
    author = "Esposito-Farese, Gilles and Polarski, D.",
    title = "{Scalar tensor gravity in an accelerating universe}",
    eprint = "gr-qc/0009034",
    archivePrefix = "arXiv",
    reportNumber = "CPT-00-PE-4053",
    doi = "10.1103/PhysRevD.63.063504",
    journal = "Phys. Rev. D",
    volume = "63",
    pages = "063504",
    year = "2001"
}

@article{Esposito-Farese:2004azw,
    author = "Esposito-Farese, Gilles",
    editor = "Martins, Carlos J. A. P. and Avelino, Pedro P. and Costa, M. S. and Mack, Katherine and Mota, M. F. and Parry, M.",
    title = "{Tests of scalar-tensor gravity}",
    eprint = "gr-qc/0409081",
    archivePrefix = "arXiv",
    doi = "10.1063/1.1835173",
    journal = "AIP Conf. Proc.",
    volume = "736",
    number = "1",
    pages = "35--52",
    year = "2004"
}

@article{Faraoni:2004qd,
    author = "Faraoni, Valerio",
    title = "{Singularities in scalar tensor gravity}",
    eprint = "gr-qc/0403020",
    archivePrefix = "arXiv",
    doi = "10.1103/PhysRevD.70.047301",
    journal = "Phys. Rev. D",
    volume = "70",
    pages = "047301",
    year = "2004"
}

@article{Gannouji:2006jm,
    author = "Gannouji, Radouane and Polarski, David and Ranquet, Andre and Starobinsky, Alexei A.",
    title = "{Scalar-Tensor Models of Normal and Phantom Dark Energy}",
    eprint = "astro-ph/0606287",
    archivePrefix = "arXiv",
    doi = "10.1088/1475-7516/2006/09/016",
    journal = "JCAP",
    volume = "09",
    pages = "016",
    year = "2006"
}

@article{Jarv:2008eb,
    author = "Jarv, Laur and Kuusk, Piret and Saal, Margus",
    title = "{Scalar-tensor cosmologies: Fixed points of the Jordan frame scalar field}",
    eprint = "0807.2159",
    archivePrefix = "arXiv",
    primaryClass = "gr-qc",
    doi = "10.1103/PhysRevD.78.083530",
    journal = "Phys. Rev. D",
    volume = "78",
    pages = "083530",
    year = "2008"
}

@article{Adam:2025kve,
    author = "Adam, Husam and Hertzberg, Mark P. and Jim{\'e}nez-Aguilar, Daniel and Khan, Iman",
    title = "{Comparing Minimal and Non-Minimal Quintessence Models to 2025 DESI Data}",
    eprint = "2509.13302",
    archivePrefix = "arXiv",
    primaryClass = "astro-ph.CO",
    month = "9",
    year = "2025"
}

@article{Lu:2019qgk,
    author = "Lu, Jianbo and Wang, Yan and Zhao, Xin",
    title = "{Linearized modified gravity theories and gravitational waves physics in the GBD theory}",
    eprint = "1904.01734",
    archivePrefix = "arXiv",
    primaryClass = "gr-qc",
    doi = "10.1016/j.physletb.2019.05.051",
    journal = "Phys. Lett. B",
    volume = "795",
    pages = "129--134",
    year = "2019"
}

@article{Lu:2020eux,
    author = "Lu, Jianbo and Li, Jiachun and Guo, Hui and Zhuang, Zhitong and Zhao, Xin",
    title = "{Linearized physics and gravitational-waves polarizations in the Palatini formalism of GBD theory}",
    eprint = "2012.02343",
    archivePrefix = "arXiv",
    primaryClass = "gr-qc",
    doi = "10.1016/j.physletb.2020.135985",
    journal = "Phys. Lett. B",
    volume = "811",
    pages = "135985",
    year = "2020"
}

@article{Dyadina:2019yon,
    author = "Dyadina, P. I. and Labazova, S. P. and Alexeyev, S. O.",
    title = "{Post-Newtonian limit of hybrid metric-Palatini f(R)-gravity}",
    eprint = "1907.06919",
    archivePrefix = "arXiv",
    primaryClass = "gr-qc",
    doi = "10.1134/S0044451019110087",
    month = "7",
    year = "2019"
}

@article{Fan:2015rha,
    author = "Fan, Yize and Wu, Puxun and Yu, Hongwei",
    title = "{Cosmological perturbations of non-minimally coupled quintessence in the metric and Palatini formalisms}",
    doi = "10.1016/j.physletb.2015.05.005",
    journal = "Phys. Lett. B",
    volume = "746",
    pages = "230--236",
    year = "2015"
}

@article{Palatini:1919ffw,
    author = "Palatini, Attilio",
    title = "{Deduzione invariantiva delle equazioni gravitazionali dal principio di Hamilton}",
    doi = "10.1007/BF03014670",
    journal = "Rend. Circ. Mat. Palermo",
    volume = "43",
    number = "1",
    pages = "203--212",
    year = "1919"
}

@article{Ferraris:1982wci,
    author = "Ferraris, M. and Francaviglia, M. and Reina, C.",
    title = "{Variational formulation of general relativity from 1915 to 1925 {\textquotedblleft}Palatini's method{\textquotedblright} discovered by Einstein in 1925}",
    doi = "10.1007/BF00756060",
    journal = "Gen. Rel. Grav.",
    volume = "14",
    number = "3",
    pages = "243--254",
    year = "1982"
}

@article{Brans:1961sx,
    author = "Brans, C. and Dicke, R. H.",
    editor = "Hsu, Jong-Ping and Fine, D.",
    title = "{Mach's principle and a relativistic theory of gravitation}",
    doi = "10.1103/PhysRev.124.925",
    journal = "Phys. Rev.",
    volume = "124",
    pages = "925--935",
    year = "1961"
}

@article{Antoniadis:2022cqh,
    author = "Antoniadis, Ignatios and Guillen, Anthony and Tamvakis, Kyriakos",
    title = "{Late time acceleration in Palatini gravity}",
    eprint = "2207.13732",
    archivePrefix = "arXiv",
    primaryClass = "gr-qc",
    doi = "10.1007/JHEP11(2022)144",
    journal = "JHEP",
    volume = "11",
    pages = "144",
    year = "2022"
}

@article{Verner:2020gfa,
    author = "Verner, Sarunas",
    title = "{Quintessential Inflation in Palatini Gravity}",
    eprint = "2010.11201",
    archivePrefix = "arXiv",
    primaryClass = "gr-qc",
    doi = "10.1088/1475-7516/2021/04/001",
    journal = "JCAP",
    volume = "04",
    year = "2021"
}

@article{SupernovaSearchTeam:1998fmf,
    author = "Riess, Adam G. and others",
    collaboration = "Supernova Search Team",
    title = "{Observational evidence from supernovae for an accelerating universe and a cosmological constant}",
    eprint = "astro-ph/9805201",
    archivePrefix = "arXiv",
    doi = "10.1086/300499",
    journal = "Astron. J.",
    volume = "116",
    pages = "1009--1038",
    year = "1998"
}

@article{SupernovaCosmologyProject:1998vns,
    author = "Perlmutter, S. and others",
    collaboration = "Supernova Cosmology Project",
    title = "{Measurements of $\Omega$ and $\Lambda$ from 42 High Redshift Supernovae}",
    eprint = "astro-ph/9812133",
    archivePrefix = "arXiv",
    reportNumber = "LBNL-41801, LBL-41801",
    doi = "10.1086/307221",
    journal = "Astrophys. J.",
    volume = "517",
    pages = "565--586",
    year = "1999"
}

@article{Adelberger:2003zx,
    author = "Adelberger, E. G. and Heckel, Blayne R. and Nelson, A. E.",
    title = "{Tests of the gravitational inverse square law}",
    eprint = "hep-ph/0307284",
    archivePrefix = "arXiv",
    doi = "10.1146/annurev.nucl.53.041002.110503",
    journal = "Ann. Rev. Nucl. Part. Sci.",
    volume = "53",
    pages = "77--121",
    year = "2003"
}

@article{Bertotti:2003rm,
    author = "Bertotti, B. and Iess, L. and Tortora, P.",
    title = "{A test of general relativity using radio links with the Cassini spacecraft}",
    doi = "10.1038/nature01997",
    journal = "Nature",
    volume = "425",
    pages = "374--376",
    year = "2003"
}

@article{Fienga:2019uds,
    author = "Fienga, A. and Bigot, L. and Mary, D. and Deram, P. and Di Ruscio, A. and Bernus, L. and Gastineau, M. and Laskar, J.",
    title = "{Evolution of INPOP planetary ephemerides and Bepi-Colombo simulations}",
    eprint = "2111.04499",
    archivePrefix = "arXiv",
    primaryClass = "astro-ph.EP",
    doi = "10.1017/S1743921321001277",
    journal = "IAU Symp.",
    volume = "364",
    pages = "31--51",
    year = "2019"
}

@article{Genova:2018mjp,
    author = "Genova, Antonio and Mazarico, Erwan and Goossens, Sander and Lemoine, Frank G. and Neumann, Gregory A. and Smith, David E. and Zuber, Maria T.",
    title = "{Solar system expansion and strong equivalence principle as seen by the NASA MESSENGER mission}",
    doi = "10.1038/s41467-017-02558-1",
    journal = "Nature Commun.",
    volume = "9",
    number = "1",
    pages = "289",
    year = "2018"
}

@phdthesis{TerenteDiaz:2024uxb,
    author = "Terente D{\'\i}az, Jos{\'e} Jaime",
    title = "{Aspects of Modified Gravity in the Early and Late Universe}",
    school = "Madrid U.",
    year = "2024"
}

@article{BeltranJimenez:2016wxw,
    author = "Beltran Jimenez, Jose and Heisenberg, Lavinia and Koivisto, Tomi S.",
    title = "{Cosmology for quadratic gravity in generalized Weyl geometry}",
    eprint = "1602.07287",
    archivePrefix = "arXiv",
    primaryClass = "hep-th",
    reportNumber = "NORDITA-2016-10",
    doi = "10.1088/1475-7516/2016/04/046",
    journal = "JCAP",
    volume = "04",
    pages = "046",
    year = "2016"
}

@article{Shimada:2018lnm,
    author = "Shimada, Keigo and Aoki, Katsuki and Maeda, Kei-ichi",
    title = "{Metric-affine Gravity and Inflation}",
    eprint = "1812.03420",
    archivePrefix = "arXiv",
    primaryClass = "gr-qc",
    reportNumber = "WU-AP/1808/18",
    doi = "10.1103/PhysRevD.99.104020",
    journal = "Phys. Rev. D",
    volume = "99",
    number = "10",
    pages = "104020",
    year = "2019"
}

@article{Iosifidis:2023eom,
    author = "Iosifidis, Damianos and Hehl, Friedrich W.",
    title = "{Motion of test particles in spacetimes with torsion and nonmetricity}",
    eprint = "2310.15595",
    archivePrefix = "arXiv",
    primaryClass = "gr-qc",
    doi = "10.1016/j.physletb.2024.138498",
    journal = "Phys. Lett. B",
    volume = "850",
    pages = "138498",
    year = "2024"
}

@article{Vitagliano:2010sr,
    author = "Vitagliano, Vincenzo and Sotiriou, Thomas P. and Liberati, Stefano",
    title = "{The dynamics of metric-affine gravity}",
    eprint = "1008.0171",
    archivePrefix = "arXiv",
    primaryClass = "gr-qc",
    doi = "10.1016/j.aop.2011.02.008",
    journal = "Annals Phys.",
    volume = "326",
    pages = "1259--1273",
    year = "2011",
    note = "[Erratum: Annals Phys. 329, 186--187 (2013)]"
}

@inbook{Iosifidis:2021pta,
    author = "Iosifidis, Damianos and Saridakis, Emmanuel N.",
    title = "{Metric-Affine Gravity}",
    doi = "10.1007/978-3-030-83715-0_10",
    year = "2021"
}

@phdthesis{JimenezCano:2021rlu,
    author = "Jim{\'e}nez Cano, Alejandro",
    title = "{Metric-affine Gauge theories of gravity. Foundations and new insights}",
    eprint = "2201.12847",
    archivePrefix = "arXiv",
    primaryClass = "gr-qc",
    school = "Granada U., Theor. Phys. Astrophys.",
    year = "2021"
}

@book{Faraoni:2004pi,
    author = "Faraoni, Valerio",
    title = "{Cosmology in scalar tensor gravity}",
    doi = "10.1007/978-1-4020-1989-0",
    isbn = "978-1-4020-1988-3",
    year = "2004"
}

@article{Carroll:1997ar,
    author = "Carroll, Sean M.",
    title = "{Lecture notes on general relativity}",
    eprint = "gr-qc/9712019",
    archivePrefix = "arXiv",
    reportNumber = "NSF-ITP-97-147",
    month = "12",
    year = "1997"
}

@article{Yuan:2013cv,
    author = "Yuan, Fang-Fang and Huang, Yong-Chang",
    title = "{A modified variational principle for gravity in the modified Weyl geometry}",
    eprint = "1301.1316",
    archivePrefix = "arXiv",
    primaryClass = "gr-qc",
    doi = "10.1088/0264-9381/30/19/195008",
    journal = "Class. Quant. Grav.",
    volume = "30",
    pages = "195008",
    year = "2013"
}

@article{BeltranJimenez:2014iie,
    author = "Beltran Jimenez, Jose and Koivisto, Tomi S.",
    title = "{Extended Gauss-Bonnet gravities in Weyl geometry}",
    eprint = "1402.1846",
    archivePrefix = "arXiv",
    primaryClass = "gr-qc",
    reportNumber = "NORDITA-2014-14",
    doi = "10.1088/0264-9381/31/13/135002",
    journal = "Class. Quant. Grav.",
    volume = "31",
    pages = "135002",
    year = "2014"
}

@article{Dabrowski:2008kx,
    author = "Dabrowski, Mariusz P. and Garecki, Janusz and Blaschke, David B.",
    title = "{Conformal transformations and conformal invariance in gravitation}",
    eprint = "0806.2683",
    archivePrefix = "arXiv",
    primaryClass = "gr-qc",
    doi = "10.1002/andp.200810331",
    journal = "Annalen Phys.",
    volume = "18",
    pages = "13--32",
    year = "2009"
}

@article{Dadhich:2012htv,
    author = "Dadhich, Naresh and Pons, Josep M.",
    title = "{On the equivalence of the Einstein-Hilbert and the Einstein-Palatini formulations of general relativity for an arbitrary connection}",
    eprint = "1010.0869",
    archivePrefix = "arXiv",
    primaryClass = "gr-qc",
    doi = "10.1007/s10714-012-1393-9",
    journal = "Gen. Rel. Grav.",
    volume = "44",
    pages = "2337--2352",
    year = "2012"
}

@article{Bernal:2016lhq,
    author = "Bernal, Antonio N. and Janssen, Bert and Jimenez-Cano, Alejandro and Orejuela, Jose Alberto and Sanchez, Miguel and Sanchez-Moreno, Pablo",
    title = "{On the (non-)uniqueness of the Levi-Civita solution in the Einstein{\textendash}Hilbert{\textendash}Palatini formalism}",
    eprint = "1606.08756",
    archivePrefix = "arXiv",
    primaryClass = "gr-qc",
    doi = "10.1016/j.physletb.2017.03.001",
    journal = "Phys. Lett. B",
    volume = "768",
    pages = "280--287",
    year = "2017"
}

@book{Weinberg:1972kfs,
    author = "Weinberg, Steven",
    title = "{Gravitation and Cosmology}: {Principles and Applications of the General Theory of Relativity}",
    isbn = "978-0-471-92567-5, 978-0-471-92567-5",
    publisher = "John Wiley and Sons",
    address = "New York",
    year = "1972"
}

@article{Toniato:2017wmk,
    author = "Toniato, J{\'u}nior D. and Rodrigues, Davi C. and de Almeida, {\'A}lefe O. F. and Bertini, Nicolas",
    title = "{Will-Nordtvedt PPN formalism applied to renormalization group extensions of general relativity}",
    eprint = "1706.09032",
    archivePrefix = "arXiv",
    primaryClass = "gr-qc",
    doi = "10.1103/PhysRevD.96.064034",
    journal = "Phys. Rev. D",
    volume = "96",
    number = "6",
    pages = "064034",
    year = "2017"
}

@article{Reula:1998ty,
    author = "Reula, Oscar A.",
    title = "{Hyperbolic methods for Einstein's equations}",
    doi = "10.12942/lrr-1998-3",
    journal = "Living Rev. Rel.",
    volume = "1",
    pages = "3",
    year = "1998"
}

@article{Choquet-Bruhat:2014hta,
    author = "Choquet-Bruhat, Yvonne",
    title = "{Beginnings of the Cauchy problem for Einstein{\textquoteright}s field equations}",
    eprint = "1410.3490",
    archivePrefix = "arXiv",
    primaryClass = "gr-qc",
    doi = "10.4310/sdg.2015.v20.n1.a1",
    journal = "Surveys Diff. Geom.",
    volume = "20",
    number = "1",
    pages = "1--16",
    year = "2015"
}

@article{Copeland:2021qby,
    author = "Copeland, Edmund J. and Millington, Peter and Mu{\~n}oz, Sergio Sevillano",
    title = "{Fifth forces and broken scale symmetries in the Jordan frame}",
    eprint = "2111.06357",
    archivePrefix = "arXiv",
    primaryClass = "hep-th",
    doi = "10.1088/1475-7516/2022/02/016",
    journal = "JCAP",
    volume = "02",
    number = "02",
    pages = "016",
    year = "2022"
}

@article{Steinwachs:2011zs,
    author = "Steinwachs, Christian F. and Kamenshchik, Alexander Yu.",
    title = "{One-loop divergences for gravity non-minimally coupled to a multiplet of scalar fields: calculation in the Jordan frame. I. The main results}",
    eprint = "1101.5047",
    archivePrefix = "arXiv",
    primaryClass = "gr-qc",
    doi = "10.1103/PhysRevD.84.024026",
    journal = "Phys. Rev. D",
    volume = "84",
    pages = "024026",
    year = "2011"
}

@book{Fujii:2003pa,
    author = "Fujii, Y. and Maeda, K.",
    title = "{The scalar-tensor theory of gravitation}",
    doi = "10.1017/CBO9780511535093",
    isbn = "978-0-521-03752-5, 978-0-521-81159-0, 978-0-511-02988-2",
    publisher = "Cambridge University Press",
    series = "Cambridge Monographs on Mathematical Physics",
    month = "7",
    year = "2007"
}

@book{CANTATA:2021asi,
    author = "Akrami, Yashar and others",
    editor = "Saridakis, Emmanuel N. and Lazkoz, Ruth and Salzano, Vincenzo and Vargas Moniz, Paulo and Capozziello, Salvatore and Beltr{\'a}n Jim{\'e}nez, Jose and De Laurentis, Mariafelicia and Olmo, Gonzalo J.",
    collaboration = "CANTATA",
    title = "{Modified Gravity and Cosmology. An Update by the CANTATA Network}",
    eprint = "2105.12582",
    archivePrefix = "arXiv",
    primaryClass = "gr-qc",
    doi = "10.1007/978-3-030-83715-0",
    isbn = "978-3-030-83714-3, 978-3-030-83717-4, 978-3-030-83715-0",
    publisher = "Springer",
    year = "2021"
}

@article{Barreiro:1999zs,
    author = "Barreiro, T. and Copeland, Edmund J. and Nunes, N. J.",
    title = "{Quintessence arising from exponential potentials}",
    eprint = "astro-ph/9910214",
    archivePrefix = "arXiv",
    reportNumber = "SUSX-TH-016",
    doi = "10.1103/PhysRevD.61.127301",
    journal = "Phys. Rev. D",
    volume = "61",
    pages = "127301",
    year = "2000"
}

@article{Anber:2009qp,
    author = "Anber, Mohamed M. and Aydemir, Ufuk and Donoghue, John F.",
    title = "{Breaking Diffeomorphism Invariance and Tests for the Emergence of Gravity}",
    eprint = "0911.4123",
    archivePrefix = "arXiv",
    primaryClass = "gr-qc",
    doi = "10.1103/PhysRevD.81.084059",
    journal = "Phys. Rev. D",
    volume = "81",
    pages = "084059",
    year = "2010"
}

@article{Jarv:2025qgo,
    author = {J{\"a}rv, Laur and Kraiko, Dmitri},
    title = "{Global portraits of inflation in nonsingular variables}",
    eprint = "2503.07544",
    archivePrefix = "arXiv",
    primaryClass = "gr-qc",
    doi = "10.1140/epjc/s10052-025-14411-7",
    journal = "Eur. Phys. J. C",
    volume = "85",
    number = "6",
    pages = "715",
    year = "2025"
}

@article{Ballardini:2023mzm,
    author = "Ballardini, Mario and Ferrari, Angelo Giuseppe and Finelli, Fabio",
    title = "{Phantom scalar-tensor models and cosmological tensions}",
    eprint = "2302.05291",
    archivePrefix = "arXiv",
    primaryClass = "astro-ph.CO",
    doi = "10.1088/1475-7516/2023/04/029",
    journal = "JCAP",
    volume = "04",
    pages = "029",
    year = "2023"
}

@article{Planck:2018vyg,
	archiveprefix = {arXiv},
	author = {Aghanim, N. and others},
	collaboration = {Planck},
	doi = {10.1051/0004-6361/201833910},
	eprint = {1807.06209},
	journal = {Astron. Astrophys.},
	note = {[Erratum: Astron.Astrophys. 652, C4 (2021)]},
	pages = {A6},
	primaryclass = {astro-ph.CO},
	title = {{Planck 2018 results. VI. Cosmological parameters}},
	volume = {641},
	year = {2020}}

@article{Wolf:2024stt,
    author = "Wolf, William J. and Ferreira, Pedro G. and Garc{\'\i}a-Garc{\'\i}a, Carlos",
    title = "{Matching current observational constraints with nonminimally coupled dark energy}",
    eprint = "2409.17019",
    archivePrefix = "arXiv",
    primaryClass = "astro-ph.CO",
    doi = "10.1103/PhysRevD.111.L041303",
    journal = "Phys. Rev. D",
    volume = "111",
    number = "4",
    pages = "L041303",
    year = "2025"
}

@article{Ye:2024ywg,
    author = "Ye, Gen and Martinelli, Matteo and Hu, Bin and Silvestri, Alessandra",
    title = "{Hints of Nonminimally Coupled Gravity in DESI 2024 Baryon Acoustic Oscillation Measurements}",
    eprint = "2407.15832",
    archivePrefix = "arXiv",
    primaryClass = "astro-ph.CO",
    doi = "10.1103/PhysRevLett.134.181002",
    journal = "Phys. Rev. Lett.",
    volume = "134",
    number = "18",
    pages = "181002",
    year = "2025"
}

@article{Ye:2024zpk,
    author = "Ye, Gen",
    title = "{Bridge the Cosmological Tensions with Thawing Gravity}",
    eprint = "2411.11743",
    archivePrefix = "arXiv",
    primaryClass = "astro-ph.CO",
    month = "11",
    year = "2024"
}

@article{Ferrari:2025egk,
    author = "Ferrari, Angelo G. and Ballardini, Mario and Finelli, Fabio and Paoletti, Daniela",
    title = "{Scalar-tensor gravity and DESI 2024 BAO data}",
    eprint = "2501.15298",
    archivePrefix = "arXiv",
    primaryClass = "astro-ph.CO",
    doi = "10.1103/PhysRevD.111.083523",
    journal = "Phys. Rev. D",
    volume = "111",
    number = "8",
    pages = "083523",
    year = "2025"
}

@article{Pan:2025psn,
    author = "Pan, Jiaming and Ye, Gen",
    title = "{Non-minimally coupled gravity constraints from DESI DR2 data}",
    eprint = "2503.19898",
    archivePrefix = "arXiv",
    primaryClass = "astro-ph.CO",
    month = "3",
    year = "2025"
}

@article{Tiwari:2024gzo,
    author = "Tiwari, Yashi and Upadhyay, Ujjwal and Jain, Rajeev Kumar",
    title = "{Exploring cosmological imprints of phantom crossing with dynamical dark energy in Horndeski gravity}",
    eprint = "2412.00931",
    archivePrefix = "arXiv",
    primaryClass = "astro-ph.CO",
    doi = "10.1103/PhysRevD.111.043530",
    journal = "Phys. Rev. D",
    volume = "111",
    number = "4",
    pages = "043530",
    year = "2025"
}

@article{Myrzakulov:2025jpk,
    author = "Myrzakulov, Yerlan and Hussain, Saddam and Shahalam, M.",
    title = "{Phase space and Data analyses of a non-minimally coupled scalar field system with decaying dark energy model}",
    eprint = "2506.11755",
    archivePrefix = "arXiv",
    primaryClass = "gr-qc",
    month = "6",
    year = "2025"
}

@article{Wang:2025znm,
    author = "Wang, Jia-Qi and Cai, Rong-Gen and Guo, Zong-Kuan and Wang, Shao-Jiang",
    title = "{Resolving the Planck-DESI tension by non-minimally coupled quintessence}",
    eprint = "2508.01759",
    archivePrefix = "arXiv",
    primaryClass = "astro-ph.CO",
    month = "8",
    year = "2025"
}

@article{Park:2017zgd,
    author = "Park, Ryan S. and Folkner, William M. and Konopliv, Alexander S. and Williams, James G. and Smith, David E. and Zuber, Maria T.",
    title = "{Precession of Mercury{\textquoteright}s Perihelion from Ranging to the MESSENGER Spacecraft}",
    doi = "10.3847/1538-3881/aa5be2",
    journal = "Astron. J.",
    volume = "153",
    number = "3",
    pages = "121",
    year = "2017"
}

@article{
doi:10.1126/science.1218809,
author = {David E. Smith  and Maria T. Zuber  and Roger J. Phillips  and Sean C. Solomon  and Steven A. Hauck  and Frank G. Lemoine  and Erwan Mazarico  and Gregory A. Neumann  and Stanton J. Peale  and Jean-Luc Margot  and Catherine L. Johnson  and Mark H. Torrence  and Mark E. Perry  and David D. Rowlands  and Sander Goossens  and James W. Head  and Anthony H. Taylor },
title = "{Gravity Field and Internal Structure of Mercury from MESSENGER}",
journal = "Science",
volume = "336",
number = "6078",
pages = "214-217",
year = "2012",
doi = "10.1126/science.1218809"
}

@book{Will:2018bme,
    author = "Will, Clifford M.",
    title = "{Theory and Experiment in Gravitational Physics}",
    isbn = "978-1-108-67982-4, 978-1-107-11744-0",
    publisher = "Cambridge University Press",
    month = "9",
    year = "2018"
}

@article{Huang:2024gvi,
    author = "Huang, Li and Deng, Xue-Mei",
    title = "{On the (un)testability of the general free scalar{\textendash}tensor gravity for the Solar System tests}",
    doi = "10.1140/epjc/s10052-024-12969-2",
    journal = "Eur. Phys. J. C",
    volume = "84",
    number = "6",
    pages = "615",
    year = "2024"
}

@article{Alsing:2011er,
    author = "Alsing, Justin and Berti, Emanuele and Will, Clifford M. and Zaglauer, Helmut",
    title = "{Gravitational radiation from compact binary systems in the massive Brans-Dicke theory of gravity}",
    eprint = "1112.4903",
    archivePrefix = "arXiv",
    primaryClass = "gr-qc",
    doi = "10.1103/PhysRevD.85.064041",
    journal = "Phys. Rev. D",
    volume = "85",
    pages = "064041",
    year = "2012"
}

@article{Dyadina:2021paa,
    author = "Dyadina, P. I. and Labazova, S. P.",
    title = "{On Shapiro time delay in massive scalar-tensor theories}",
    eprint = "2111.13900",
    archivePrefix = "arXiv",
    primaryClass = "gr-qc",
    doi = "10.1088/1475-7516/2022/01/029",
    journal = "JCAP",
    volume = "01",
    number = "01",
    pages = "029",
    year = "2022"
}

@article{Lambert:2009xy,
    author = "Lambert, S. B. and Le Poncin-Lafitte, C.",
    title = "{Determination of the relativistic parameter gamma using very long baseline interferometry}",
    eprint = "0903.1615",
    archivePrefix = "arXiv",
    primaryClass = "gr-qc",
    doi = "10.1051/0004-6361/200911714",
    journal = "Astron. Astrophys.",
    volume = "499",
    pages = "331",
    year = "2009"
}

@article{Titov:2010zn,
    author = "Titov, O. and Lambert, S. B. and Gontier, A. -M.",
    title = "{VLBI measurement of the secular aberration drift}",
    eprint = "1009.3698",
    archivePrefix = "arXiv",
    primaryClass = "astro-ph.CO",
    doi = "10.1051/0004-6361/201015718",
    journal = "Astron. Astrophys.",
    volume = "529",
    pages = "A91",
    year = "2011"
}

@article{Sotiriou:2008rp,
    author = "Sotiriou, Thomas P. and Faraoni, Valerio",
    title = "{f(R) Theories Of Gravity}",
    eprint = "0805.1726",
    archivePrefix = "arXiv",
    primaryClass = "gr-qc",
    doi = "10.1103/RevModPhys.82.451",
    journal = "Rev. Mod. Phys.",
    volume = "82",
    pages = "451--497",
    year = "2010"
}

@article{Toniato:2019rrd,
    author = "Toniato, J{\'u}nior D. and Rodrigues, Davi C. and Wojnar, Aneta",
    title = "{Palatini $f(R)$ gravity in the solar system: post-Newtonian equations of motion and complete PPN parameters}",
    eprint = "1912.12234",
    archivePrefix = "arXiv",
    primaryClass = "gr-qc",
    doi = "10.1103/PhysRevD.101.064050",
    journal = "Phys. Rev. D",
    volume = "101",
    number = "6",
    pages = "064050",
    year = "2020"
}

@article{Alves:2023cuo,
    author = "Alves, Matheus F. S. and Toniato, J{\'u}nior D. and Rodrigues, Davi C.",
    title = "{Detailed first-order post-Newtonian analysis of massive Brans-Dicke theories: Numerical constraints and the meaning of the {\ensuremath{\beta}} parameter}",
    eprint = "2307.11883",
    archivePrefix = "arXiv",
    primaryClass = "gr-qc",
    doi = "10.1103/PhysRevD.109.044045",
    journal = "Phys. Rev. D",
    volume = "109",
    number = "4",
    pages = "044045",
    year = "2024"
}

@article{Fienga:2023ocw,
    author = "Fienga, Agn{\`e}s and Minazzoli, Olivier",
    title = "{Testing theories of gravity with planetary ephemerides}",
    eprint = "2303.01821",
    archivePrefix = "arXiv",
    primaryClass = "gr-qc",
    doi = "10.1007/s41114-023-00047-0",
    journal = "Living Rev. Rel.",
    volume = "27",
    number = "1",
    pages = "1",
    year = "2024"
}

@article{Jarv:2017npl,
    author = {J{\"a}rv, Laur},
    title = "{Effective Gravitational {\textquotedblleft}Constant{\textquotedblright} in Scalar-(Curvature)Tensor and Scalar-Torsion Gravities}",
    doi = "10.3390/universe3020037",
    journal = "Universe",
    volume = "3",
    number = "2",
    pages = "37",
    year = "2017"
}

@article{DESI:2025zgx,
    author = "Abdul Karim, M. and others",
    collaboration = "DESI",
    title = "{DESI DR2 results. II. Measurements of baryon acoustic oscillations and cosmological constraints}",
    eprint = "2503.14738",
    archivePrefix = "arXiv",
    primaryClass = "astro-ph.CO",
    reportNumber = "FERMILAB-PUB-25-0169-PPD",
    doi = "10.1103/tr6y-kpc6",
    journal = "Phys. Rev. D",
    volume = "112",
    number = "8",
    pages = "083515",
    year = "2025"
}

@article{Wetterich:1987fm,
    author = "Wetterich, C.",
    title = "{Cosmology and the Fate of Dilatation Symmetry}",
    eprint = "1711.03844",
    archivePrefix = "arXiv",
    primaryClass = "hep-th",
    reportNumber = "PRINT-87-0756, DESY-87-123",
    doi = "10.1016/0550-3213(88)90193-9",
    journal = "Nucl. Phys. B",
    volume = "302",
    pages = "668--696",
    year = "1988"
}

@article{Ratra:1987rm,
    author = "Ratra, Bharat and Peebles, P. J. E.",
    title = "{Cosmological Consequences of a Rolling Homogeneous Scalar Field}",
    reportNumber = "PUPT-1072",
    doi = "10.1103/PhysRevD.37.3406",
    journal = "Phys. Rev. D",
    volume = "37",
    pages = "3406",
    year = "1988"
}

@article{Huterer:2017buf,
    author = "Huterer, Dragan and Shafer, Daniel L.",
    title = "{Dark energy two decades after: observables, probes, consistency tests}",
    eprint = "1709.01091",
    archivePrefix = "arXiv",
    primaryClass = "astro-ph.CO",
    doi = "10.1088/1361-6633/aa997e",
    journal = "Rept. Prog. Phys.",
    volume = "81",
    number = "1",
    pages = "016901",
    year = "2018"
}

@article{SanchezLopez:2023ixx,
    author = "S{\'a}nchez L{\'o}pez, Samuel and Dimopoulos, Konstantinos and Karam, Alexandros and Tomberg, Eemeli",
    title = "{Observable gravitational waves from hyperkination in Palatini gravity and beyond}",
    eprint = "2305.01399",
    archivePrefix = "arXiv",
    primaryClass = "gr-qc",
    doi = "10.1140/epjc/s10052-023-12332-x",
    journal = "Eur. Phys. J. C",
    volume = "83",
    number = "12",
    pages = "1152",
    year = "2023"
}

@article{Lopez:2025gfu,
    author = "L{\'o}pez, Samuel S{\'a}nchez and Terente D{\'\i}az, Jos{\'e} Jaime",
    title = "{Scalar-Induced Gravitational Waves in Palatini f(R) gravity}",
    eprint = "2505.13420",
    archivePrefix = "arXiv",
    primaryClass = "astro-ph.CO",
    doi = "10.1088/1475-7516/2025/12/029",
    journal = "JCAP",
    volume = "12",
    pages = "029",
    year = "2025"
}

@article{Benisty:2022txp,
    author = "Benisty, David",
    title = "{Testing modified gravity via Yukawa potential in two body problem: Analytical solution and observational constraints}",
    eprint = "2207.08235",
    archivePrefix = "arXiv",
    primaryClass = "gr-qc",
    doi = "10.1103/PhysRevD.106.043001",
    journal = "Phys. Rev. D",
    volume = "106",
    number = "4",
    pages = "043001",
    year = "2022"
}

@article{Benisty:2021cmq,
    author = "Benisty, David and Davis, Anne-Christine",
    title = "{Dark energy interactions near the Galactic Center}",
    eprint = "2108.06286",
    archivePrefix = "arXiv",
    primaryClass = "astro-ph.CO",
    doi = "10.1103/PhysRevD.105.024052",
    journal = "Phys. Rev. D",
    volume = "105",
    number = "2",
    pages = "024052",
    year = "2022"
}

@article{Benisty:2023dkn,
    author = "Benisty, David and Brax, Philippe and Davis, Anne-Christine",
    title = "{Multiscale constraints on scalar-field couplings to matter: The geodetic and frame-dragging effects}",
    eprint = "2305.19977",
    archivePrefix = "arXiv",
    primaryClass = "gr-qc",
    doi = "10.1103/PhysRevD.108.063031",
    journal = "Phys. Rev. D",
    volume = "108",
    number = "6",
    pages = "063031",
    year = "2023"
}

@article{Khoury:2003aq,
    author = "Khoury, Justin and Weltman, Amanda",
    title = "{Chameleon fields: Awaiting surprises for tests of gravity in space}",
    eprint = "astro-ph/0309300",
    archivePrefix = "arXiv",
    doi = "10.1103/PhysRevLett.93.171104",
    journal = "Phys. Rev. Lett.",
    volume = "93",
    pages = "171104",
    year = "2004"
}

@article{Khoury:2003rn,
    author = "Khoury, Justin and Weltman, Amanda",
    title = "{Chameleon cosmology}",
    eprint = "astro-ph/0309411",
    archivePrefix = "arXiv",
    doi = "10.1103/PhysRevD.69.044026",
    journal = "Phys. Rev. D",
    volume = "69",
    pages = "044026",
    year = "2004"
}

@article{Hinterbichler:2010es,
    author = "Hinterbichler, Kurt and Khoury, Justin",
    title = "{Symmetron Fields: Screening Long-Range Forces Through Local Symmetry Restoration}",
    eprint = "1001.4525",
    archivePrefix = "arXiv",
    primaryClass = "hep-th",
    doi = "10.1103/PhysRevLett.104.231301",
    journal = "Phys. Rev. Lett.",
    volume = "104",
    pages = "231301",
    year = "2010"
}

@article{Hinterbichler:2011ca,
    author = "Hinterbichler, Kurt and Khoury, Justin and Levy, Aaron and Matas, Andrew",
    title = "{Symmetron Cosmology}",
    eprint = "1107.2112",
    archivePrefix = "arXiv",
    primaryClass = "astro-ph.CO",
    doi = "10.1103/PhysRevD.84.103521",
    journal = "Phys. Rev. D",
    volume = "84",
    pages = "103521",
    year = "2011"
}

@article{Vainshtein:1972sx,
    author = "Vainshtein, A. I.",
    title = "{To the problem of nonvanishing gravitation mass}",
    doi = "10.1016/0370-2693(72)90147-5",
    journal = "Phys. Lett. B",
    volume = "39",
    pages = "393--394",
    year = "1972"
}

\end{document}